# An X-ray view of 82 LINERs with *Chandra* and *XMM-Newton* data

O. González-Martín[1,*], J. Masegosa[2], I. Márquez[2], M. Guainazzi[3] and E. Jiménez-Bailón[4]

[1] X-ray Astronomy Group, Department of Physics and Astronomy, Leicester University, Leicester LE1 7RH, UK e-mail: gmo4@star.le.ac.uk
[2] Instituto de Astrofísica de Andalucía, CSIC, Granada, SPAIN
[3] European Space Astronomy Centre of ESA, P.O. Box 78, Villanueva de la Canada, E-28691 Madrid, SPAIN
[4] Instituto de Astronomía, Universidad Nacional Autónoma de Mexico, Apartado Postal 70-264, 04510 Mexico DF, MEXICO

**ABSTRACT**

We present the results of an homogeneous X-ray analysis for 82 nearby LINERs selected from the catalogue of Carrillo et al. (1999). All sources have available *Chandra* (68 sources) and/or *XMM-Newton* (55 sources) observations. This is the largest sample of LINERs with X-ray spectral data (60 out of the 82 objects) and significantly improves our previous analysis based on *Chandra* data for 51 LINERs (Gonzalez-Martin et al. 2006). It increases both the sample size and adds *XMM-Newton* data. New models permit the inclusion of double absorbers in the spectral fits. Nuclear X-ray morphology is inferred from the compactness of detected nuclear sources in the hard band (4.5-8.0 keV). Sixty per cent of the sample shows a compact nuclear source and are classified as AGN candidates. The spectral analysis indicates that best fits involve a composite model: absorbed primary continuum and (2) soft spectrum below 2 keV described by an absorbed scatterer and/or a thermal component. The resulting median spectral parameters and their standard deviations are: $<\Gamma>=2.11\pm 0.52$, $<kT>=0.54\pm 0.30$ keV, $<\log(NH1)>=21.32\pm 0.71$ and $<\log(NH2)>=21.93\pm 1.36$. We complement our X-ray results with our analysis of HST optical images and literature data on emission lines, radio compactness and stellar population. Adding all these multiwavelength data, we conclude that evidence do exist supporting the AGN nature of their nuclear engine for 80% of the sample (66 out of 82 objects).

## 1. Introduction

The term *Active Galactic Nucleus* (AGN) generally refers to the galaxies that show energetic phenomena in their nuclei which cannot be unambiguously attributed to starlight. The 2-10 keV X-ray luminosity has been used as a reasonably reliable measure of AGN power allowing one to extract information about the central engine from the spectral fit and to quantify the absorbing material ($N_H < 1.5 \times 10^{24}$cm$^{-2}$ from current missions), i.e. the dusty torus. Nearby examples of both Seyfert 2 (Maiolino et al., 1998; Risaliti et al., 1999; Page et al., 2003; Cappi et al., 2006; Panessa et al., 2006), and Seyfert 1 (Nandra et al., 2007) or PG-QSOs (Piconcelli et al., 2005; Jimenez-Bailon et al., 2005) have been studied. The X-ray regime can also be used to search for AGN in heavily obscured objects. This is important for high redshift surveys (Comastri and Brusa, 2008) and starburst samples (Tzanavaris and Georgantopoulos, 2007).

These studies are especially important for *low luminosity* AGN (LLAGN) with bolometric luminosities $L_B < 10^{44}$erg s$^{-1}$ because more than 40% of nearby galaxies show evidence for low-power AGN (see the review by Ho, 2008). Even at lower luminosities Zhang et al. (2009) have recently studied the X-ray nuclear activity of 187 nearby galaxies, most of them classed as non-active, with *Chandra* data, finding evidences for AGN in 46% of their sample (60 % when considering ellipticals and early-type spirals).

We focused our attention on *low ionisation narrow emission-line regions* (LINERs), originally defined as a subclass of LLAGN by Heckman (1980). They show optical spectra dominated by emission lines of moderate intensities arising from gas in lower ionisation states than classical AGN. Previous studies of LINERs reached different conclusions about the ionisation mechanism responsible for the LINER emission. Possibilities included: 1) shock heating (Dopita and Sutherland, 1995), 2) Wolf-Rayet or OB stars in compact near-nuclear star clusters (Terlevich and Melnick, 1985; Filippenko and Terlevich, 1992) and 3) low luminosity AGN (Ho et al., 1997; Eracleous and Halpern, 2001).

It is therefore important to isolate and study the nature of the source hosted by most of the LINERs. Satyapal et al. (2005) (see also Ho et al., 2001; Satyapal et al., 2004; Dudik et al., 2005) estimated 2-10 keV luminosities for a sample of 41 LINERs using *Chandra* data. They found that AGN are very frequent among LINERs. Eddington ratio considerations led them to conclude that LINERs represent the faint end of the fundamental correlation between mass accretion and star formation rates (e.g. LINERs very inefficient accreting systems). Flohic et al. (2006) studied a sample of 19 LINERs using archival *Chandra* data and found that the median AGN contribution to the 0.5–10 keV luminosity is 60%). They suggest that AGN power is not sufficient to produce the observed optical emission lines and invoke shocks and/or stellar processes. Our previous analysis (Gonzalez-Martin et al., 2006b, hereinafter GM+06) presented results from an homogeneous analysis of 51 LINER galaxies observed with *Chandra*. Morphological classification together with spectral analysis (when possible) led us to conclude that at least 60% of LINERs may host an AGN. Following Ho et al. (1997), we consider a LINER to host an AGN when the nuclear regions showing an unresolved source at high X-ray energies.

---

[*] E-mail: gmo4@star.le.ac.uk (University of Leicester); omaira@iaa.es (IAA)

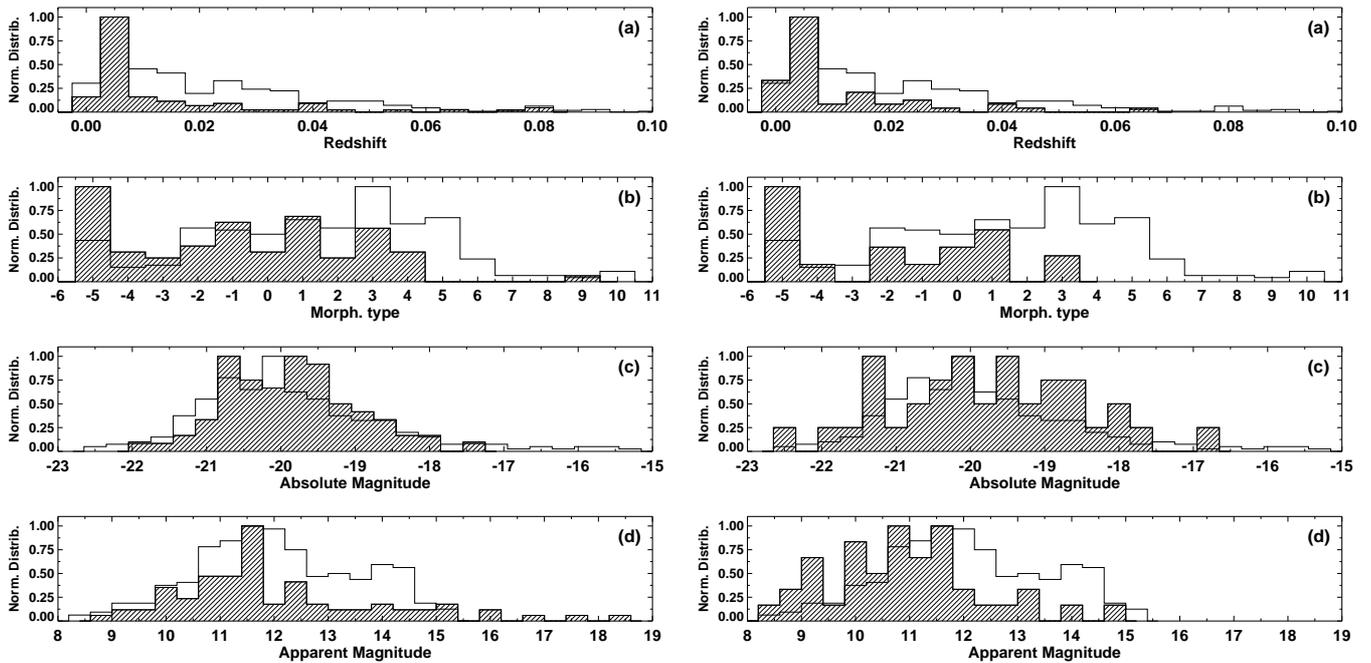

**Fig. 1.** (*Left*): total sample of LINERs in MCL (empty histogram) versus X-ray sample (dashed histogram), normalised to the peak. (*Right*): X-ray sample (empty histogram) versus the sample reported by GM+06, normalised to the peak. (a) Redshift, (b) morphological types (from the RC3 catalog: t<0 are for ellipticals, t=0 for S0, t=1 for Sa, t=3 for Sb, t=5 for Sc, t=7 for Sd, and t>8 for irregulars), (c) absolute magnitudes, and (d) apparent magnitudes distributions.

This paper presents a study of 82 LINERs which is the largest sample so far analyzed at X-ray wavelengths. In contrast to GM+06 we include objects observed with *XMM-Newton* yielding 68 sources with *Chandra* and 55 with *XMM-Newton* data. The high spatial resolution of the *Chandra* optics is optimally suited to disentangle the different components of the often complex X-ray morphology (GM+06) found in our sample. Analysis of X-ray images is improved by using a different smoothing process. On the other hand, the addition of *XMM-Newton* data complements *Chandra* by allowing more detailed spectral analysis in 60 of the sources. In contrast to GM+06, we now use a different baseline model with the main improvement being the use of an additional power-law component and two neutral absorbers. We present a detailed study of the X-ray LINER properties added to multiwavelength information. In summary, compared to GM+06, the current paper presents 33 new LINERs, 19 new *Chandra* data, presents *XMM-Newton* information for the 41 objects with both *Chandra* and *XMM-Newton* data, and proves new X-ray spectral baseline models.

Section 2 introduces the sample and the data. Section 3 explains the data reduction. The data analysis is presented in Section 4. Section 5 includes the results and the discussion while section 6 includes summary and conclusions.

This paper is complemented by a second paper (González-Martín et al., in preparation, hereinafter Paper II) which contains a detailed analysis of LINER obscuring matterial and its *Compton-thickness* (objects with a primary source completely suppressed below 10 keV).

## 2. The sample and the data

Our sample was extracted from the multi-wavelength LINER catalogue compiled by Carrillo et al. (1999) (hereinafter MCL[1]). We updated the sample by including all the galaxies in MCL with available *Chandra* data up to 2007-06-30 and *XMM-Newton* data up to 2007-04-30. We recall that only *Chandra* data were used in GM+06. The search has been done using the HEASARC[2] archive. The sample includes 108 LINERs with *Chandra* data and 107 LINERs with *XMM-Newton* data. Seventy six objects are present in both archives yielding a total of 139 LINERs.

LINER identifications were revised using standard diagnostic diagrams (Baldwin et al., 1981; Veilleux and Osterbrock, 1987). We obtained emission line fluxes for all but 18 objects from the literature (Ho et al., 2001; Veilleux et al., 1995; Moustakas and Kennicutt, 2006; Wu et al., 1998). The 18 objects were excluded to ensure a sample of bona-fide LINERs. In the same vein, another four objects, lacking [OIII] measured fluxes (NGC 3189, NGC 4414, NGC 5350 and NGC 6503) were excluded. NGC 4013 was also rejected because its [OIII] measurement is affected by a 100% error. Fifty one out of the 116 objects were already classified as LINERs in our previous work (GM+06). However, later reanalysis showed that the classification given by GM+06 for NGC 4395 and NGC 5194 was not correct. Thus, they are most likely Seyfert galaxies. This yields 49 LINERs in common with GM+06.

After optical re-identification we ended up with a final sample of 83 sources including 68 observed with *Chandra* and 55 with *XMM-Newton*. *Chandra* data for 19 new LINERs are provided here. Forty LINERs are found in both datasets. Observations for one of these objects showed strong pile-up effects leaving us with a final sample of 82 objects (see Sect. 3).

---

[1] MCL includes most LINER galaxies known through 1999. It provides information on broad band and monochromatic emission from radio to X-ray frequencies for 476 objects classified as LINERs.

[2] $http://heasarc.gsfc.nasa.gov/$

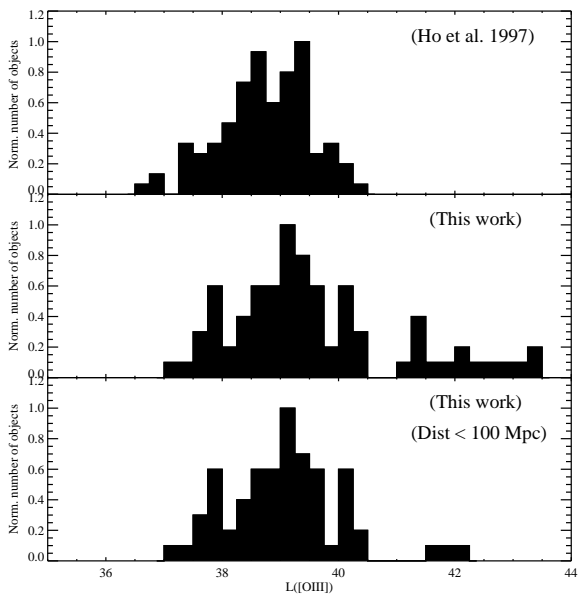

**Fig. 2.** [OIII] luminosity distributions using Ho et al. (1997) sample (*top*), this sample (*centre*) and this sample excluding objects with distances larger than 100 Mpcs (*bottom*).

Table 1 includes the properties of the host galaxies[3]. Distances have been taken form Tonry et al. (2001), Ferrarese et al. (2000), Tully (1998) and otherwise assuming $H_0 = 75$ km s$^{-1}$Mpc, in this priority order.

Fig. 1 shows, from top to bottom, redshift, morphological type, absolute B magnitude and apparent B magnitude distributions. The distributions for the 476 LINERs in the MCL catalogue are shown with empty histograms. To the left (right) panels, the corresponding distribution for our final (GM+06) sample are shown with dashed histograms. All distributions are normalised to their maximum values.

Comparing with our previous work (GM+06) we now have later morphological types. The lack of faint LINERs present in our previous analysis (GM+06) now disappears. The median total absolute magnitude is now $M_B = -19.9 \pm 0.9$ which is consistent with that in the MCL catalogue ($M_B = -20.0 \pm 1.5$). The median redshift is $z = 0.059 \pm 0.018$, higher to that for MCL ($z = 0.015 \pm 0.040$) but still within one sigma deviation. Table 2 provides the observational details of the current X-ray LINER sample.

MCL catalogue contains all the known LINERs until 1999. Therefore, it is not biased to any kind of LINERs. However, the X-ray selection relies on the availability of X-ray archival data, and consequently might introduce a bias, whose nature is not obvious to quantify. Nevertheless, looking at the different goals of the individual proposals, a wide range of scientific aims is covered: AGN, Superwinds, radio galaxies, ULIRGs, LINERs, ULX, LMXBs, clusters, merger systems, starburst nuclei, star forming galaxies or SNe. Moreover, there are a few objects that were not the target, but were observed by chance (e.g. NGC 3226 is in the field of view of NGC 3227). It is worth noticing that the study of LINERs itself is the main goal for 8 of the sources and the study of the host galaxy (ULXs, LMXBs and/or diffuse emission) is the principal topic in 38 objects in our sample.

---

[3] Main properties extracted from the RC3 catalogue.

To evaluate possible biases towards bright AGN LINERs we have used [OIII] luminosity as it can be considered as an isotropic indicator of AGN power. Fig. 2 plots the distribution of L([OIII]) for our LINERs (middle panel) and for those in the volume limited sample of nearby AGN reported by Ho et al. (1997) (top panel). Our sample is similarly distributed as that by Ho et al. (1997) but it shows an extension towards the most luminous objects. However, when selecting the objects in our sample within the same volume as those in Ho's (bottom panel), the discrepancy disappears (K-S probability of 90% for both samples coming from the same parent population). Therefore, our sample seems to be representative of the level of nuclear activity of LINERs in the local universe, with X-ray observations not introducing any significant bias in favour of AGN-type objects. Another interesting parameter to consider is the ratio L([OIII])/L(X-rays), which we defer for a full discussion in Paper II. Here we only anticipate that the values for this ratio may indicate a high percentage of LINERs hosting Compton-thick AGN (GM+09).

## 3. Data reduction

ACIS instrument level 2 event data have been extracted from *Chandra* archive. The data products have been analysed in a uniform, self-consistent manner using *CXC Chandra Interactive Analysis of Observations* (CIAO) software version 3.4. *Chandra* data have been reduced following the prescriptions in GM+06. In the following subsections we include the details on *Chandra* reduction only if different to that reported in GM+06.

*XMM-Newton* data have been reduced with SAS v7.0.0, using the most updated calibration files available. In this paper, only data from the EPIC pn camera (Struder et al., 2001) will be discussed. The spectral analysis has been performed with *XSPEC* (version 12.3.1).

Images could be dominated by the background if time intervals affected by "flares" are not excluded. For *Chandra* data see GM+06. For *XMM-Newton* data these time intervals of quiescent particle background have been determined through an algorithm that maximises the signal-to-noise ratio of the net source spectrum by applying different constant count rate threshold on the single-events, $E > 10$ keV field-of-view background light curve. At the same time, the algorithm calculates the optimal source extraction region size yielding the maximum number of net source counts for a given background threshold.

Pileup affects both flux measurements and spectral characterisation of bright sources (Ballet, 2001). The pileup estimation has been performed using PIMMS software. To evaluate the importance of pileup for each source we have used the 0.5-2 keV and 2-10 keV flux, the best-fit model and the redshift. The resulting pileup fraction in 0.5-2 keV and 2-10 keV are reported in Cols. (12) and (15) in Table 2. We notice that these effects are unimportant in our sample, being in most cases below 10%. For *Chandra* data six galaxies, NGC 3998, NGC 4486, NGC 4494, NGC 4579, NGC 4594 and NGC 5813, show pileup fractions between 10% and 20%.

An inspection of the final *XMM-Newton* spectral data of MCG -5-23-16, NGC 4486 and NGC 4696 showed a wavy structure that cannot be fitted properly. It might be attributed to the high pileup fraction. We have hence decided not to use these *XMM-Newton* data that results in a final sample of 82 galaxies. We reject MCG -5-23-16 since it is the only case with *XMM-*

*Newton* data strongly affected by pileup and with no *Chandra* data.

### 3.1. Image data reduction

In order to gain insight into the emission mechanisms for the LINER sample, we have studied the X-ray morphology of the sources in six energy bands: 0.6–0.9, 0.9–1.2, 1.2–1.6, 1.6–2.0, 2.0–4.5, and 4.5–8.0 keV (see GM+06). In the last energy band (4.5–8.0 keV), the range from 6.0 to 7.0 keV was excluded to avoid possible contamination due to the FeK emission line[4] (the corresponding band will be called (4.5–8.0)$^*$ keV hereafter). All the bands have been shifted according to the redshift although the effect is low because of the low redshift of the sample. An example of the images in the four bands 0.6–0.9, 1.6–2.0, 4.5–8.0$^*$, and 6–7 keV are given in Fig. 4. See Appendix C of the catalogue of images that provides this information for the whole sample. *Chandra* data have been used for image analysis when available; *XMM-Newton* data have been used otherwise. Fourteen *XMM-Newton* images have been included in the image catalogue, namely NGC 0410, NGC 2639, NGC 2655, NGC 2685, NGC 3185, NGC 3226, NGC 3623, NGC 3627, IRAS 12112+0305, NGC 5005, NGC 5363, IC 4395, NGC 7285 and NGC 7743. They have been flagged in Tab. 7 with asterisks and their X-ray morphological classification is obviously not considered as robust as that from *Chandra* data (see Sect. 5).

In Fig. 4 in Appendix C (Bottom-right) we also provide the processed optical images from archival *HST* data available for 67 out of the 82 galaxies; the images have been processed following the sharp dividing method (Marquez et al., 2003, and references therein) in order to better show the internal structure of the galaxy. The observational details of the *HST* data are included in Table 2 (Cols. 16, 17 and 18). The same figure provides the 2MASS image in the Ks band (Bottom-center).

We have employed smoothing techniques which enhance weak structures, to have a conservative estimate of the morphological compactness of each X-ray source. We applied the adaptive smoothing CIAO tool CSMOOTH to the *Chandra* data, based on the algorithm developed by Ebeling et al. (2006). CSMOOTH is an adaptive smoothing tool for images containing multi-scale complex structures, and it preserves the spatial signatures and the associated counts as well as significance estimates. We have used CSMOOTH task with a minimum and maximum significance signal-to-noise ratio between 3 and 4, respectively, smaller than the value used by GM+06. It allows the enhancement of small structures while avoiding the detection of extended features. Given that our main goal is the study of the central engine, it is better suited for our purposes. Only in a few cases differences on the morphological classification are found due to this improvement (see Section 5.1.3 and Appendix B, for a discussion of the particular cases).

*XMM-Newton* smoothed images have been generated using the ASMOOTH SAS task applied to the pn images. We have applied the adaptive convolution technique, designed for Poissonian images, with S/N ≥ 5.

### 3.2. Nuclear identification and extraction region

The inner parts of the galaxies hosting a LINER show a complex morphology, with several sources surrounded by diffuse emission (GM+06, see the catalogue of images presented in Appendix C). This complex structure makes the nuclear identification a decisive issue.

The extraction region of the nuclear source for *Chandra* data is in most cases around 2″ and always smaller than 8″ (see Table 2). This corresponds to less than 100 pc in 5 cases (NGC 2787, NGC 2841, NGC 4594, NGC 4736 and NGC 5055) and to a median value of 300 pc for the whole sample. The use of such small radii rules out a significant contamination of extra-nuclear sources in the extraction regions in our sample (see also the discussion in Sect. 5.2). All but two sources are within the extraction region, namely NGC 4696 and MRK 0848. NGC 4696 was already reported as having extended morphology without a nuclear component in GM+06. None of the point like sources in MRK 0848 corresponds to the 2MASS identified nuclear position.

For *XMM-Newton* data, the nuclear positions have been retrieved from NED and circular regions with 25″ radii (500 pixels) have been automatically used as the extraction regions. This 25″ radius is between 80% (85%) of the PSF at 1.5 keV (9.0 keV) for an on-axis source with EPIC pn instrument. These extraction regions range between 630 pc (NGC 4736) and 40 kpc (IRAS 14348-1447) at the distances of our sample. The extraction radius is hence large enough to include the nucleus together with the circumnuclear central region of the galaxy, or even the whole galaxy (see Appendix F). In four objects (IRAS 14348-1447, IRAS 17208-0014, MRK 0848 and NPM1G -12.0625) the size provided by hyperleda[5] is smaller than 25″.

### 3.3. Spectral data reduction

Only objects with more than 200 total of counts in the 0.5-10 keV energy range have been considered for the spectral fitting. It allows to have enough bins to make a realible fit, after spectral binning of at least 20 counts per bin (required to the use of $\chi^2$ statistics). The number of counts have been computed by using DMEXTRACT and DMLIST tasks (CIAO package) for *Chandra* data and EVSELECT (SAS package) for *XMM-Newton* data.

For *Chandra* data the source and background regions have been selected following GM+06 (no contaminating sources included in either regions). For *XMM-Newton* data the background was extracted from a circular region (between 30″ and 75″ of radius) in the same chip as for the source region and excluding point sources. The regions were extracted by using EVSELECT task and pn redistribution matrix and effective areas were calculated with RMFGEN and ARFGEN tasks, respectively. For *Chandra* data the background regions have been selected within the corresponding galaxy, so that the host contribution is subtracted from the nuclear spectrum. For *XMM-Newton* data the background region have been located as close as possible to the extraction region. In the four objects with sizes smaller than 25″ (IRAS 14348-1447, IRAS 17208-0014, MRK 0848 and NPM1G -12.0625) the background region is taken out of the host galaxies. However, all these objects have also *Chandra* data what allows us to get an insight of the posible galaxy contribution. Sixty out of the 82 resulting nuclear spectra have at least 200 counts (see above) and are therefore used for the spectral fitting.

---

[4] The most common emission features in the 2-10 keV band of AGN spectra are those of iron between 6.4–6.97 keV.

[5] http://leda.univ-lyon1.fr/

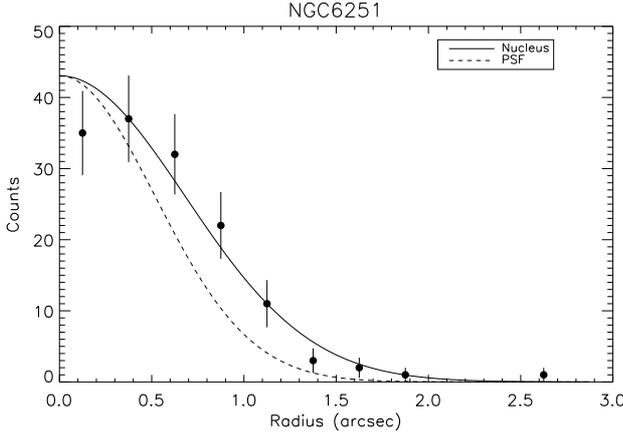

**Fig. 3.** Radial profile of NGC 6251 (black circles). Gaussian fit of this radial profile is shown as a continuous line and the Guassian fit of the *Chandra* PSF at the same position is shown as a dashed line.

## 4. Data analysis

### 4.1. Image analysis

Since we focus our attention on the nuclear sources, no attempt has been made to fully characterise the flux and spectral properties of extra-nuclear sources. As a first insight into the nature of LINERs, we have taken the presence of an unresolved compact nuclear source in the hard band (4.5–8.0* keV) as evidence of an AGN.

We have searched for all sources in the (4.5-8.0)* keV energy band within 10″ of the NED position in *Chandra* data using WAVEDETECT algorithm (Freeman et al., 2002, see also CIAO software). Therefore, we have selected the closest source to the NED position. No sources within the 15″ of the NED position were found in 21 out of the 68 objects with *Chandra* data. Furthermore, 4 objects (NGC 3379, NGC 3628, NGC 5866 and NGC 7331) shows an identification too far (8, 12, 13 and 12″) to be the nuclear source. If the nucleus is detected in the 4.5-8 keV energy range, it is always coincident with the nuclear region in the 0.5-10 keV energy range selected for the spectral analysis.

In order to determine quantitatively whether a nuclear source is resolved, its radial profile has to be studied. For this purpose, a minimum of 100 counts in the (4.5-8.0)* keV energy band is needed to perform the PSF analysis. Only nine objects (NGC 315, 3C 218, NGC 3998, NGC 4261, NGC 4594, UGC 08696, NGC 6251, NGC 6240 and IC 1459) accomplish this condition. In these cases we have extracted the Point Spread Function (PSF) from the PSF *Chandra* library at the same position (MKPSF task within CIAO software). Fig. 3 shows the radial profile of NGC 6251 (continuous line) and the PSF of *Chandra* (dashed line). All but NGC 6240 are consistent with the PSF of the instrument (< FWHM(Nucleus) – FWHM(PSF) >= 0.2″). NGC 6240 is a well studied binary AGN, explaining such broadened profile (FWHM(Nucleus) – FWHM(PSF) = 1.7″).

The sample has been grouped into 2 main categories (same as GM+06[6]):

– **AGN candidates:** include all galaxies with a clearly identified unresolved nuclear source in the (4.5-8.0)* keV energy band. Classification was based on a visual inspection of each image carried out independently by three co-authors of the paper. In Fig. 4(i) we show NGC 4594, as an exampleof AGN candidate, where a clear point-like source exists in the hardest band (*centre-right*).

[6] Note that the Non-AGN class corresponds to SB class in GM+06.

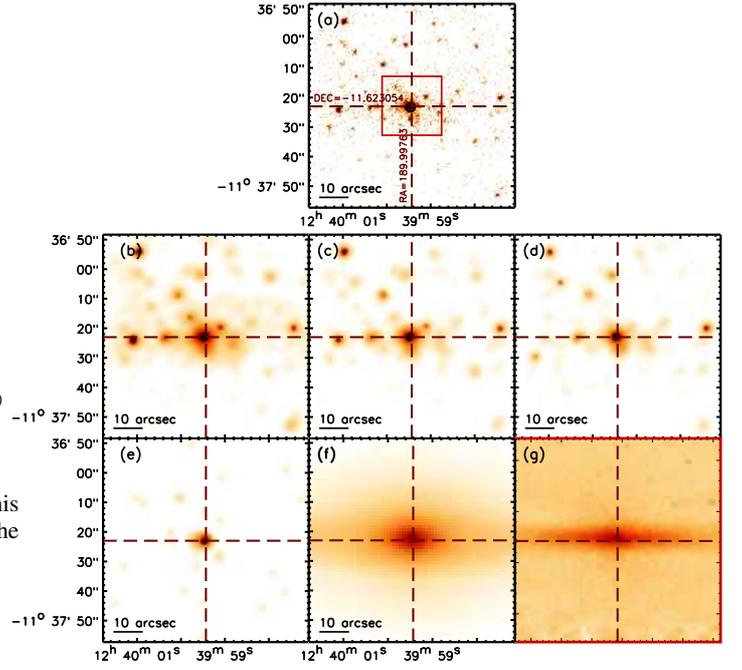

(i) NGC 4594

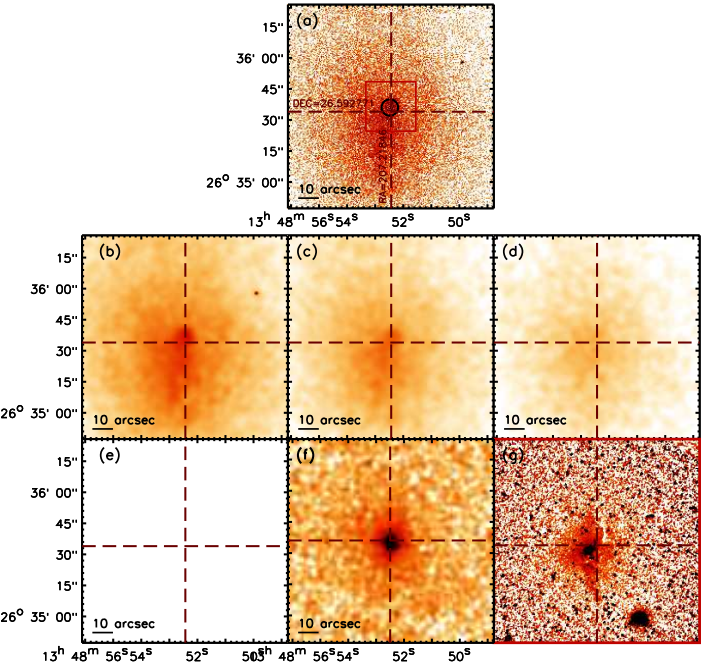

(ii) CGCG 162-010

**Fig. 4.** Images of (i) the AGN candidate NGC 4594 and (ii) of the Non-AGN candidate CGCG 162-010. The top image corresponds to the 0.6-8.0 keV band without smoothing. The extraction region is plotted with a black circle. The following 4 images correspond to the X-ray bands 0.6–0.9 (*Centre-left*), 1.6–2.0 (*Centre-Centre*), 4.5–8.0* (*Centre-Right*) and 6.0–7.0 keV (*Bottom-Left*). The 2MASS image in Ks band is plotted in the *Centre-Bottom* box. The sharp-divided HST/F814W image (*Bottom-Right*) shows the red box region shown in the top image. Note that this figure plots the same example than that reported in GM+06. However the smoothing treat is different (see Section 3.1).

– **Non-AGN candidates:** include all objects without a clearly identifiable nuclear source in the hard band. However we note that the non existence of a point source is not evidence for alack of AGN activity. In Fig. 4(ii) we show the images of CGCG 162-010 as an example of these systems. Note that there does not appear to be a nuclear source in the hardest energy band (*centre-right*).

The PSF profile classification of the nine objects in which it is possible agree with the on a eye based classification.

### 4.2. Spectral fit

The spectra in the 0.5-8.0 keV energy range have been fitted using XSPEC v12.3.1. To be able to use the $\chi^2$ statistics, the spectra were binned in order to obtain a minimum of 20 counts per spectral bin before background subtraction. The GRPPHA task included in FTOOLS software has been used for this purpose.

We have selected five models in order to parametrise five scenarios. We want to stress that each model could have more than a physical interpretation. For instance, a single power-law model could be interpreted as the AGN continuum emission, the emission of X-ray binaries, or the scattering component of the AGN intrinsic continuum. Therefore, the spectral fit is another evidence of the AGN nature but it is not conclusive by itself (see the discussion in Sect. 5.2). Here, we report the five models and the the scenario for which they have been chosen:

1. *Power-law model*: (for simplicity hereinafter **PL**). This is the simplest scenario of an AGN. The column density is added as a free parameter, to take into account the absorption by matter between our galaxy and the target nucleus.
2. *MEKAL model*: (for simplicity hereinafter **ME**). In this case the thermal emission (from unresolved binaries or supernovae remnants) is the responsible for the bulk of the X-ray energy distribution.
3. *MEKAL plus power-law model*: (for simplicity hereinafter **MEPL**). The AGN dominates the hard X-rays although it cannot explain the soft X-rays (< 2 keV) that require an additional thermal contribution.
4. *Power-law plus power-law model*: (for simplicity hereinafter **2PL**). This is the general scenario for which the bulk of the hard X-rays is due to a primary continuum described by a power-law and the soft X-ray spectrum is due to a scattering component also described by a power-law with the same spectral index.
5. *MEKAL plus power-law and power-law model*: (for simplicity hereinafter **ME2PL**). As model (iv), but including the plausible contribution of thermal emission at soft X-rays. This is the *Compton-thin* (obscured with NH < $1 \times 10^{24}$ cm$^{-2}$ and above the Galactic value) Seyfert 2 *baseline* model used by Guainazzi et al. (2005).

For models (iii), (iv) and (v), two absorbing column densities are used, which will be called hereinafter NH1 and NH2. In the most complex model, (v), NH2 is assumed to cover the hardest power-law component and NH1 covers MEKAL plus power-law components. Moreover, Galactic absorption has been fixed to the predicted value (Col. (3) in Table 2) using NH tool within FTOOLS (Dickey and Lockman, 1990; Kalberla et al., 2005).

We have searched for the presence of the neutral iron fluorescence emission line, adding a narrow Gaussian with centroid energy fixed at the observed energy corresponding to a rest frame at 6.4 keV. Two Gaussians were also included to model recombination lines from FeXXV at 6.7 keV and FeXXVI at 6.95 keV.

### 4.3. Best-fit selection criteria

We have chosen the best fit model as the simplest model that gives a good fit with acceptable parameters. To estimate whether the inclusion of a more complex model improves significantly the fit, the F-statistics test (F-test task within XSPEC software) has been applied. A standard threshold for selecting the more complex model is a significance lower than 0.05 (95% confidence). Tables 3 and 4 show the F-test results for *Chandra* and *XMM-Newton* data, respectively.

We have considered it to be a *reliable model* when $\chi^2$ reduced ($\chi^2_\nu$) is in the range from 0.6 to 1.5, its null hypothesis probability is higher than 0.01 and the resulting parameters are within an acceptable range of values, that we consider kT=0-2 keV or $\Gamma$=0-3. An upper limit of kT=2 keV has been considered to take into account the characteristic temperatures observed in central cluster galaxies (Kaastra et al., 2008, and references therein). An upper limit of $\Gamma$=3 has been assumed since this is the upper value obtained in Starburst (Grimes et al., 2005) and LINERs (GM+06). Those values corresponding to unphysical parameters and/or with a bad $\chi^2_\nu$ are marked with 'U' in Tables 3 and 4.

Within the *reliable models*, we have determined the *best-fit model* as the simplest model for which the quality of the fit is not improved by more complex models at the 95% confidence. We have chosen the model with a $\chi^2_\nu$ closest to the unity only when two models with the same number of components agree with the *best-fit model* definition above.

### 4.4. Luminosities

Soft (0.5-2.0 keV) and hard (2-10 keV) luminosities have been computed using the best-fit in the subsample of 60 objects available with spectral fitting (16 objects with *Chandra* data, 16 objects with *XMM-Newton* data and 28 objects with *Chandra* and *XMM-Newton* data). In order to get a luminosity estimation for the remaining 22 galaxies we have used the procedure developed in GM+06, which was based on that proposed by Ho et al. (2001). We obtained a count rate to flux conversion factor for 2.0–10.0 keV and 0.5–2.0 keV energy ranges respectively, assuming a power-law model with a spectral index of 1.8 and the Galactic interstellar absorption (NH task within FTOOLS) when the spectral fit is not available. Tables 5 and 6 show the resulting X-ray fluxes and luminosities, for *Chandra* and *XMM-Newton* data respectively

To validate the approximation, we have computed the soft and hard X-ray luminosity assuming a spectral index of 1.8 and the Galactic interstellar absorption in the subsample of objects with spectral fit. We assume that the faint objects (i.e. those without enough counts to make spectral analysis) show the same spectral shape than bright objects (i.e. those with spectral fit). Color-color diagrams as reported in GM+06 (Fig. 7) show that faint objects are located in the same locus than bright objects (ruling out the thermal model), reinforcing this assumption.

The median values and standard deviation of the soft luminosities are log(L(0.5-2 keV))=39.70±1.00 (log(L(0.5-2 keV))=40.34±1.26) using the best-fit model and log(L(0.5-2 keV))=40.86±1.18 (log(L(0.5-2 keV))=41.06±1.41) using the fixed power-law assumption for *Chandra* (*XMM-Newton*) data.

The soft luminosity is consistent with the assumption of fixed power-law although it tends to be underestimated (median standard deviation from the expected value of 0.65). Seven objects have large differences only for the soft X-ray luminosity calculated assuming a fixed power-law model (NGC 0315, NGC 3898, NGC 4111, NGC 4261, NGC 4696, NGC 5813, and NGC 7130). For NGC 0315, NGC 3898, NGC 4111 we can attribute such differences to the differing column densities, but this is not the case for the remaining objects.

The 2-10 keV luminosity using the best-fit model is log(L(2-10 keV))=39.87±1.00 (log(L(2-10 keV))=40.29±1.47) while the estimated luminosity assuming a single power-law is log(L(2-10 keV))=40.51±1.33 (log(L(2-10 keV))=40.54±1.26) for *Chandra* (*XMM-Newton*) data. Therefore, both hard X-ray luminosities agree with a median deviation of 0.10. NGC 0833, NGC 0835, NGC 1052, UGC 05101, UGC 08696 and NGC 6240 show high discrepancies between estimated and computed luminosities, mainly because the spectral model suggests that high column densities ($NH=1-5\times10^{23}$ cm$^{-2}$) and/or strong iron emission lines are required, whereas we are assuming a Galactic column density value in our estimation.

It therefore appears that while soft X-ray luminosities calculated with a fixed power-law model have to be taken with some reserves, the estimation of the hard X-ray luminosity calculated in the same way is reasonably good, except for very obscured galaxies.

## 5. Results and Discussion

We present the largest sample of LINERs ever analysed with X-ray data, including the spectral analysis of 60 of them.

### 5.1. X-ray characterisation of LINERs

#### 5.1.1. Imaging

Provided the complex morphology of LINERs with surrounding point-like sources (e.g. NGC 4594) and diffuse emission (e.g. CGCG 162-010) *Chandra* data are better suited for imaging purposes. For completeness we have added the results of the whole sample including *XMM-Newton* data, but as explained above, the classification from *XMM-Newton* imaging is indicative and the corresponding column in Table 7 is marked with an asterisk.

- **AGN candidates:** Sixty three per cent (43/68) of our *Chandra* sample galaxies have been classified as AGN-like nuclei, almost sixty per cent (48/82) including *XMM-Newton* data. This fraction increases up to an 80% when only objects with more than 200 counts are considered (35 out of the 44 objects). Our results might represent a lower limit to the true fraction of AGN candidates in our sample.
- **Non-AGN candidates:** Thirty seven per cent (25/68) of the *Chandra* sample of LINERs falls into this category, 40% (34/82) including *XMM-Newton* data.

Among the 14 objects with X-ray imaging from *XMM-Newton*, six are classified as AGN-like objects (NGC 2655, NGC 2685, NGC 3226, NGC 5005, NGC 5363 and NGC 7285). We note that HETG *Chandra* data have been reported by George et al. (2001) for NGC 3226 and a nuclear point source has been detected using the zero order data; for NGC 3623, NGC 3627 and NGC 5005. Snapshot ACIS-S data were shown in Dudik et al. (2005) who, following Ho et al. (2001), classified the first two galaxies as class IV (no nuclear source) and the last one as class III (a hard nuclear point source embedded in soft diffuse emission). All agree with our classifications.

#### 5.1.2. Best-fit

Spectral analysis was possible for 44 (44) out of 68 (55) LINERs with *Chandra* (*XMM-Newton*) data. Fig. 5 shows an example of the nuclear *XMM-Newton* spectrum of NGC 2655. The five panels show the results for each of the five models used in the spectral fitting (see the caption).

For the full-sample analysis we grouped our sources into those best-fit with "Simple Models" (one component, i.e. ME or PL) and those fit with "Composite Models" (more than a single component required, as MEPL, 2PL and ME2PL models). We get:

1. *Simple Models: Chandra* data: (see Table 3) a PL model is reliable in 25 cases however 2PL and/or MEPL models result in an improvement in all but 7 (NGC 2787, NGC 3414, NGC 3945, NGC 4594, NGC 5055, NGC 5746 and IRAS 17208-0014). ME model is reliable in 14 galaxies however PL model, or the inclusion of a composite model, result in an improvement in all but 3 cases (NGC 3507, CGCG 162-010 and NGC 6482).
   *XMM-Newton* data: (see Table 4), PL model is reliable in 13 cases but 2PL and/or MEPL models result in an improvement in all but 4 objects (NGC 3628, NGC 3998, NGC 4494 and MRK 0848). ME model is reliable in 7 galaxies, but PL model or the inclusion of a composite MEPL model result in an improvement in all but 5 cases (NGC 2639, UCG 04881, IRAS 14348-1447, IRAS 17208-0014 and NGC 6482).

2. *Composite Models:* Excluding cases where ME or PL models are required the complex model is needed in 34 out of the 44 objects with *Chandra* data and 36 out of the 44 objects *XMM-Newton* data.
   *Chandra* data: 2PL model is better than MEPL model in 12 cases (NGC 0833, NGC 1052, UGC 05101, NGC 3998, NGC 4374, NGC 4486, NGC 4736, MRK 266NE, UGC 08696, NGC 6251, NGC 6240 and IC 1459). MEPL model is better than 2PL model in 19 objects and in three cases (NGC 0835, NGC 4111 and NGC 4261) no good fit has been found with neither MEPL nor 2PL models.
   *XMM-Newton* data: 2PL model is better than MEPL model in 3 cases (NGC 1052, NGC 3226 and NGC 4594). MEPL model is better than 2PL model in 29 objects. Within these 29 objects, none of the models provides a statistically acceptable fit for NGC 7743, being a MEPL the one providing the best fit ($\chi^2_\nu = 1.76$). Moreover, in UGC 05101, NGC 4261, UGC 08696 and NGC 7743 neither MEPL nor 2PL models are good representations of the data-sets.
   For these 12 (3) objects with *Chandra* (*XMM-Newton*) data reliably fitted with 2PL model, the F-test demonstrates that the use of ME2PL model is an improvement in all the objects (all but NGC 3226) with *Chandra* (*XMM-Newton*) data. Within the objects reliably fitted with MEPL model, the F-test demonstrates that the use of ME2PL model improves the results in two objects (NGC 2681 and NGC 3898) from the *Chandra* dataset and in 12 of the cases with *XMM-Newton* data. Three objects (NGC 0835, NGC 4111 and NGC 4261) observed with *Chandra* and four objects with *XMM-Newton* (UGC 05101, NGC 4261, UGC 08696 and NGC 7743) statistically require the ME2PL model. For the interested readers, full detail on results of the fit are given in Gonzalez-Martin et al. (2008).

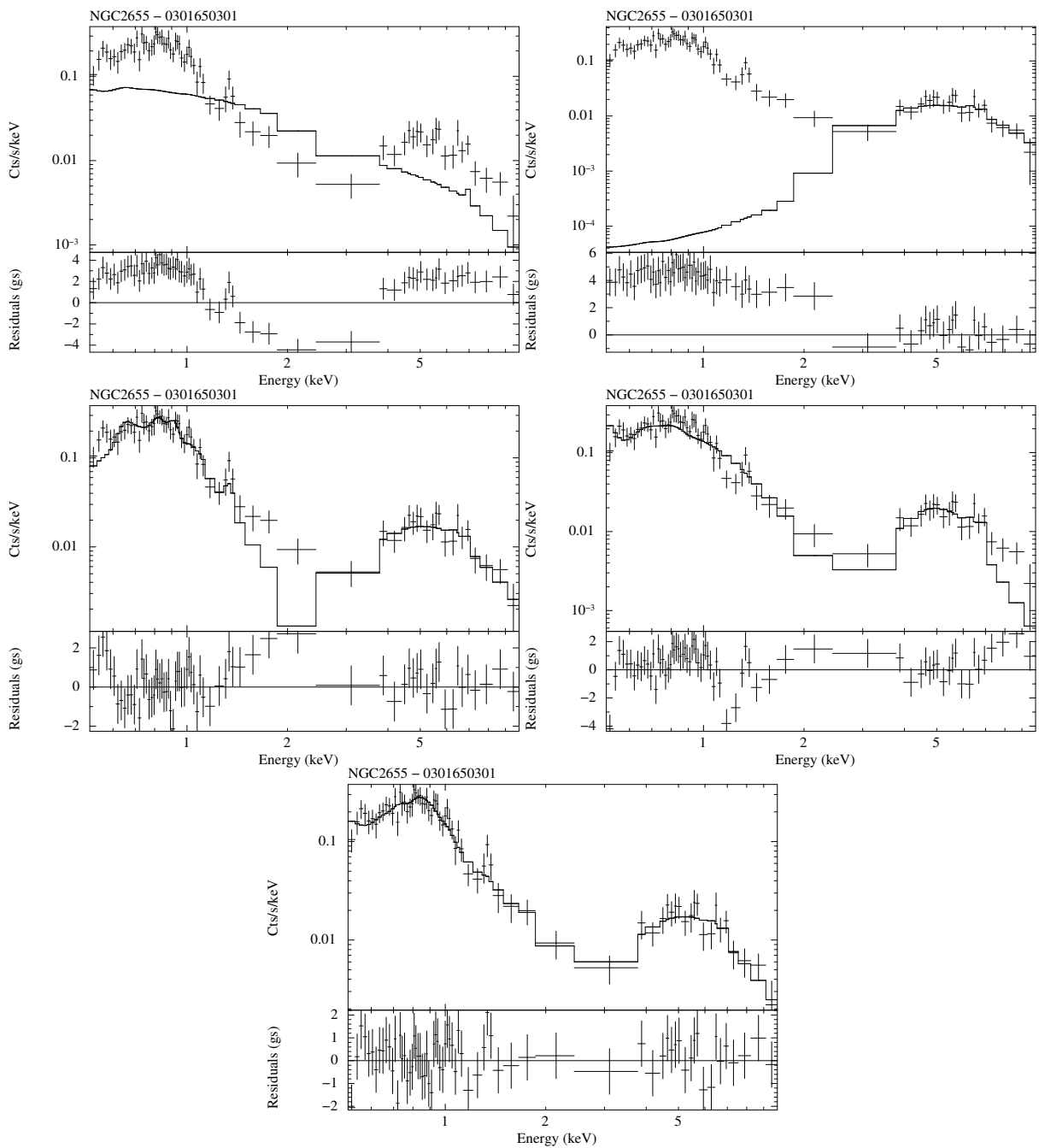

**Fig. 5.** Spectral fits (*top panels*) and residuals (*bottom panels*) for the nuclear spectrum of NGC 2655 (*XMM-Newton* data). (*Top-left*): Thermal model (ME), (*Top-right*): Power-law model (PL), (*Centre-left*): Power-law plus thermal model (MEPL), (*Centre-right*): Two power-law model (2PL), (*Bottom*): Two power-law plus thermal model (ME2PL). The best-fit for this object is a ME2PL model (see Table 7). Figures of spectral fits of the LINER sample are in the electronic edition in Appendix D and E.

A comparative analysis of the spectral fitting of the 40 sources observed with both by *Chandra* and *XMM-Newton* is presented in Appendix A. This analysis consists on: (a) A comparison of the the X-ray properties of the 40 objects in observed by both satellites, (b) the same comparison using the same extraction region and, (c) a statistical comparison of the two samples. The main results of this study are: (1) The spectral index is a robust parameter, independent of the extraction radius of the source, (2) *XMM-Newton* luminosities are about a factor of 5-10 brighter in the 0.5-2 keV range and 2.5 times in the 2-10 keV range larger than *Chandra* luminosities, (3) NH2 column density is independent on the selected aperture and appears to be strongly linked to the best-fit model and, (4) the discrepancies in NH1 are due to aperture effects.

Having in mind all the considerations when comparing *Chandra* and *XMM-Newton* data, we have constructed the final sample for spectral analysis by using *XMM-Newton* data only when *Chandra* data are not available. The results are shown in Tables 7 and 8.

Regarding the best model for LINERs, we found that MEPL or ME2PL models are the best representation for 73% of the sample (24 galaxies have been fitted with MEPL model and 20 with ME2PL model) while 2PL model is a poor representation of the data (only 1 case). Moreover, when the number counts are taken into consideration, all the objects with a simple model best-fit are those with low number counts in their spectra while composite models cover a wide range of number counts. This suggests that the requirement of a composite model seems to be the most choice for our LINER sample.

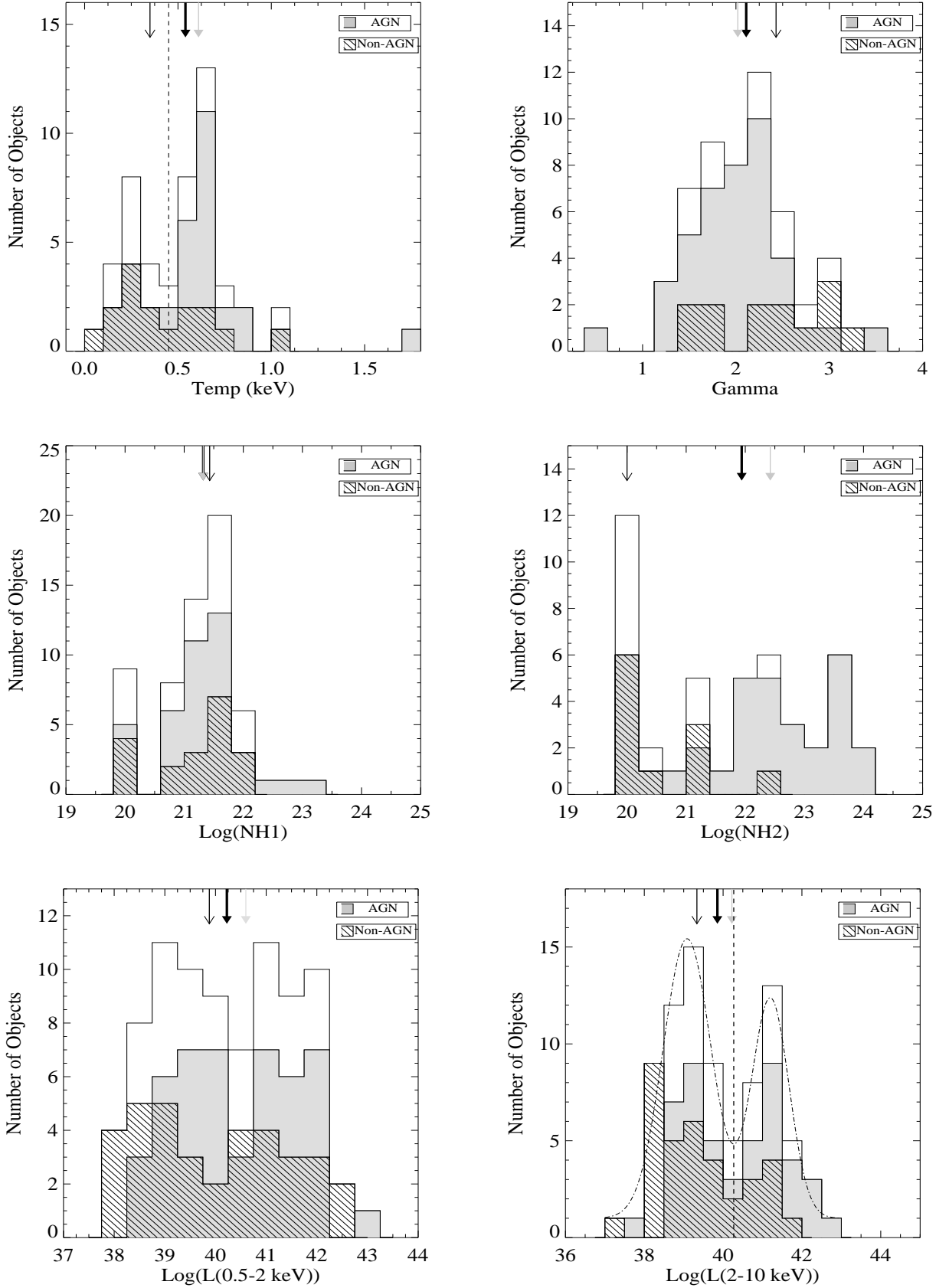

**Fig. 6.** Distributions of temperature (*Top-Left*), spectral index (*Bottom-Left*), NH1 (*Top-Centre*), NH2 (*Bottom-Centre*), soft (0.5-2 keV, *Top-Right*) and hard (2-10 keV, *Bottom-Right*) luminosities. The median values are marked with arrows (bold-face arrows for the whole LINER sample, grey arrows the AGN-like sample and thin arrows the Non-AGN sample, see Table 10). Empty distributions show the whole LINER sample results, grey filled distributions show the subsample of AGN candidates and dashed-filled distributions show the subsample of Non-AGN candidates. Dashed lines show the minima between the two peaks found in the distribution of temperature (kT=0.45 keV) and hard luminosity ($L(2-10\ \text{keV}) = 1.9 \times 10^{40}\ \text{erg s}^{-1}$). The dot-dashed line in the L(2-10 keV) histogram shows the best fit of the distribution to a two Gaussians model. IRAS 14348-1447 has been excluded from the temperature histogram because it shows a temperature above 2 keV.

We have also computed the relative contribution coming from each of the spectral fit components to the 0.5-2 keV and 2-10 keV fluxes (Table 8, Cols. 9, 10 and 11). The thermal component dominates the soft emission (> %50) in 31 LINERs and its contribution is higher than 20% in 23 objects above 2 keV, although the main contribution (> %50) above 2 keV comes from the power-law components.

Tables 9 and 10 show the statistic of best-fits and the median parameters of the sample. Fig. 6 shows the distributions of temperature (*Top-Left*), spectral index (*Bottom-Left*), NH1 (*Top-Centre*), NH2 (*Bottom-Centre*), soft (0.5-2 keV, *Top-Right*) and hard (2-10 keV, *Bottom-Right*) luminosities. Table 10 also includes the K-S test probability to investigate whether AGN and non-AGN LINERs are consistent with deriving from the same parent population.

A deeper understanding on the nature of LINER nuclei can be obtained based on their spectral characteristics. Within our expectations for the AGN-like population, ME model is not found as a good fit for any of the galaxies.

Within the family of Non-AGN nuclei, a proper spectral fitting have been possible for 9 galaxies (NGC 3507, NGC 3898, NGC 4321, NGC 4696, CGCG 162-010, NGC 5813, NGC 5846, NGC 6482 and NPM1G -12.0625). In four cases (NGC 4321, NGC 4696, CGCG 162-010 and NPM1G -12.0625) the resulting (2-10 keV) X-ray luminosities seem to be too high ($> 10^{40}$ erg s$^{-1}$) to be interpreted as due to star forming processes; additional sources for such an excess could be either an obscured AGN or an additional component like that coming from the hot gas observed in galaxy clusters. On that respect, it is worthwhile to notice that 5 out of the 9 Non-AGN galaxies (NGC 4696, CGCG 162-010, NGC 5846, NGC 6482 and NPM1G -12.0625) are confirmed brightest galaxies in clusters; all the remaining, excepting NGC 3507, are known to be members of clusters. Thus, the inclusion of cluster emission could be a possible explanation for their high luminosity. A proper model of the hot gas from the underlying cluster, out of the scope of this paper, needs to be done indeed before any conclusion is drawn. Moreover, regarding the model that best fits their spectra, none of them need two power-law components. In three cases (NGC 3507, CGCG 162-010 and NGC 6482) a simple ME model is the best representation. For the remaining 6 galaxies, MEPL model is shown as the best fit.

The spectral index shows an asymmetrical distribution with a median value ($<\Gamma> = 2.11 \pm 0.52$), consistent with other AGN as reported by Guainazzi et al. (2005) (see Table 10 and Fig. 6). This is an important clue in favour of the AGN nature. No differences are found between AGN-like LINERs and Non-AGN LINERs.

### 5.1.3. Comparison with GM+06

With respect to GM+06, this paper presents the analysis of X-ray data on 33 new LINERs, includes *XMM-Newton* information of 41 objects with both *Chandra* and *XMM-Newton* data, uses a different smoothing process for the X-ray images and proves new X-ray spectral baseline models. We recall that only *Chandra* data were used in GM+06. The morphological types of this enlarged sample, with a total of 82 LINERs, cover later morphological types than in GM+06, but the latest morphological types (t>4) appear still lacking. Concerning optical luminosities, it seems that the lack of faint LINERs in GM+06's sample now disappears.

The differences in the smoothing process used in this paper lead to a different morphological classification on four sources. NGC 3245, NGC 4438 and NGC 4676B are now classed as AGN candidate. NGC 3628 has changed the classification from AGN to a Non-AGN source. Although a source is present in the hard band for NGC 3628 it is not point-like (see Appendix C). Using the same 46 objects than in GM+06, we end up now with 29 AGN candidates instead of 27 in GM+06.

We present the spectral analysis for 60 LINER nuclei, 36 more objects than in GM+06. This improvement comes from the inclusion of both new *Chandra* data and *XMM-Newton* data.

In addition to this, the spectral analysis itself shows differences with the previous reported analysis. The main differences between the current spectral analysis and that in GM+06 are: (1) We have excluded the Raymond Smith model, since MEKAL model has been proven to be significantly better, in particular with respect to the Fe-L emission line forest; (2) two absorptions have been included in composite models; (3) we have introduced 2PL and ME2PL models, as some of the sample sources could not be fitted with the models used by GM+06; and (4) we have explicitly included Fe emission lines in all the fits. Thus, the comparison between our current spectral analysis with that in GM+06 is not straightforward and the results there cannot be simply added up to those for the new objects here. However, if we assume that the previous RS+PL and ME+PL models are the same than the current MEPL model reported here, the final best-fit does not agree with that obtained by GM+06 in 10 cases (NGC 0315, NGC 3690B, NGC 4374, NGC 4410A, NGC 4696, UGC 08696, CGCG 162-010, NGC 6240, NGC 7130 and IC 1459). NGC 0315, NGC 3690B, UGC 08696, NGC 6240, NGC 7130 and IC 1459 are better fitted now with a more complex model (ME2PL). NGC 4374, NGC 4410A and NGC 4696 are better fitted with MEPL model instead of PL model. This can be explained for the inclusion of two absorptions in our current MEPL model. CGCG 162-010 is better fitted with ME model, instead of RS+PL model. Statistically speaking, the reduced $\chi^2$ for the objects in common shows a median value and standard deviation $<\chi_r^2> = 0.88 \pm 0.13$ in this paper and $<\chi_r^2> = 1.04 \pm 0.21$ in GM+06. The median value is lower in this work because more complicated models with a large set of free parameters have been used. However, the standard deviation is lower, hinting that we have found a better fit in more cases.

Temperature (spectral index) here and in GM+06 show a correlation coefficient of r=0.7 (r=0.5). The inclusion of two absorptions and additional power-law components are for sure responsible for the poor correlation between the two sets of spectral indices. Column densities cannot be compared because we use here two column densities instead of the one used in GM+06.

### 5.1.4. Iron emission lines

Within the *Chandra* sample, FeK$\alpha$ emission line is detected in 7 objects (NGC 0833, UGC 05101, NGC 4486, NGC 4579, NGC 6240, UGC 08696 and NGC 7130). For three of them (NGC 0833, NGC 6240 and NGC 7130) the detections are compatible with high EW (EW(FeK$\alpha$)>500 eV). FeXXV emission line is detected in NGC 4486 and NGC 6240. FeXXVI emission line is detected in four cases (NGC 4486, NGC 4579, NGC 4736 and NGC 6240).

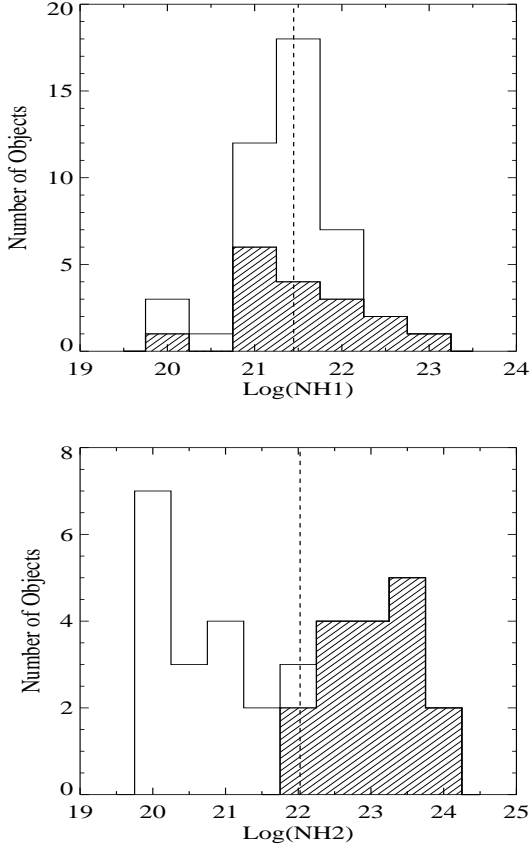

**Fig. 7.** Histogram of column densities obtained with *Chandra* data (emptied histogram). The filled histogram is the sub-sample of objects fitted to ME2PL and 2PL models. (*Top*): NH1 column density histogram and (*Bottom*): NH2 column density histogram. Dashed lines show the locus of the median value for the *Chandra* sample.

For the *XMM-Newton* sample, the FeKα emission line is measured in 10 objects (NGC 0315, NGC 0835, NGC 1052, UGC 05101, NGC 3690B, NGC 4579, MRK 266NE, UGC 08696, NGC 6240 and NGC 7285). Four of them (NGC 0835, UGC 05101, MRK 266NE and NGC 6240) are compatible with high EW (EW>500 eV). Nine out of these 10 cases have *Chandra* observations (except NGC 7285); however, only in 3 objects we detected the FeKα emission lines, whose width is coincident with that from *XMM-Newton* data. FeXXV emission line is detected in 12 cases (NGC 0315, NGC 0410, NGC 1052, UGC 05101, 3C 218, NGC 3690B, NGC 4579, UGC 08696, CGCG 162-010, NGC 6251, NGC 6240 and NPM1G -12.0625). The equivalent widths of the ionised lines are also consistent with *Chandra* results.

*XMM-Newton* data is better suited in the range where these emission lines are placed (∼ 7 keV). Thus, *XMM-Newton* results are taken when available while *Chandra* results are taken otherwise (see Table 11). Thirteen detections of the FeKα emission line, 13 detections of the FeXXV emission line and 8 detections of the FeXXVI emission lines are reported. All the objects with detected FeKα emission line are morphologically classified as AGN-like objects. Iron emission lines, as an indication of the obscuration, will be discussed in Paper II.

### 5.1.5. Obscuring material

We focus the analysis of the obscuration on *Chandra* data because the column density results might be affected by the lower spatial resolution of *XMM-Newton* data (see Appendix A for a detailed explanation).

The histograms with the resulting column densities for *Chandra* data are given in Fig. 7 (empty histogram). While NH1 column density has a narrow distribution (see Table 10), NH2 column density shows a wide range of values. NH1 column density shows a distribution similar to that reported in GM+06 ($< NH > = 3.1 \pm 3.3 \times 10^{21} cm^{-2}$), where only one column density was used in the spectral fit.

The median value for NH1 column density in our LINERs with *Chandra* data ($< \log(NH1) > = 21.32 \pm 0.71$) is higher than that reported for type 2 Seyferts (consistent with the Galactic value, Bianchi et al. 2009). This is also shown with the median value for the whole sample (see Fig. 6, *Centre-Right*). We have also found that AGN and non-AGN populations shows different NH2 distributions (2% of probability that they come from the same parent population, see Table 10).

We have also studied how column densities vary with the used spectral model. The dashed histograms in Fig. 7 represent the sub-sample of objects with two power-laws (ME2PL and 2PL) as best fit. NH1 column density distributions are quite similar for the whole sample and the previous subset (Kolmogorov-Smirnov test probability of 93%). Nevertheless, the highest values of NH2 column density are obtained for the nuclei best fitted with ME2PL or 2PL models (Kolmogorov-Smirnov test probability of being the same distribution of 6%). A detailed analysis of the obscuring matter in LINERs, including the amount and location of the absorbers, will be presented in Paper II.

### 5.1.6. Luminosities

Our derived luminosities for the LINER sample completely overlap with those found for type 2 Seyferts, although both soft and hard luminosities in LINERs tend to cover lower values. Hard X-ray luminosities show a bimodal distribution centered at $L(2-10 keV) \simeq 1 \times 10^{39}$ erg s$^{-1}$ and $L(2-10 keV) \simeq 1 \times 10^{41}$ erg s$^{-1}$ (Fig. 6), indicating a plausible distinction between two populations of LINERs. We have tried to fit the distribution to one and two Gaussian models. The single Gaussian model gives $L_o = 40.0$ and $\sigma = 1.6$ ($\chi_\nu^2 = 4.3$). The double Gaussians are centred at $L_{o1} = 39.3$ and $L_{o2} = 41.4$, with $\sigma = 0.6$ and $\sigma = 0.5$, respectively ($\chi_\nu^2 = 0.9$). Kolmogorov-Smirnov test gives a 98% of probability of being the L(2-10 keV) histogram and the two Gaussian model the same distribution. The Gaussian model shows a 73% probability of arising from the same distribution. The hard X-ray luminosity distribution is better fit with a double Gaussian model. The minimun between the two gaussians occurs at $L(2-10 keV) \simeq 1.9 \times 10^{41}$ erg s$^{-1}$. A hint of such dual behaviour is also seen for soft luminosities (see Fig. 6). Moreover, the AGN versus non-AGN distribution of luminosities shows that non-AGN objects would tend to have lower luminosities than AGN-like objects, although the K-S test is not conclusive (21% and 29% probability, respectively, see Table 10).

As a consequence of the high obscuration obtained in ME2PL and 2PL models, the unabsorbed X-ray luminosities are among the highest values of the sample when ME2PL or 2PL models are the best fit. To illustrate the relationship between this obscuration and the luminosity Fig. 8 shows NH1 and NH2 column densities versus L(0.5-2.0 keV) and L(2-10 keV) X-ray luminosities. *XMM-Newton* data are included as open circles. Note that upper limits are not included in the plots for clarity but are consistent with the detections. NH1 column density does not show any tendency with luminosity (correlation coefficient r≃0.3) while a hint on such a correlation is seen when either soft or hard luminosity is compared with NH2 column density. This is not an artifact of the quality of the data since no trend has been found between the NH2 column density and the number counts. This correlation was not found by Panessa et al. (2005) for a sample of unobscured type 2 Seyfert galaxies (see also Risaliti et al., 1999). However, their result is perfectly consistent with ours, because their column density is closer to our NH1 column density. In Paper II we re-discuss this result including an exploration about the *Compton-thick* nature of LINER nuclei.

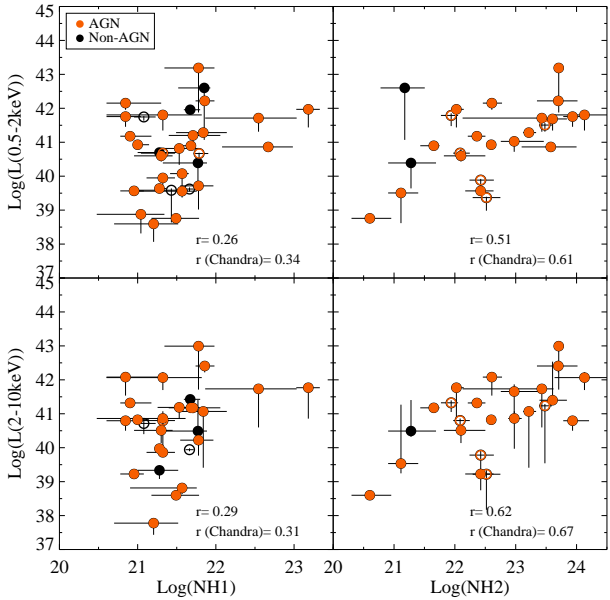

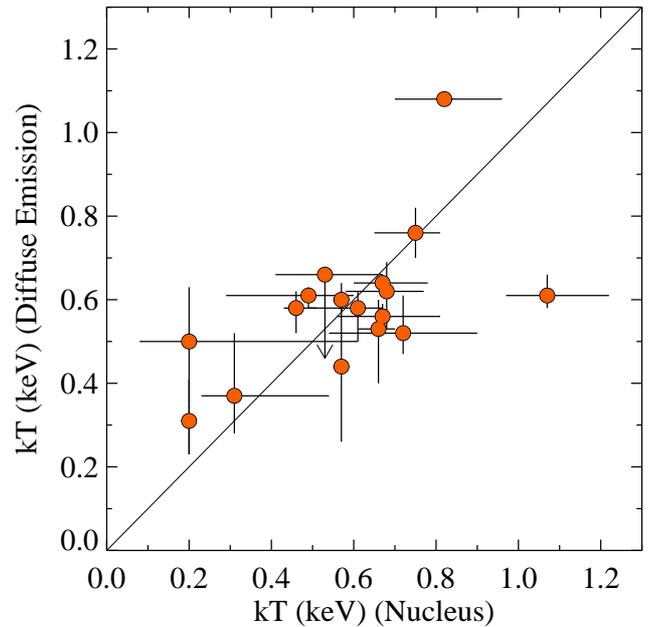

**Fig. 8.** Comparison between NH column densities and luminosities. *Top-left:* Soft (0.5-2.0 keV) luminosity versus NH1 column density; *Top-right:* Soft (0.5-2.0 keV) luminosity versus NH2 column density; *Bottom-left:* Hard (2-10 keV) luminosity versus NH1 column density and ; *Bottom-right:* Hard (2-10 keV) luminosity versus NH2 column density. *XMM-Newton* data are plotted with open circles. For clarity upper limits are not included in the plot. The correlation coeficient r (r(Chandra)) are provided for each plot for the whole sample (the subsample with *Chandra* data). AGN candidates are shown as red circles while Non-AGN candidates are shown as black circles.

**Fig. 9.** Temperature of the diffuse emission (kT (keV) (Diffuse Emission)) versus the temperature of the nuclear emission (kT (keV)(Nucleus)). Arrows are upper limits. The unity slope is shown as a continuous line. 3C 218 and CGCG 162-010 are out of the plot with coordinates (x, y)=[1.7 keV, 3.0 keV] and (x, y)=[1.0 keV, 4.1 keV], respectively (see text).

### 5.1.7. Thermal component

The thermal component is an important ingredient in LINERs, since it is needed in 53 out of the 60 objects. This contribution is a high fraction of the emission, specially at kT< 2 keV (see Table 8 Col. 9). Its asymmetrical, bimodal distribution is centred at kT=0.25 keV and kT=0.65 keV (see Fig. 6). Only IRAS 14348-0014 shows kT>2 keV. Temperatures above kT=0.45 keV are related to strong thermal processes (Strickland et al., 2002) whereas temperatures around kT=0.2 keV are typical of Seyfert 1 objects (Teng et al., 2005; Panessa et al., 2008).

To study if the thermal component comes from the host galaxy we have made the spectral analysis of the diffuse emis-sion around the nucleus using *Chandra* data. From the sample of 55 objects with spectral fitting, we have selected the 19 objects for which the spectral analysis on the diffuse emission is expected to be reliable (i.e. a clean extraction of the diffuse emission can be done). Any point-like source has been excluded from the extraction region.

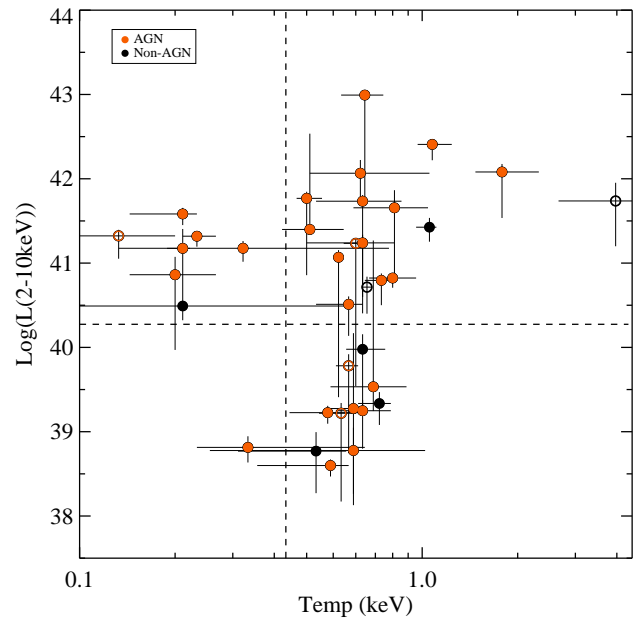

**Fig. 10.** Hard X-ray (2-10 keV) luminosity versus temperature. AGN candidates are shown as red circles while Non-AGN candidates are shown as black circles. The *XMM-Newton* data are shown with open circles. Dashed lines show the minima between the two peaks found in the distribution of temperature (kT=0.45 keV) and hard luminosity ($1.9 \times 10^{40}$ erg s$^{-1}$).

We have investigated ME, PL and MEPL models. ME model is used to reproduce thermal emission. PL model is considered to take into account the eventual contamination of unresolved point-like sources. MEPL model is included to add the possibility that unresolved point-like sources and thermal emission are contributing to the final emission[7]. The resulting parameters for the best-fit models are summarized in Table 13.

The comparison between the nuclear temperature (kT (keV) (Nucleus)) and the diffuse emission temperature (kT (keV) (Diffuse Emission)) is shown in Fig. 9. Only 4 objects (3C 218, NGC 4486, CGCG 162-010 and NGC 6240) show discrepancies. Thus, it can be safely concluded that the material responsible for the thermal component is the same than that producing the circumnuclear diffuse emission. However, the physical mechanism is still unknown.

Since both, 2-10 keV luminosities and temperatures, show a bimodal distribution, we have attempted to analyse whether a connection exists between L(2-10 keV) and kT (see Fig. 10) The dashed lines correspond to the values where the histograms reach the minimum between the two peaks: kT=0.45 keV and $L(2-10 \text{ keV}) = 1.9 \times 10^{40} \text{erg s}^{-1}$. *XMM-Newton* data are included with open circles. All the objects but one (NGC 4457[8]) with low temperatures (kT < 0.45 keV) are in the group of high luminosities ($> 1.9 \times 10^{40}$erg s$^{-1}$) (NGC 3690B, NGC 3998, NGC 4321, NGC 4410, NGC 4579, NGC 6251 and NGC 7285). Also all the objects but NGC 4457 in the low luminosity range ($< 1.9 \times 10^{40}$erg s$^{-1}$) show a temperature above kT=0.45 keV (NGC 2681, NGC 2841, NGC 4278, NGC 4374, NGC 4494, NGC 4552, NGC 4696, NGC 4736, NGC 5363, NGC 5813 and NGC 6482). There is also a mixed group of objects with high temperature and high luminosity. The same trend is found with the soft luminosity.

The temperature of thermal spectral components that we formally obtain in high-power LINERs is comparable to that obtained in type 1 AGN (Teng et al., 2005; Panessa et al., 2008). In type 1 AGN the soft excess is probably associated with atomic physics processes such as ionised absorption or disk reflection (Gierliński and Done, 2004; Crummy et al., 2006). Therefore, we suggest that in AGN candidates amongst LINERs the soft excess is not indicative of thermal processes. Only Non-AGN candidates would exhibit signatures of truly thermal processes in their X-ray spectra. Thus, the nature of this thermal component remains unclear but it is very suggestive that it is present in a high fraction of LINERs. In fact, some scattering can be seen below 2 keV in most cases, which might point out to more complex assumptions than a simple thermal model to reproduce the soft emission. For this purpose we are analysis*XMM-Newton* data from the Reflection Grating Spectrometer (RGS) (Gonzalez-Martin et al. in preparation).

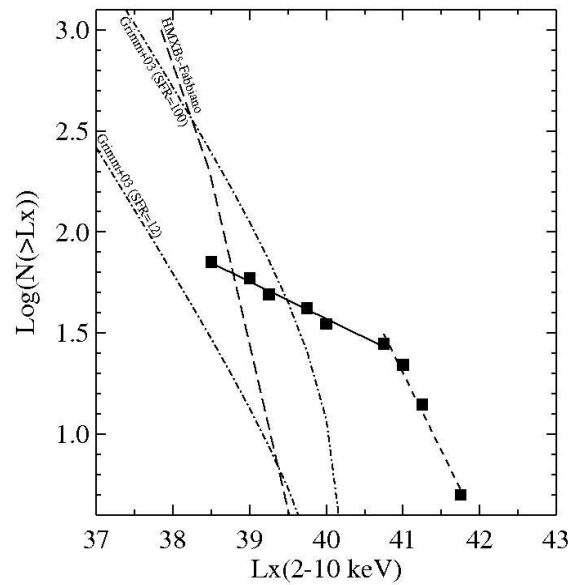

**Fig. 11.** Hard (2-10 keV) luminosity function (LF) for the whole LINER sample (continuous line). That for ULXs in Kim and Fabbiano (2004) is plotted as a dashed line. The dashed-dotted lines correspond to the LF of HMXRBs from Grimm et al. (2003) for two different star forming rates, 12 and 100M$_\odot$/yr.

### 5.2. Multiwavelength analysis

In addition to the X-ray signatures used in this paper (e.g. hard X-ray compact nuclear source, FeK$\alpha$ 6.4 keV iron emission line or the X-ray spectral analysis) evidence about the AGN nature of our LINER nuclei can be found at other wavelengths. The last part of our study deals hence with this multiwavelength analysis.

Before entering into it, we want to say a few words about the longstanding controversy (see Ho, 2008, for a full discussion) involving other alternatives to explain the X-ray emission in LINERs, including Starburst and/or ULX contamination. A starbust contribution has been invoked to explain the observed X-ray emission for some particular cases (i.e. Eracleous et al., 2002; Jimenez-Bailon et al., 2005, for the LINER NGC 4736 and for the low luminosity AGN NGC 1808, respectively). In these cases the emission from high mass X-ray binaries (HMXB) can be claimed as an important contributor. Nevertheless, in GM+06 we have already argued that HMXB cannot be considered as an important ingredient for most of the galaxies. By using the data from Cid-Fernandes et al. (2004) and Gonzalez-Delgado et al. (2004), we have found that for the 21 objects in common a young stellar population can be only claimed in three cases, NGC 3507, NGC 3998 and NGC 4321 (see Col. (8) in Table 12). NGC 3998 has a reported broad H$\alpha$ emission line (Ho et al., 1997) what reinforces its AGN nature. For the other two galaxies their X-ray morphology is that of an Non-AGN candidate and a detailed analysis needs to be done to evaluate the importance of HMXB in them. But for the remaining 18 objects young stars appear not to be important contributors to the observed emission.

On the other hand, ULX sources (see Fabbiano, 2006, for a review) can constitute an important contribution at X-rays energies. The high X-ray luminosity found for an ULX in the star-forming galaxy NGC 7424 (Soria et al., 2006), with luminosity of $L(0.5 - 10\text{keV}) = 10^{40}$ erg s$^{-1}$, proves how important the contamination produced by such objects can be if they are found at nuclear locations. This makes the analysis extremely difficult and implies that only indirect proofs can be invoked. But it has to be taken into account that the reported ULXs are mostly associated to young clusters where strong events of star formation are occurring (see for instance the data on the Antenae and NGC 1275, Wolter et al., 2006; Gonzalez-Martin et al., 2006a, respectively) which appears not to be our case. Fig. 11 shows the luminosity function of the whole LINER sample (continuous line) and the luminosity function (LF) reported for ULXs (Kim and Fabbiano, 2004, dotted line) and HMXBs (Grimm et al., 2003, dashed-dotted line) for two different star forming rates, 12 and $100 M_{\odot}/yr$. Kim and Fabbiano (2004) found that the ULX luminosity function for a sample of normal galaxies is well fitted by two power-laws with a break at $5 \times 10^{38} erg\, s^{-1}$ (see Fig. 11, dashed line). The slope before the break is around -1.8 and after the break is -2.8. However, the luminosity function we obtain for LINER nuclear sources fits to a broken power-law with a break at $\sim 10^{41} erg\, s^{-1}$ and slopes -0.2 and -0.8 before and after the break (see Fig. 11, dashed line). Hence, the LF of LINER nuclei notably differs from that of HMXBs and ULX. Thus it seems safe to conclude that neither HMXBs nor ULXs can be considered as serious alternatives to explain the X-rays emission in LINERs.

---

[7] 2PL and ME2PL were not explored since they have not a physical correspondence in the model of the diffuse emission.

[8] Note that its temperature range falls within the range of high temperature objects.

To further analyse the nature of our LINER nuclei, we compiled all the information available from the literature at frequencies from radio to UV. Table 12 lists the multiwavelength data discussed in this paper. Under the assumptions of the unified model the main ingredients expected to be common to AGN are: (1) a region of high velocity and high velocity dispersion, called the Broad Line Region (BLR); (2) variability at UV and; (3) an unresolved radio compact core associated with a flat continuum (Nagar et al., 2005; Filho et al., 2004). The signature of any of these ingredients will be taken as an AGN signature.

**(1)** Broad $H\alpha$ emission line components has been reported for 15 cases (18%) in our sample. We stress that weak broad $H\alpha$ emission lines could exist and remain undetected. Low spectral resolution could result in misclassifying the blending of narrow $H\alpha$ emission line as a broad $H\alpha$ emission line. With the exception of NGC 2639 and NGC 4636, all the 15 BLR LINER nuclei in our sample are classified as having AGN X-ray morphology (86%). Ho (2008) has shown that the type 1 LINERs in his sample host a point like source in 95% of the cases, consistent with our findings. On the contrary, among type 2 LINERs, our detection rate of X-ray AGN is 50%, lower than the reported rate by Ho (2008). This difference may be related to the less homogeneous character of our sample with respect to Ho's, who only uses sources from the Palomar Sky spectroscopic survey.

From the comparison between optical and X-ray results, the existence of a BLR in NGC 2639 and NGC 4636, with no AGN signatures at X-rays remains unclear. We could imagine two scenarios to explain it: (1) The matter which obscures the BLR at optical wavelengths is not the same than that obscuring the X-ray source, with the later possibly located in the inner parts, eventually within the BLR itself (Elvis et al., 2004; Risaliti et al., 2005) or (2) the obscuring torus is clumpy, what might explain transitions between type 1 and type 2 (Elitzur, 2007). In the later case the time evolution of the clouds around the AGN is the responsible for such discrepancies, provided that they were not observed simultaneously at optical an X-ray frequencies. In fact, a general picture can emerge in which the optical obscuring torus is a smooth continuation of the BLR clouds. However, the clumpy torus seems to dilute at bolometric luminosities lower than $10^{42} erg\, s^{-1}$ (Elitzur, 2007), which is well above the luminosities computed for NGC 4636 and NGC 2639. The fact that only a few LINERs show broad $H\alpha$ would support the hypothesis of the dissolution of the BLR in LLAGN (Nicastro et al., 2003; Martínez et al., 2008).

**(2)** In six cases (NGC 3998, NGC 4486, NGC 4552, NGC 4579, NGC 4594 and NGC 4736) UV variability has been found (Maoz et al., 2005, Table 12 Col. (10)), all of them compatible with the X-ray morphological classification as AGN. Interestingly, all the sources showing UV variability appear as *Compton-thin* at X-rays frequencies.

**(3)** Finally, observations at radio-frequencies do exist for 75 out of our 82 LINERs (Nagar et al., 2005, and references therein). Twenty eight sources have not been detected[9] at 2 cm whereas compact sources are detected in 54/75 (72%). Among the 54 compact sources, six of them show steep nuclear spectra; 25 of them are compatible with the presence of an AGN (17 show flat spectra and 8 show jet structure). Among these 25 objects, 14 have been classified as AGN candidates and 11 have been classified as Non-AGN candidates. Among the 54 sources detected at radio frequencies, 30 are detected at X-rays.

Adding together the X-ray results presented in this paper with the information on UV variability, broad $H\alpha$ emission line and radio classifications, 16 sources (19.5%) do not show any evidence of the presence of an AGN, namely NGC 0410, NGC 474, NGC 0524, UGC 4881, NGC 3379, NGC 3623, NGC 3628, NGC 3898, NGC 4314, NGC 4321, NGC 4459, NGC 4596, NGC 4676A, IC 4395, NGC 6482 and NPM1G -12.0625. The spectral analysis was possible for 10 sources of these sources and it is incompatible with a pure thermal origin (i.e. ME model) in three cases (UGC 4881, NGC 3507 and NGC 6482).

Hence, taking into account the X-ray and multiwavelength information of this large compilation of LINERs, a huge amount of LINERs (80%) seem to host an AGN. Recently, Dudik et al. (2009), combining optical, X-ray, and mid-IR diagnostics, found an AGN detection rate of 74%, close to our findings. This fraction might be a lower limit since we have not taken into account *Compton-thick* sources (a full discussion will be reported in Paper II).

## 6. Summary and conclusions

In this paper the nuclear characterisation of 82 LINER galaxies in X-rays, 68 with *Chandra* and 55 with *XMM-Newton* data is presented. They make the largest sample of LINERs with X-ray spectral fits ever analysed (60 out of the 82 objects). We also use the information on the optical morphology with *HST* data, optical emission lines, UV variability, radio compactness and stellar populations reported in the literature.

To summarise, the most relevant findings in this paper are:

1. *X-ray Imaging:* 60% (49/82) of the sample shows a compact nuclear source at 4.5-8 keV band (the so-called AGN candidates), 68% (41/60) within the sample with spectral fit.
2. *X-ray spectroscopy:* MEPL and ME2PL models are the best representation of the data in 73% (44/60) of the sample. 2PL was good enough only in one case (2%) while simple models (ME or PL) were needed in 15 cases (25%). The median and standard deviation of the parameters are: $<\Gamma>=2.11\pm 0.52$, $<kT>=0.54\pm 0.30$ keV, $<\log(NH1)>=21.32\pm 0.71$ and $<\log(NH2)>=21.93\pm 1.36$. Spectral indices are consistent with the presence of an AGN (Page et al., 2003; Piconcelli et al., 2005; Nandra et al., 2007). A single thermal model is reported as the best fit in only six cases, all of them morphologically classified at X-rays as Non-AGN candidates. The FeK iron emission line was detected in 13 cases, all of them AGN-like sources.

---

[9] Note that the non detection might be due to low S/N level of the data

3. *Luminosities:* Soft and hard luminosities range between log(L(0.5-2 keV)) = 37.5-43.5 and log(L(2-10 keV)) = 37-43 with median values of $\log(L(0.5 - 2 \text{ keV})) = 40.22 \pm 1.33$ and $\log(L(2 - 10 \text{ keV})) = 39.85 \pm 1.26$, respectively. The 2-10 keV luminosity shows a bimodal distribution centred at log(L(2-10 keV))=39 and log(L(2-10 keV))=41.5. A weak dependence of the NH2 column density with the intrinsic luminosities is found. The X-ray luminosities overlap with those found for type 2 Seyferts (Guainazzi et al., 2005; Cappi et al., 2006; Panessa et al., 2006).

4. *Multiwavelength analysis:* Adding up the multiwavelength information discussed throughout this paper, at least 66 out of the 82 objects (80%) show evidence of harbouring an AGN, although we do not rule out the presence of an AGN in the remaining ones. NGC 2639 and NGC 4636 do not show X-ray signatures of an AGN while, interestingly, the host BLR (Ho et al., 2001). A more complex scenario than the simplest version of the Unified Model is claimed to include the LINER nuclei family.

## Acknowledgements

Thanks to the anonymous referee for his/her useful comments that helped to improve the paper. JM, IM and OGM also are gratefully acknowledge the useful comments from R. Dupke and F. Panessa. We thank F. Durret for her helpfull comments on this work. We gratefully acknowledge J. Acosta, F. Carrera and E. Florido, members of OGM's PhD jury. We also thank to J. Sulentic and A. de Ugarte Postigo for the help to improve the text. This work was financed by DGICyT grants AYA 2003-00128, AYA 2006-01325, AYA 2007-62190 and the Junta de Andalucía TIC114. OGM acknowledges financial support from the Ministerio de Educación y Ciencia through the Spanish grant FPI BES-2004-5044 and research fellowship of STFC. This research has made use of data obtained from the *Chandra* Data Archive provided by the *Chandra* X-ray Center (CXC). This research has made use of data obtained from the *XMM-Newton* Data Archive provided by the *XMM-Newton* Science Archive (XSA). This research made use of the NASA/IPAC extragalactic database (NED), which is operated by the Jet Propulsion Laboratory under contract with the National Aeronautics and Space Administration. This publication makes use of data products from the Two Micron All Sky Survey, which is a joint project of the University of Massachusetts and the Infrared Processing and Analysis Center/California Institute of Technology, funded by the National Aeronautics and Space Administration and the National Science Foundation. We acknowledge the usage of the Hyperleda database. This paper is partially based on NASA/ESA *Hubble Space Telescope* observations.

## Appendices

The appendices below are provided in the on-line edition.

- Appendix A. *XMM-Newton* versus *Chandra* results.
- Appendix B. Notes on individual objects.
- Appendix C. Catalogue of X-ray images.
- Appendix D. Nuclear spectra from *Chandra* data.
- Appendix E. Nuclear spectra from *XMM-Newton* data.
- Appendix F. Large scale optical (DSS) images.

**Table 1.** Summary of the general properties of our LINER sample.

| Num | Name | Other Name | RA | Dec | Redshift | Dist. (Mpc) | Ref. | B | E(B-V) | Morph. Type |
|---|---|---|---|---|---|---|---|---|---|---|
| (1) | (2) | (3) | (4) | (5) | (6) | (7) | (8) | (9) | (10) | (11) |
| 1 | NGC 0315 | UGC 0597 | 00 57 48.88 | +00 21 08.8 | 0.01646 | 68.11 | (b) | 12.202 | 0.065 | E |
| 2 | NGC 0410 | UGC 0735 | 01 10 58.87 | +03 09 08.3 | 0.01765 | 71.87 | (b) | 12.519 | 0.059 | E(s) |
| 3 | NGC 0474 | | 01 20 06.70 | +00 24 55.0 | 0.00775 | 32.51 | (d) | 12.382 | 0.035 | S0(r) |
| 4 | IIIZW 035 | | 01 44 30.50 | +17 06 05.0 | 0.02743 | 109.67 | (b) | | 0.063 | |
| 5 | NGC 0524 | UGC 0968 | 01 24 47.72 | +00 32 19.8 | 0.00793 | 23.99 | (c) | 11.381 | 0.082 | S0-a |
| 6 | NGC 0833 | ARP 318B | 02 09 20.88 | +10 08 00.3 | 0.01357 | 49.74 | (b) | 13.697 | 0.025 | Sa |
| 7 | NGC 0835 | ARP 318A | 02 09 24.69 | −10 08 10.5 | 0.01357 | 50.45 | (b) | 13.159 | 0.025 | Sab(sr) |
| 8 | NGC 1052 | | 02 41 04.80 | +08 15 20.8 | 0.0049 | 19.41 | (c) | 11.436 | 0.027 | E |
| 9 | NGC 2639 | UGC 4544 | 08 43 38.08 | +50 12 20.0 | 0.01112 | 45.45 | (b) | 12.596 | 0.024 | Sa(r) |
| 10 | NGC 2655 | UGC 4637 | 08 55 37.73 | +78 13 23.1 | 0.00467 | 24.43 | (d) | 10.98 | 0.031 | S0-a(s) |
| 11 | NGC 2681 | UGC 4645 | 08 53 32.73 | +51 18 49.3 | 0.00230 | 17.22 | (c) | 11.148 | 0.023 | S0-a(s) |
| 12 | NGC 2685 | UGC 4666 | 08 55 34.75 | +58 44 03.9 | 0.00294 | 16.22 | (d) | 12.16 | 0.062 | S0-a(s) |
| 13 | UGC 4881 | | 09 15 55.10 | +44 19 55.0 | 0.03930 | 157.09 | (b) | | 0.017 | |
| 14 | 3C 218 | Hydra A | 09 18 05.67 | −12 05 44.0 | 0.05487 | 218.04 | (b) | 14.296 | 0.041 | E-S0 |
| 15 | NGC 2787 | UGC 4914 | 09 19 18.56 | +69 12 12.0 | 0.00232 | 7.48 | (c) | 11.602 | 0.133 | S0-a(sr) |
| 16 | NGC 2841 | UGC 4966 | 09 22 02.63 | +50 58 35.5 | 0.00212 | 11.97 | (d) | 10.063 | 0.015 | Sb(r) |
| 17 | UGC 05101 | | 09 35 51.65 | +61 21 11.3 | 0.03939 | 157.3 | (b) | 15.278 | 0.033 | |
| 18 | NGC 3185 | | 10 17 38.57 | +21 41 17.7 | 0.00406 | 21.28 | (d) | 12.936 | 0.027 | Sa(sr) |
| 19 | NGC 3226 | UGC 5617 | 10 23 27.01 | +19 53 54.7 | 0.0059 | 23.55 | (c) | 12.339 | 0.023 | E |
| 20 | NGC 3245 | UGC 5663 | 10 27 18.39 | +28 30 26.6 | 0.00522 | 20.89 | (c) | 11.655 | 0.025 | S0(r) |
| 21 | NGC 3379 | M 105 | 10 47 49.60 | +12 34 53.9 | 0.00265 | 10.57 | (c) | 10.218 | 0.024 | E |
| 22 | NGC 3414 | UGC 5959 | 10 51 16.23 | +27 58 30.0 | 0.00471 | 25.23 | (c) | 12.055 | 0.025 | S0(s) |
| 23 | NGC 3507 | UGC 6123 | 11 03 25.39 | +18 08 07.4 | 0.00495 | 19.77 | (d) | 12.084 | 0.024 | SBb(s) |
| 24 | NGC 3607 | UGC 6297 | 11 16 54.66 | +18 03 06.5 | 0.0057 | 22.80 | (c) | 10.933 | 0.021 | E-S0 |
| 25 | NGC 3608 | UGC 6299 | 11 16 58.96 | +18 08 54.9 | 0.00572 | 22.91 | (c) | 11.567 | 0.021 | E |
| 26 | NGC 3623 | M 65 | 11 18 55.96 | +13 05 32.0 | 0.00269 | 7.28 | (d) | 10.139 | 0.025 | SABa(s) |
| 27 | NGC 3627 | M 66 | 11 20 15.03 | +12 59 29.6 | 0.00242 | 10.28 | (a) | 9.735 | 0.033 | SABb(s) |
| 28 | NGC 3628 | UGC 6350 | 11 20 17.01 | +13 35 22.9 | 0.00192 | 7.73 | (d) | 9.972 | 0.027 | Sb |
| 29 | NGC 3690B | MRK 171 | 11 28 32.20 | +58 33 44.0 | 0.0104 | 41.60 | (b) | 13.369 | 0.017 | SBm(s) |
| 30 | NGC 3898 | UGC 6787 | 11 49 15.37 | +56 05 03.7 | 0.00392 | 21.98 | (d) | 11.257 | 0.020 | Sab |
| 31 | NGC 3945 | UGC 6860 | 11 53 13.73 | +60 40 32.0 | 0.0042 | 22.49 | (d) | 11.7 | 0.028 | S0-a(sr) |
| 32 | NGC 3998 | UGC 6946 | 11 57 56.12 | +55 27 12.7 | 0.00346 | 14.13 | (c) | 11.409 | 0.016 | S0(r) |
| 33 | NGC 4036 | UGC 7005 | 12 01 26.75 | +61 53 44.8 | 0.00482 | 24.55 | (d) | 11.536 | 0.024 | E-S0 |
| 34 | NGC 4111 | UGC 7103 | 12 07 03.13 | +43 03 55.4 | 0.00375 | 15.00 | (c) | 11.67 | 0.014 | S0-a(r) |
| 35 | NGC 4125 | UGC 7118 | 12 08 06.02 | +65 10 26.9 | 0.00597 | 23.88 | (c) | 10.633 | 0.019 | E |
| 36 | IRAS 12112+0305 | | 12 13 46.00 | +02 48 38.0 | 0.07334 | 291.12 | (b) | 18.272 | 0.021 | S? |
| 37 | NGC 4261 | UGC 7360 | 12 19 23.22 | +05 49 30.8 | 0.00737 | 31.62 | (c) | 11.35 | 0.018 | E |
| 38 | NGC 4278 | M 98 | 12 20 06.83 | +29 16 50.7 | 0.00216 | 16.07 | (c) | 11.042 | 0.029 | E |
| 39 | NGC 4314 | UGC 7443 | 12 22 31.99 | +29 53 43.3 | 0.00242 | 9.68 | (d) | 11.416 | 0.025 | Sa(sr) |
| 40 | NGC 4321 | M 100 | 12 22 54.90 | +15 49 20.6 | 0.00524 | 16.14 | (a) | 10.022 | 0.026 | SABb(s) |
| 41 | NGC 4374 | M 84 | 12 25 03.74 | +12 53 13.1 | 0.0046 | 18.37 | (c) | 10.114 | 0.041 | E |
| 42 | NGC 4410A | UGC 7535 | 12 26 28.86 | +09 01 10.8 | 0.02512 | 100.30 | (d) | 13.948 | 0.025 | Sab |
| 43 | NGC 4438 | UGC 7574 | 12 27 45.59 | +13 00 31.8 | 0.0042 | 16.83 | (d) | 10.927 | 0.028 | Sa |
| 44 | NGC 4457 | UGC 7609 | 12 28 59.01 | +03 34 14.1 | 0.00435 | 17.38 | (d) | 11.711 | 0.022 | S0-a(sr) |
| 45 | NGC 4459 | UGC 7614 | 12 29 00.03 | +13 58 42.8 | 0.00402 | 16.14 | (d) | 11.454 | 0.045 | S0-a(r) |
| 46 | NGC 4486 | M 87 | 12 30 49.42 | +12 23 28.0 | 0.00402 | 16.07 | (c) | 9.587 | 0.023 | E |
| 47 | NGC 4494 | UGC 7662 | 12 31 24.03 | +25 46 29.9 | 0.00427 | 17.06 | (c) | 10.681 | 0.021 | E |
| 48 | NGC 4552 | M 89 | 12 35 39.81 | +12 33 22.8 | 0.0038 | 15.35 | (c) | 10.67 | 0.041 | E |
| 49 | NGC 4589 | UGC 7797 | 12 37 25.03 | +74 11 30.8 | 0.00660 | 21.98 | (c) | 11.703 | 0.028 | E |
| 50 | NGC 4579 | M 58 | 12 37 43.52 | +11 49 05.5 | 0.0042 | 16.83 | (d) | 10.478 | 0.041 | SABb(s) |
| 51 | NGC 4596 | UGC 7828 | 12 39 55.94 | +10 10 33.9 | 0.0042 | 16.83 | (d) | 11.437 | 0.022 | S0-a(sr) |
| 52 | NGC 4594 | M 104 | 12 39 59.43 | −11 37 23.0 | 0.00245 | 9.77 | (c) | 9.155 | 0.051 | Sa |
| 53 | NGC 4636 | UGC 7878 | 12 42 49.87 | +02 41 16.0 | 0.00365 | 14.66 | (c) | 10.429 | 0.029 | E |
| 54 | NGC 4676A | | 12 46 10.08 | +30 43 55.2 | 0.022 | 88.00 | (b) | 14.4 | 0.016 | S0-a(s) |
| 55 | NGC 4676B | | 12 46 11.23 | +30 43 21.6 | 0.022 | 88.00 | (b) | 14.954 | 0.016 | S0-a(s) |
| 56 | NGC 4698 | UGC 7970 | 12 48 22.92 | +08 29 14.3 | 0.0042 | 16.83 | (d) | 11.554 | 0.026 | Sa |
| 57 | NGC 4696 | Abell 3526 | 12 48 49.28 | −41 18 40.0 | 0.00887 | 35.48 | (c) | 11.684 | 0.112 | E |
| 58 | NGC 4736 | M 94 | 12 50 53.06 | +41 07 13.6 | 0.0013 | 5.20 | (c) | 8.707 | 0.018 | Sab(r) |
| 59 | NGC 5005 | UGC 8256 | 13 10 56.23 | +37 03 33.1 | 0.00315 | 21.28 | (d) | 10.539 | 0.014 | SABb(s) |
| 60 | NGC 5055 | M 63 | 13 15 49.33 | +42 01 45.4 | 0.0018 | 7.14 | (d) | 9.323 | 0.018 | Sbc |
| 61 | MRK 266NE | NGC 5256 | 13 38 17.80 | +48 16 41.2 | 0.02806 | 112.00 | (b) | 14.10 | 0.013 | |
| 62 | UGC 08696 | MRK 273 | 13 44 42.11 | +55 53 12.7 | 0.03778 | 150.9 | (b) | 15.106 | 0.008 | |
| 63 | CGCG 162-010 | Abell 1795 | 13 48 52.43 | +26 35 34.0 | 0.06315 | 252.6 | (b) | 15.197 | 0.013 | |
| 64 | NGC 5363 | | 13 56 07.24 | +05 15 17.0 | 0.00379 | 22.39 | (d) | 11.084 | 0.027 | S0-a |
| 65 | IC 4395 | UGC 9141 | 14 17 21.08 | +26 51 26.7 | 0.03651 | 148.50 | (b) | 14.698 | 0.017 | Sb |
| 66 | IRAS 14348-1447 | | 14 37 38.37 | −15 00 21.0 | 0.08273 | 329.65 | (b) | 16.623 | 0.123 | S? |
| 67 | NGC 5746 | UGC 9499 | 14 44 55.92 | +01 57 18.0 | 0.00735 | 29.38 | (d) | 11.342 | 0.040 | SABb(s) |
| 68 | NGC 5813 | | 15 01 11.26 | +01 42 07.1 | 0.00657 | 32.21 | (c) | 11.483 | 0.057 | E |
| 69 | NGC 5838 | UGC 9692 | 15 05 26.26 | +02 05 57.6 | 0.00453 | 28.44 | (d) | 11.785 | 0.053 | E-S0 |
| 70 | NGC 5846 | UGC 9705 | 15 06 29.29 | +01 36 20.2 | 0.00622 | 24.89 | (c) | 11.074 | 0.056 | E |
| 71 | NGC 5866 | UGC 9723 | 15 06 29.50 | +55 45 47.6 | 0.00385 | 15.35 | (c) | 10.73 | 0.013 | S0-a |
| 72 | MRK 0848 | IZW 107 | 15 18 06.35 | +42 44 36.7 | 0.04019 | 164.48 | (b) | 15.99 | 0.026 | S0 |
| 73 | NGC 6251 | UGC 10501 | 16 32 31.97 | +82 32 16.4 | 0.02297 | 91.90 | (b) | 13.768 | 0.087 | E |
| 74 | NGC 6240 | UGC 10592 | 16 52 58.89 | +02 24 03.4 | 0.02445 | 97.80 | (b) | 13.816 | 0.076 | S0-a |
| 75 | IRAS 17208-0014 | | 17 23 21.96 | +00 17 00.9 | 0.04275 | 171.0 | (b) | 17.53 | 0.344 | S? |
| 76 | NGC 6482 | UGC 11009 | 17 51 48.81 | +23 04 19.0 | 0.01315 | 52.60 | (b) | 12.225 | 0.101 | E |
| 77 | NGC 7130 | IC 5135 | 21 48 19.50 | −34 57 04.7 | 0.01612 | 64.50 | (b) | 12.927 | 0.029 | Sa |
| 78 | NGC 7285 | | 22 28 38.00 | −24 50 26.8 | 0.01442 | 57.54 | (b) | 12.963 | 0.025 | SBa(s) |
| 79 | NGC 7331 | UGC 12113 | 22 37 04.09 | +34 24 56.3 | 0.00377 | 13.12 | (c) | 10.199 | 0.091 | Sbc |
| 80 | IC 1459 | IC 5265 | 22 57 10.60 | −36 27 44.0 | 0.0055 | 29.24 | (c) | 10.961 | 0.016 | E |
| 81 | NPM1G -12.0625 | Abell 2567 | 23 25 19.82 | −12 07 26.4 | 0.08299 | 327.93 | (b) | 16.177 | 0.030 | S? |

**Table 1.** Continuation

| Num | Name | Other Name | RA | Dec | Redshift | Dist. (Mpc) | Ref. | B | E(B-V) | Morph. Type |
|---|---|---|---|---|---|---|---|---|---|---|
| (1) | (2) | (3) | (4) | (5) | (6) | (7) | (8) | (9) | (10) | (11) |
| 82 | NGC 7743 ....... | UGC 12759 | 23 44 21.14 | +09 56 02.7 | 0.00570 | 20.70 | (c) | 12.392 | 0.068 | S0-a(s) |

NOTES.–Distances from the following references: (a) Ferrarese et al. (2000); (b) from cosmology assuming $H_o = 75$ Km Mpc$^{-1}$ s$^{-1}$; (c) Tonry et al. (2001); and (d) Tully (1998)

**Table 2.** Observational details*

| | | | | Chandra data | | | | | | | | XMM-Newton data | | | HST data | | |
|---|---|---|---|---|---|---|---|---|---|---|---|---|---|---|---|---|---|
| Num | Name | NH(Gal) (cm$^{-2}$) | Scale (pc/″) | ObsID | Expt. (ks) | RA | Dec | Radii (pc) | Radii (arcs) | Offset (arcs) | Pileup** (%) | ObsID | Expt. (ksec) | Pileup** (%) | Filter | ObsID | Expt. (sec) |
| (1) | (2) | (3) | (4) | (5) | (6) | (7) | (8) | (9) | (10) | (11) | (12) | (13) | (14) | (15) | (16) | (17) | (18) |
| 1 | NGC 0315 | 0.0588 | 330.2 | 4156 | 52.3 | 00 57 48.870 | +30 21 08.75 | 649.8 | 1.968 | 0.271 | 4.917/ 5.078 | 305290201 | 43.7 | 0.154/ 0.156 | F814W | 6673 | 230 |
| 2 | NGC 0410 | 0.0540 | 348.4 | ... | ... | ... | ... | ... | ... | ... | .../ ... | 203610201 | 19.6 | 0.179/ 0.183 | ... | ... | ... |
| 3 | NGC 0474 | 0.0332 | 157.6 | 7144 | 14.9 | 01 20 06.799 | +03 24 57.06 | 897.7 | 5.696 | 2.217 | 0.066/ 0.082 | 200780101 | 16.3 | <0.001/<0.001 | ... | ... | ... |
| 4 | IIIZW 035 | 0.0505 | 531.7 | 6855 | 14.5 | 01 24 47.751 | +09 32 19.71 | 1129.3 | 2.124 | 0.593 | 0.248/ 0.118 | 203390201 | 13.8 | <0.001/<0.001 | F814W | 10592 | 720 |
| 5 | NGC 0524 | 0.0485 | 116.3 | 6778 | 13.7 | 01 44 30.500 | +17 06 07.53 | 295.2 | 2.538 | 2.520 | 0.000/ 0.470 | ... | ... | .../ ... | F814W | 5999 | 160 |
| 6 | NGC 0833 | 0.0221 | 241.1 | 923 | 12.2 | 02 09 24.662 | -10 08 10.44 | 1897.9 | 7.872 | 0.746 | 1.632/ 1.432 | 115810301 | 44.8 | 0.001/ 0.002 | ... | ... | ... |
| 7 | NGC 0835 | 0.0221 | 244.6 | 923 | 12.1 | 02 09 20.797 | -10 07 58.62 | 1684.8 | 6.888 | 1.017 | 0.249/ 0.203 | 115810301 | 44.8 | 0.001/ 0.002 | ... | ... | ... |
| 8 | NGC 1052 | 0.0307 | 94.1 | 385 | 10.9 | 02 41 04.833 | -08 15 20.57 | 138.9 | 1.476 | 1.552 | 5.999/ 6.875 | 306230101 | 47.2 | 0.267/ 0.268 | F555W | 3639 | 500 |
| 9 | NGC 2639 | 0.0304 | 220.3 | ... | ... | ... | ... | ... | ... | ... | .../ ... | 301651101 | 7.4 | 0.014/ 0.016 | F606W | 5479 | 500 |
| 10 | NGC 2655 | 0.0209 | 118.4 | ... | ... | ... | ... | ... | ... | ... | .../ ... | 301650301 | 8.9 | 0.116/ 0.113 | F547M | 5419 | 300 |
| 11 | NGC 2681 | 0.0245 | 83.5 | 2060 | 70.7 | 08 53 32.783 | +51 18 48.85 | 205.4 | 2.46 | 0.770 | 1.081/ 1.083 | ... | ... | .../ ... | F555W | 4854 | 300 |
| 12 | NGC 2685 | 0.0414 | 78.6 | ... | ... | ... | ... | ... | ... | ... | .../ ... | 085030101 | 8.6 | 0.002/ 0.002 | ... | ... | ... |
| 13 | UGC 4881 | 0.0153 | 761.6 | 6857 | 14.6 | 09 15 55.556 | +44 19 57.90 | 2421.1 | 3.179 | 0.554 | 0.292/ 0.309 | 203390401 | 19.0 | <0.001/ 0.001 | F814W | 10592 | 760 |
| 14 | 3C 218 | 0.0494 | 1057.1 | 4969 | 79.9 | 09 18 05.675 | -12 05 44.06 | 2323.5 | 2.198 | 0.672 | 3.212/ 3.488 | 109980301 | 24.0 | 3.787/ 3.960 | ... | ... | ... |
| 15 | NGC 2787 | 0.0432 | 36.3 | 4689 | 29.4 | 09 19 18.570 | +69 12 11.59 | 95.8 | 2.639 | 0.388 | 9.968/ 0.132 | 200250101 | 32.9 | 0.018/ 0.019 | F814W | 6633 | 365 |
| 16 | NGC 2841 | 0.0145 | 58.0 | 6096 | 26.7 | 09 22 02.687 | +50 58 35.89 | 99.4 | 1.714 | 0.267 | 0.726/ 0.733 | 201440101 | 29.9 | 0.054/ 0.054 | F814W | 9788 | 120 |
| 17 | UGC 05101 | 0.0267 | 762.6 | 2033 | 48.2 | 09 35 51.676 | +61 21 11.06 | 2251.2 | 2.952 | 1.322 | 0.763/ 0.775 | 085640201 | 26.5 | 0.003/ 0.004 | F814W | 6346 | 400 |
| 18 | NGC 3185 | 0.0212 | 103.2 | ... | ... | ... | ... | ... | ... | ... | .../ ... | 112552001 | 8.4 | 0.003/ 0.003 | F814W | 9042 | 160 |
| 19 | NGC 3226 | 0.0214 | 114.2 | ... | ... | ... | ... | ... | ... | ... | .../ ... | 101040301 | 30.9 | 0.062/ 0.063 | ... | ... | ... |
| 20 | NGC 3245 | 0.0208 | 101.3 | 2926 | 9.6 | 10 27 18.339 | +28 30 27.21 | 199.4 | 1.968 | 1.031 | 0.610/ 0.617 | ... | ... | .../ ... | F702W | 7403 | 140 |
| 21 | NGC 3379 | 0.0275 | 51.2 | 1587 | 31.3 | 10 47 49.611 | +12 34 53.75 | 100.8 | 1.968 | 0.367 | 0.329/ 0.333 | ... | ... | .../ ... | F814W | 5512 | 335 |
| 22 | NGC 3414 | 0.0212 | 122.3 | 6779 | 13.5 | 10 51 16.216 | +27 58 30.14 | 240.7 | 1.968 | 0.133 | 0.821/ 0.701 | ... | ... | .../ ... | ... | ... | ... |
| 23 | NGC 3507 | 0.0163 | 95.8 | 3149 | 38.0 | 11 03 25.313 | +18 08 07.32 | 235.7 | 2.46 | 0.845 | 0.655/ 0.660 | ... | ... | .../ ... | F606W | 5446 | 80 |
| 24 | NGC 3607 | 0.0148 | 110.5 | 2073 | 38.4 | 11 16 54.612 | +18 03 04.39 | 326.2 | 2.952 | 2.119 | 0.307/ 0.313 | ... | ... | .../ ... | F814W | 5999 | 160 |
| 25 | NGC 3608 | 0.0149 | 111.1 | 2073 | 38.4 | 11 16 58.891 | +18 08 53.71 | 218.6 | 1.968 | 1.840 | 0.078/ 0.094 | ... | ... | .../ ... | F814W | 5454 | 230 |
| 26 | NGC 3623 | 0.0216 | 35.3 | ... | ... | ... | ... | ... | ... | ... | .../ ... | 082140301 | 28.9 | 0.017/ 0.005 | F814W | 9042 | 230 |
| 27 | NGC 3627 | 0.0243 | 49.8 | ... | ... | ... | ... | ... | ... | ... | .../ ... | 093641101 | 5.3 | 0.033/ 0.006 | F814W | 8602 | 350 |
| 28 | NGC 3628 | 0.0222 | 37.5 | 2039 | 54.7 | 11 20 16.919 | +13 35 22.75 | 110.7 | 2.952 | 1.381 | 0.159/ 0.162 | 110980101 | 41.6 | 0.028/ 0.029 | F606W | 5446 | 80 |
| 29 | NGC 3690B | 0.0098 | 201.7 | 1641 | 24.3 | 11 28 30.946 | +58 33 40.46 | 645.0 | 3.198 | 1.177 | 1.760/ 2.183 | 112810101 | 15.5 | 0.378/ 0.053 | F814W | 8602 | 350 |
| 30 | NGC 3898 | 0.0108 | 106.6 | 4740 | 56.9 | 11 49 15.353 | +56 05 03.62 | 494.4 | 4.638 | 1.364 | 0.797/ 1.050 | ... | ... | .../ ... | F814W | 9042 | 230 |
| 31 | NGC 3945 | 0.0166 | 109.0 | 6780 | 13.8 | 11 53 13.647 | +60 40 32.17 | 254.6 | 2.336 | 0.652 | 0.602/ 0.008 | ... | ... | .../ ... | F814W | 6633 | 125 |
| 32 | NGC 3998 | 0.0122 | 68.5 | 6781 | 13.6 | 11 57 56.123 | +55 27 13.37 | 123.9 | 1.809 | 0.577 | 19.463/ 19.717 | 090020101 | 9.0 | 5.226/ 5.242 | F658N | 5924 | 553 |
| 33 | NGC 4036 | 0.0189 | 119.0 | 6783 | 13.7 | 12 01 26.787 | +61 53 44.33 | 189.9 | 1.596 | 0.818 | 0.229/ 0.039 | ... | ... | .../ ... | F702W | 6785 | 400 |
| 34 | NGC 4111 | 0.0140 | 72.7 | 1578 | 14.8 | 12 07 03.142 | +43 03 57.34 | 214.6 | 2.952 | 1.000 | 1.747/ 1.760 | ... | ... | .../ ... | F702W | 6785 | 600 |
| 35 | NGC 4125 | 0.0184 | 115.8 | 2071 | 61.9 | 12 08 05.990 | +65 10 27.60 | 284.9 | 2.46 | 1.703 | 0.672/ 0.061 | 141570201 | 35.3 | 0.071/ 0.071 | F814W | 6587 | 700 |
| 36 | IRAS 12112+0305 | 0.0179 | 1411.4 | ... | ... | ... | ... | ... | ... | ... | .../ ... | 081340801 | 18.0 | <0.001/<0.001 | ... | ... | ... |
| 37 | NGC 4261 | 0.0155 | 153.3 | 834 | 31.3 | 12 19 23.235 | +05 49 30.39 | 528.0 | 3.444 | 0.897 | 5.238/ 5.266 | 056340101 | 27.0 | 0.214/ 0.210 | F702W | 5476 | 140 |
| 38 | NGC 4278 | 0.0177 | 77.9 | 7077 | 11.0 | 12 20 06.850 | +29 16 50.64 | 216.3 | 2.777 | 0.450 | 9.207/ 9.432 | 205010101 | 30.5 | 1.129/ 1.117 | F814W | 5454 | 230 |
| 39 | NGC 4314 | 0.0178 | 46.9 | 2062 | 14.5 | 12 22 32.122 | +29 53 44.76 | 276.9 | 5.904 | 2.076 | 0.563/ 0.570 | 201690301 | 22.3 | 0.009/ 0.009 | F814W | 6265 | 300 |
| 40 | NGC 4321 | 0.0239 | 78.2 | 6727 | 37.6 | 12 22 54.820 | +15 49 17.34 | 384.7 | 4.92 | 3.387 | 1.967/ 4.035 | 106860201 | 26.9 | 0.106/ 0.106 | F814W | 9776 | 100 |
| 41 | NGC 4374 | 0.0260 | 89.1 | 0803 | 28.2 | 12 25 03.719 | +12 53 13.24 | 131.5 | 1.476 | 0.386 | 2.317/ 2.423 | ... | ... | .../ ... | F814W | 6094 | 260 |
| 42 | NGC 4410A | 0.0170 | 486.3 | 2982 | 34.8 | 12 26 28.236 | +09 01 10.81 | 957.0 | 1.968 | 1.079 | 2.472/ 2.489 | ... | ... | .../ ... | F606W | 5479 | 500 |
| 43 | NGC 4438 | 0.0266 | 81.6 | 2883 | 24.6 | 12 27 45.592 | +13 00 33.01 | 240.9 | 2.952 | 1.786 | 2.113/ 9.916 | ... | ... | .../ ... | F814W | 6791 | 350 |
| 44 | NGC 4457 | 0.0182 | 84.3 | 3150 | 36.1 | 12 28 59.039 | +03 34 14.29 | 248.9 | 2.952 | 0.453 | 2.083/ 1.969 | ... | ... | .../ ... | ... | ... | ... |
| 45 | NGC 4459 | 0.0267 | 78.2 | 2927 | 9.8 | 12 29 00.038 | +13 58 41.73 | 153.9 | 1.968 | 1.046 | 0.246/ 0.249 | ... | ... | .../ ... | F814W | 5999 | 160 |
| 46 | NGC 4486 | 0.0254 | 77.9 | 2707 | 98.7 | 12 30 49.423 | +12 23 28.31 | 191.6 | 2.46 | 0.439 | 26.338/ 26.013 | 200920101 | 72.9 | >10/ >10 | F814W | 6775 | 246 |
| 47 | NGC 4494 | 0.0152 | 82.7 | 2079 | 12.7 | 12 31 24.098 | +25 46 29.80 | 162.8 | 1.968 | 0.913 | 26.771/ 0.054 | 071340301 | 24.5 | 0.015/ 0.015 | F814W | 6554 | 900 |
| 48 | NGC 4552 | 0.0257 | 74.4 | 2072 | 54.3 | 12 35 39.813 | +12 33 23.35 | 183.0 | 2.46 | 0.689 | 3.231/ 3.267 | 141570101 | 32.1 | 0.365/ 0.362 | F814W | 6099 | 500 |

**Table 2.** Continuation

| | | | | Chandra data | | | | | | | | XMM-Newton data | | | HST data | | |
|---|---|---|---|---|---|---|---|---|---|---|---|---|---|---|---|---|---|
| Num | Name | NH(Gal) (cm$^{-2}$) | Scale (pc/") | ObsID | Expt. (ks) | RA | Dec | Radii (pc) | Radii (arcs) | Offset (arcs) | Pileup** (%) | ObsID | Expt. (ksec) | Pileup** (%) | Filter | ObsID | Expt. (sec) |
| (1) | (2) | (3) | (4) | (5) | (6) | (7) | (8) | (9) | (10) | (11) | (12) | (13) | (14) | (15) | (16) | (17) | (18) |
| 49 | NGC 4589 | 0.0199 | 106.6 | 6785 | 13.5 | 12 37 24.832 | +74 11 30.62 | 251.6 | 2.360 | 2.118 | 0.057/ 0.051 | ... | ... | ... / ... | F814W | 5454 | 230 |
| 50 | NGC 4579 | 0.0247 | 81.6 | 0807 | 29.8 | 12 37 43.498 | +11 49 05.49 | 160.6 | 1.968 | 0.363 | 15.607/ 15.803 | 112840101 | 19.6 | 1.652/ 5.148 | F791W | 6436 | 300 |
| 51 | NGC 4596 | 0.0198 | 81.6 | 2928 | 9.2 | 12 39 55.971 | +10 10 35.04 | 301.1 | 3.69 | 0.698 | 0.287/ 0.291 | ... | ... | ... / ... | F606W | 5446 | 80 |
| 52 | NGC 4594 | 0.0377 | 47.4 | 1586 | 18.5 | 12 39 59.447 | -11 37 23.05 | 70.0 | 1.476 | 0.497 | 12.404/ 12.459 | 084030101 | 15.8 | 0.338/ 0.341 | F814W | 5512 | 245 |
| 53 | NGC 4636 | 0.0181 | 71.1 | 4415 | 73.8 | 12 42 49.822 | +02 41 15.87 | 909.4 | 12.79 | 0.657 | 1.869/ 2.733 | 111190201 | 16.4 | 1.523/<0.001 | F814W | 8686 | 100 |
| 54 | NGC 4676A | 0.0128 | 426.6 | 2043 | 27.9 | 12 46 10.118 | +30 43 55.61 | 1679.1 | 3.936 | 1.018 | 0.256/ 0.266 | ... | ... | ... / ... | F814W | 8669 | 160 |
| 55 | NGC 4676B | 0.0128 | 426.6 | 2043 | 27.9 | 12 46 11.147 | +30 43 23.13 | 1679.1 | 3.936 | 2.044 | 0.495/ 0.505 | ... | ... | ... / ... | F814W | 8669 | 160 |
| 56 | NGC 4698 | 0.0187 | 81.6 | 3008 | 29.4 | 12 48 22.908 | +08 29 14.63 | 180.7 | 2.214 | 0.243 | 0.312/ 0.311 | 112551101 | 10.3 | 0.004/ 0.004 | F814W | 9042 | 230 |
| 57 | NGC 4696 | 0.0806 | 172.0 | 1560 | 54.6 | 12 48 48.830 | -41 18 43.30 | 677.0 | 3.936 | 7.450 | 0.652/ 0.558 | 046340101 | 40.2 | 1.482/ >10 | F814W | 8683 | 500 |
| 58 | NGC 4736 | 0.0144 | 25.2 | 808 | 46.3 | 12 50 53.120 | +41 07 13.23 | 37.2 | 1.476 | 0.748 | 1.345/ 1.349 | 094360601 | 16.8 | 0.750/ 0.750 | F814W | 9042 | 230 |
| 59 | NGC 5005 | 0.0108 | 103.2 | ... | ... | ... | ... | ... | ... | ... | .../... | 110930501 | 8.5 | 0.127/ 0.127 | F814W | 9042 | 230 |
| 60 | NGC 5055 | 0.0132 | 34.6 | 2197 | 27.7 | 13 15 49.244 | +42 01 45.86 | 51.1 | 1.476 | 0.440 | 1.787/ 0.028 | ... | ... | ... / ... | F814W | 9042 | 230 |
| 61 | MRK 266NE | 0.0168 | 543.0 | 2044 | 17.4 | 13 38 17.881 | +48 16 40.82 | 1068.6 | 1.968 | 1.216 | 1.856/ 1.902 | 055990501 | 16.2 | ... / ... | F606W | 5479 | 500 |
| 62 | UGC 08696 | 0.0109 | 731.6 | 809 | 41.5 | 13 44 42.112 | +55 53 13.13 | 1439.8 | 1.968 | 0.453 | 2.535/ 2.655 | 101640401 | 18.0 | 0.032/ 0.030 | F814W | 6346 | 400 |
| 63 | CGCG 162-010 | 0.0119 | 1224.6 | 493 | 19.6 | 13 48 52.461 | +26 35 36.26 | 4820.0 | 3.936 | 2.333 | 4.118/ 3.478 | 097820101 | 42.3 | 5.514/ 5.695 | F814W | 7281 | 500 |
| 64 | NGC 5363 | 0.0209 | 108.5 | ... | ... | ... | ... | ... | ... | ... | .../... | 201670201 | 19.3 | 0.033/ 0.034 | ... | ... | ... |
| 65 | IC 4395 | 0.0155 | 719.9 | ... | ... | ... | ... | ... | ... | ... | .../... | 150480401 | 18.3 | 0.007/ 0.007 | ... | ... | ... |
| 66 | IRAS 14348-1447 | 0.0783 | 1598.2 | 6861 | 14.7 | 14 37 38.324 | -15 00 24.88 | 3711.0 | 2.322 | 0.992 | 0.228/ 0.251 | 081341401 | 18.0 | 0.002/ 0.002 | F814W | 6346 | 400 |
| 67 | NGC 5746 | 0.0327 | 142.4 | 3929 | 36.8 | 14 44 55.995 | +01 57 18.11 | 210.2 | 1.476 | 0.549 | 1.465/ 1.481 | ... | ... | ... / ... | F814W | 9046 | 400 |
| 68 | NGC 5813 | 0.0421 | 156.2 | 5907 | 48.4 | 15 01 11.195 | +01 42 07.24 | 646.0 | 4.136 | 0.553 | 15.795/ 0.041 | 302460101 | 28.1 | 0.654/<0.001 | F814W | 5454 | 230 |
| 69 | NGC 5838 | 0.0419 | 137.9 | 6788 | 13.6 | 15 05 26.268 | +02 05 57.92 | 286.3 | 2.076 | 0.402 | 0.061/ 0.049 | ... | ... | ... / ... | F814W | 7450 | 140 |
| 70 | NGC 5846 | 0.0426 | 120.7 | 4009 | 23.0 | 15 06 29.190 | +01 36 19.65 | 593.6 | 4.92 | 1.283 | 1.437/ 0.190 | 021540101 | 25.2 | 0.546/ 0.541 | F814W | 5920 | 150 |
| 71 | NGC 5866 | 0.0146 | 74.4 | 2879 | 31.9 | 15 06 29.477 | +55 45 46.08 | 329.4 | 4.428 | 1.563 | 0.235/ 0.237 | ... | ... | ... / ... | ... | ... | ... |
| 72 | MRK 0848 | 0.0190 | 797.4 | 6858 | 14.5 | 15 18 06.109 | +42 44 44.94 | 2019.0 | 2.532 | 8.980 | 0.835/ 0.870 | 203390601 | 19.8 | 0.004/ 0.004 | F814W | 10592 | 760 |
| 73 | NGC 6251 | 0.0549 | 445.5 | 847 | 25.4 | 16 32 31.850 | +82 32 15.67 | 876.7 | 1.968 | 1.948 | 9.701/ 9.886 | 056340201 | 41.1 | 1.287/ 1.314 | F814W | 6653 | 500 |
| 74 | NGC 6240 | 0.0578 | 474.1 | 1590 | 36.7 | 16 52 58.905 | +02 24 03.27 | 1516.2 | 3.198 | 0.455 | 9.609/ 9.998 | 147420201 | 20.0 | 0.230/ 0.263 | F814W | 6430 | 400 |
| 75 | IRAS 17208-0014 | 0.0999 | 829.0 | 2035 | 48.4 | 17 23 22.008 | -00 17 00.64 | 2039.3 | 2.46 | 0.682 | 0.466/ 0.507 | 081340601 | 13.9 | <0.001/<0.001 | F814W | 6346 | 400 |
| 76 | NGC 6482 | 0.0788 | 255.0 | 3218 | 19.1 | 17 51 48.831 | +23 04 18.92 | 627.3 | 2.46 | 0.000 | 2.673/ 2.742 | 304160401 | 6.7 | 0.329/ 0.336 | ... | ... | ... |
| 77 | NGC 7130 | 0.0198 | 312.7 | 2188 | 36.9 | 21 48 19.532 | -34 57 05.04 | 615.4 | 1.968 | 0.431 | 3.439/ 2.221 | ... | ... | ... / ... | F606W | 5479 | 500 |
| 78 | NGC 7285 | 0.0179 | 279.0 | ... | 29.5 | ... | ... | ... | ... | ... | .../... | 206490301 | 17.2 | 0.019/ 0.021 | ... | ... | ... |
| 79 | NGC 7331 | 0.0861 | 63.6 | 2198 | ... | 22 37 04.027 | +34 24 55.96 | 125.2 | 1.968 | 0.945 | 0.208/ 0.217 | 103861301 | 0.8 | 0.016/ 0.012 | F814W | 7450 | 185 |
| 80 | IC 1459 | 0.0118 | 141.8 | 2196 | 53.5 | 22 57 10.604 | -36 27 43.57 | 418.6 | 2.952 | 0.604 | 7.933/ 7.996 | 135980201 | 26.9 | 0.202/ 0.205 | F814W | 5454 | 230 |
| 81 | NPM1G -12.0625 | 0.0249 | 1589.8 | 7329 | 58.9 | 23 25 19.726 | -12 07 27.11 | 2953.8 | 1.858 | 0.153 | 1.219/ 4.865 | 147330101 | 89.6 | 3.147/ 3.433 | F702W | 6228 | 1050 |
| 82 | NGC 7743 | 0.0525 | 100.4 | ... | ... | ... | ... | ... | ... | ... | .../... | 301651001 | 9.6 | 0.005/ 0.006 | F606W | 8597 | 280 |

(*)Column density NH(Galactic) in units of 10$^{22}$ cm$^{-2}$. (**) Pileup estimated in (0.5-2 keV)/(2-10 keV) ranges.

**Table 3.** F-test applied to the *Chandra* data fits.

| Num | Name | (2PL) vs(PL) | | (MEPL) vs(ME) | | (MEPL) vs(PL) | | (ME2PL) vs(MEPL) | | (ME2PL) vs(2PL) | | Class. |
|---|---|---|---|---|---|---|---|---|---|---|---|---|
| (1) | (2) | (3) | | (4) | | (5) | | (6) | | (7) | | (8) |
| 1 | NGC 0315 | 3.40e-02 | | 2.20e-16 | U | 3.20e-18 | | 2.00e-02 | | 9.90e-19 | | ME2PL |
| 6 | NGC 0833 | 9.00e-05 | U | 9.00e-09 | U | 1.60e-07 | U | 1.40e-01 | U | 1.40e-04 | | ME2PL |
| 7 | NGC 0835 | 3.80e-04 | U | 7.20e-04 | U | 3.80e-04 | U | 9.60e-03 | U | 9.20e-03 | U | ME2PL |
| 8 | NGC 1052 | 1.90e-02 | | 5.60e-07 | | 1.10e-02 | | 1.90e-01 | | 8.00e-02 | | ME2PL |
| 11 | NGC 2681 | 1.50e-09 | U | 1.90e-24 | U | 1.20e-17 | U | 1.00e+00 | | 6.50e-09 | U | MEPL |
| 14 | 3C 218 | 3.60e-04 | | 8.80e-06 | U | 6.60e-06 | | 2.90e-04 | | 6.00e-06 | | ME2PL |
| 15 | NGC 2787 | 9.01e-01 | | 2.80e-01 | | 6.92e-01 | | 1.00e+00 | U | 5.89e-01 | U | PL |
| 17 | UGC 05101 | 2.00e-02 | | 3.10e-01 | U | 3.50e-03 | U | 1.00e+00 | | 6.20e-02 | | ME2PL |
| 22 | NGC 3414 | 7.60e-01 | U | 2.00e-01 | | 4.15e-01 | | 1.00e+00 | U | 7.87e-01 | U | PL |
| 23 | NGC 3507 | 9.50e-01 | U | 9.70e-01 | | 1.00e-02 | U | 1.90e-02 | U | 2.80e-04 | U | ME |
| 29 | NGC 3690B | 1.75e-13 | | 3.70e-13 | U | 4.30e-11 | | 1.80e-04 | | 3.20e-02 | | ME2PL |
| 30 | NGC 3898 | 1.80e-02 | U | 2.00e-04 | U | 4.70e-03 | | 1.00e+00 | U | 2.10e-01 | U | MEPL |
| 31 | NGC 3945 | 1.27e-01 | | 1.70e-08 | U | 5.92e-01 | U | 1.00e+00 | U | 1.00e+00 | U | PL |
| 32 | NGC 3998 | 0.00e+00 | | 2.20e-10 | U | 1.20e-04 | | 2.15e-05 | | 3.79e-09 | | ME2PL |
| 34 | NGC 4111 | 1.60e-03 | U | 8.80e-05 | U | 5.60e-04 | U | 5.47e-05 | U | 2.77e-05 | | ME2PL |
| 35 | NGC 4125 | 1.60e-01 | U | 3.00e-02 | | 3.60e-02 | | 1.00e+00 | | 1.20e-01 | | MEPL |
| 37 | NGC 4261 | 0.00e+00 | U | 5.10e-37 | U | 5.50e-67 | U | 1.81e-05 | | 0.00e+00 | U | ME2PL |
| 38 | NGC 4278 | 2.43e-08 | U | 6.00e-79 | U | 3.00e-10 | | 1.50e-01 | | 7.70e-04 | U | ME2PL |
| 40 | NGC 4321 | 8.96e-05 | U | 2.00e-09 | U | 2.60e-04 | | 2.90e-02 | U | 7.10e-02 | U | MEPL |
| 41 | NGC 4374 | 2.50e-01 | | 2.00e-05 | U | 7.60e-04 | | 1.00e+00 | U | 9.30e-04 | U | MEPL |
| 42 | NGC 4410A | 1.70e-02 | | 5.50e-07 | | 1.93e-01 | | 3.15e-02 | | 5.25e-01 | U | ME2PL |
| 43 | NGC 4438 | 5.17e-06 | U | 1.90e-06 | | 6.30e-18 | | 1.00e+00 | U | 2.20e-11 | U | MEPL |
| 44 | NGC 4457 | 5.55e-06 | U | 4.10e-15 | U | 4.10e-11 | U | 4.20e-01 | U | 1.10e-06 | U | MEPL |
| 46 | NGC 4486 | 1.80e-38 | U | 2.50e-18 | U | 2.60e-16 | U | 7.20e-26 | U | 1.40e-03 | | ME2PL |
| 47 | NGC 4494 | 8.70e-03 | | 1.70e-02 | | 2.10e-03 | | 2.60e-01 | | 4.20e-02 | U | MEPL |
| 48 | NGC 4552 | 2.60e-05 | U | 5.70e-21 | U | 1.10e-09 | | 1.00e+00 | | 8.53e-06 | U | MEPL |
| 50 | NGC 4579 | 9.90e-01 | U | 2.00e-07 | | 4.50e-08 | | 1.00e+00 | U | 1.50e-05 | U | MEPL |
| 52 | NGC 4594 | 3.60e-01 | | 9.39e-01 | U | 6.20e-01 | U | 1.20e-01 | U | 3.48e-01 | | PL |
| 57 | NGC 4696 | 2.20e-03 | | 2.50e-02 | | 8.70e-05 | | 1.00e+00 | | 1.10e-02 | | MEPL |
| 58 | NGC 4736 | 0.00e+00 | U | 2.30e-31 | U | 0.00e+00 | | 1.00e+00 | | 1.47e-05 | U | MEPL |
| 60 | NGC 5055 | 7.10e-01 | | 9.80e-02 | | 3.10e-01 | | 1.00e+00 | U | 4.53e-01 | U | PL |
| 61 | MRK 266NE | 2.55e-05 | | 4.30e-08 | U | 7.10e-08 | | 3.00e-01 | | 3.50e-04 | | ME2PL |
| 62 | UGC 08696 | 2.52e-33 | U | 4.50e-30 | U | 1.20e-22 | U | 6.10e-13 | | 2.20e-02 | | ME2PL |
| 63 | CGCG 162-010 | 7.20e-01 | U | 9.65e-01 | U | 1.74e-01 | U | 2.90e-04 | U | 9.90e-06 | U | ME |
| 67 | NGC 5746 | 8.20e-01 | U | 9.62e-01 | U | 6.51e-01 | U | 1.00e+00 | U | 1.00e+00 | U | PL |
| 68 | NGC 5813 | 0.00e+00 | U | 1.60e-05 | U | 2.10e-08 | U | 3.60e-01 | | 4.50e-04 | U | MEPL |
| 70 | NGC 5846 | 9.30e-01 | U | 1.50e-06 | | 6.70e-18 | U | 1.00e+00 | U | 1.00e+00 | U | MEPL |
| 73 | NGC 6251 | 9.64e-16 | | 8.90e-19 | U | 2.00e-12 | | 2.10e-04 | | 3.10e-01 | | MEPL |
| 74 | NGC 6240 | 4.80e+00 | | 4.10e-17 | U | 0.00e+00 | U | 8.20e-11 | U | 4.02e-06 | | ME2PL |
| 75 | IRAS 17208-0014 | 6.80e-01 | | 4.60e-02 | U | 6.10e-02 | | 1.00e+00 | | 2.53e-01 | | PL |
| 76 | NGC 6482 | 1.00e+00 | U | 5.68e-01 | | 2.20e-09 | U | 1.00e+00 | U | 2.90e-08 | U | ME |
| 77 | NGC 7130 | 7.73e-07 | U | 1.10e-32 | U | 9.70e-19 | U | 1.60e-02 | | 1.10e-14 | U | ME2PL |
| 80 | IC 1459 | 2.30e-05 | | 5.70e-35 | U | 7.70e-10 | | 9.85e-04 | | 2.25e-07 | | ME2PL |
| 81 | NPM1G -12.0625 | 1.00e+00 | U | 1.50e-10 | | 3.20e-04 | U | 1.00e+00 | U | 5.40e-04 | U | MEPL |

**Table 4.** F-test applied to the *XMM-Newton* data fits.

| Num | Name | (2PL) vs(PL) | | (MEPL) vs(ME) | | (MEPL) vs(PL) | | (ME2PL) vs(MEPL) | | (ME2PL) vs(2PL) | | Class. |
|---|---|---|---|---|---|---|---|---|---|---|---|---|
| (1) | (2) | (3) | | (4) | | (5) | | (6) | | (7) | | (8) |
| 1 | NGC 0315 | 9.80e-27 | U | 2.20e-150 | U | 3.30e-132 | U | 4.33e-04 | | 0.00e+00 | U | ME2PL |
| 2 | NGC 0410 | 9.99e-01 | U | 2.30e-07 | | 2.20e-61 | | 1.00e+00 | U | 0.00e+00 | U | MEPL |
| 7 | NGC 0835 | 8.70e-04 | U | 6.30e-11 | U | 1.20e-08 | U | 3.37e-06 | | 6.03e-11 | U | ME2PL |
| 8 | NGC 1052 | 2.30e-175 | U | 5.10e-26 | U | 2.60e-86 | U | 1.24e-121 | | 1.07e-32 | U | ME2PL |
| 9 | NGC 2639 | 5.22e-01 | U | 1.00e+00 | U | 1.00e+00 | U | 1.57e-03 | U | 1.43e-02 | U | ME |
| 10 | NGC 2655 | 1.40e-17 | U | 4.20e-34 | U | 7.10e-24 | U | 6.14e-07 | | 2.28e-13 | U | ME2PL |
| 13 | UGC 4881 | 6.43e-01 | U | 9.60e-01 | U | 6.13e-01 | U | 1.20e-01 | U | 1.85e-01 | U | ME |
| 14 | 3C 218 | 1.00e+00 | U | 1.70e-135 | U | 1.60e-94 | U | 1.00e+00 | | 0.00e+00 | U | MEPL |
| 15 | NGC 2787 | 2.30e-02 | | 4.50e-10 | | 3.00e-03 | | 3.95e-02 | | 5.11e-03 | | ME2PL |
| 16 | NGC 2841 | 5.30e-07 | U | 8.30e-29 | U | 2.40e-10 | U | 1.47e-02 | | 4.18e-06 | | ME2PL |
| 17 | UGC 05101 | 3.60e-03 | U | 3.60e-08 | U | 2.00e-03 | U | 1.08e-01 | U | 5.10e-02 | U | ME2PL |
| 19 | NGC 3226 | 8.90e-10 | | 3.20e-06 | U | 4.60e-02 | U | 9.85e-09 | U | 1.00e+00 | U | 2PL |
| 28 | NGC 3628 | 5.42e-01 | U | 7.10e-03 | U | 7.32e-01 | U | 2.88e-02 | U | 8.84e-02 | U | PL |
| 29 | NGC 3690B | 8.30e-19 | U | 3.20e-75 | U | 5.99e-55 | U | 1.00e+00 | | 6.97e-37 | U | MEPL |
| 32 | NGC 3998 | 2.61e-01 | | 1.80e-193 | U | 5.46e-01 | | 2.59e-02 | | 1.10e-01 | | PL |
| 35 | NGC 4125 | 9.86e-01 | U | 1.10e-22 | | 1.07e+00 | U | 7.96e-05 | | 0.00e+00 | U | MEPL |
| 37 | NGC 4261 | 3.40e-23 | U | 2.30e-115 | U | 1.06e-82 | U | 1.02e-30 | U | 0.00e+00 | U | ME2PL |
| 38 | NGC 4278 | 6.50e-05 | | 1.20e-198 | U | 3.70e-04 | | 5.06e-02 | | 2.31e-01 | | MEPL |
| 39 | NGC 4314 | 9.10e-04 | U | 8.10e-17 | U | 3.16e-08 | U | 2.07e-02 | | 6.37e-07 | U | MEPL |
| 40 | NGC 4321 | 3.00e-12 | U | 1.20e-56 | U | 9.98e-18 | | 8.82e-02 | | 1.09e-07 | U | MEPL |
| 47 | NGC 4494 | 1.35e-01 | | 1.30e-19 | U | 8.97e-01 | | 8.08e-03 | U | 1.47e-01 | | PL |
| 48 | NGC 4552 | 3.00e-33 | U | 9.20e-236 | U | 1.39e-125 | U | 3.43e-01 | | 0.00e+00 | U | MEPL |
| 50 | NGC 4579 | 8.00e-24 | | 1.70e-223 | U | 1.62e-42 | | 3.41e-06 | | 3.92e-25 | | ME2PL |
| 52 | NGC 4594 | 1.00e-02 | | 6.60e-34 | U | 1.00e+00 | U | 2.35e-09 | | 1.69e-06 | | ME2PL |
| 53 | NGC 4636 | 1.00e+00 | U | 1.60e-19 | U | 2.42e-110 | U | 1.00e+00 | U | 0.00e+00 | U | MEPL |
| 58 | NGC 4736 | 3.30e-49 | U | 6.50e-122 | U | 3.29e-68 | U | 5.96e-02 | | 6.83e-21 | U | MEPL |
| 59 | NGC 5005 | 1.00e-10 | U | 6.90e-24 | U | 3.09e-18 | U | 3.14e-01 | | 2.95e-09 | U | MEPL |
| 61 | MRK 266NE | 2.10e-10 | U | 8.30e-37 | U | 6.24e-18 | U | 1.17e-07 | | 3.56e-15 | U | ME2PL |
| 62 | UGC 08696 | 1.30e-09 | U | 2.50e-13 | U | 1.23e-09 | U | 9.40e-09 | U | 8.68e-09 | U | ME2PL |

**Table 4.** Continuation

| Num | Name | (2PL) vs(PL) | | (MEPL) vs(ME) | | (MEPL) vs(PL) | | (ME2PL) vs(MEPL) | | (ME2PL) vs(2PL) | | Class. |
|---|---|---|---|---|---|---|---|---|---|---|---|---|
| (1) | (2) | (3) | | (4) | | (5) | | (6) | | (7) | | (8) |
| 63 | CGCG 162-010.. | 3.00e-32 | U | 8.00e-221 | U | 7.99e-193 | U | 1.14e-01 | | 0.00e+00 | U | MEPL |
| 64 | NGC 5363 ...... | 9.00e-07 | U | 2.80e-20 | U | 7.57e-14 | U | 4.74e-03 | | 2.97e-10 | U | ME2PL |
| 65 | IC 4395 ......... | 1.00e+00 | U | 2.90e-03 | U | 4.15e-03 | U | 1.02e-01 | U | 3.47e-07 | U | MEPL |
| 66 | IRAS 14348-1447 | 3.27e-01 | U | 5.40e-03 | U | 4.21e-01 | U | 5.65e-01 | U | 6.30e-01 | U | ME |
| 68 | NGC 5813 ...... | 9.70e-03 | U | 1.70e-09 | U | 4.20e-129 | U | 7.91e-01 | | 0.00e+00 | U | MEPL |
| 70 | NGC 5846 ...... | 2.80e-05 | U | 8.90e-32 | U | 7.60e-121 | U | 3.53e-02 | | 0.00e+00 | U | ME2PL |
| 72 | MRK 0848 ...... | 2.73e-01 | | 1.30e-06 | U | 5.67e-01 | | 7.89e-03 | U | 3.63e-02 | U | PL |
| 73 | NGC 6240 ...... | 8.50e-07 | U | 6.80e-56 | U | 1.00e+00 | | 4.26e-01 | | 4.04e-04 | U | ME2PL |
| 74 | NGC 6251 ...... | 5.20e-21 | U | 4.30e-237 | U | 1.00e+00 | U | 2.91e-10 | | 3.29e-24 | U | ME2PL |
| 75 | IRAS 17208-0014 | 2.46e-01 | U | 6.29e-01 | | 1.31e-01 | | 3.82e-01 | U | 1.74e-01 | U | ME |
| 76 | NGC 6482 ...... | 4.40e-03 | U | 1.80e-03 | U | 2.33e-35 | U | 7.24e-02 | U | 2.69e-34 | U | ME |
| 78 | NGC 7285 ...... | 6.50e-05 | | 1.40e-01 | | 1.80e-04 | | 4.30e-01 | | 5.01e-01 | | MEPL |
| 80 | IC 1459 ......... | 2.80e-18 | U | 2.20e-70 | U | 1.01e-32 | U | 1.00e+00 | U | 7.44e-16 | U | MEPL |
| 81 | NPM1G -12.0625 | 2.70e-34 | U | 2.10e-283 | U | 1.96e-225 | U | 4.72e-05 | | 0.00e+00 | U | ME2PL |
| 82 | NGC 7743 ...... | 1.71e-01 | U | 1.50e-03 | U | 5.35e-02 | U | 2.40e-01 | U | 6.84e-02 | U | MEPL |

**Table 5.** Observed fluxes and absorption corrected luminosities with *Chandra* data[*].

| Num | Name | Flux(0.5-2.0keV) ($10^{-13}$ erg s$^{-1}$ cm$^{-2}$) | Flux(2.0-10.0keV) ($10^{-13}$ erg s$^{-1}$ cm$^{-2}$) | Log(Lum(0.5-2.0keV)) | Log(Lum(2.0-10.0keV)) |
|---|---|---|---|---|---|
| (1) | (2) | (3) | (4) | (5) | (6) |
| 1 | NGC 0315 ........ | 1.52 [ 0.46 , 1.80] | 6.50 [ 0.23 , 7.76] | 42.0 [ 41.4 , 42.0] | 41.8 [ 40.9 , 41.8] |
| 3 | NGC 0474 ...... (*) | 0.01 [ 0.00 , 0.04] | 0.02 [ 0.00 , 0.06] | 38.3 [ 0.0 , 38.7] | 38.5 [ 0.0 , 38.9] |
| 4 | IIIZW 035 ...... (*) | 0.04 [ 0.01 , 0.08] | 0.07 [ 0.02 , 0.13] | 39.9 [ 39.2 , 40.1] | 40.0 [ 39.4 , 40.3] |
| 5 | NGC 0524 ...... (*) | 0.03 [ 0.00 , 0.06] | 0.05 [ 0.00 , 0.11] | 38.4 [ 36.1 , 38.7] | 38.6 [ 37.1 , 38.9] |
| 6 | NGC 0833 ........ | 0.11 [ 0.05 , 0.15] | 5.52 [ 0.42 , 6.42] | 41.7 [ 41.3 , 41.8] | 41.7 [ 40.6 , 41.8] |
| 7 | NGC 0835 ........ | 0.41 [ 0.20 , 0.52] | 1.37 [ 1.21 , 17.58] | 41.7 [ 41.3 , 41.8] | 41.4 [ 41.4 , 42.5] |
| 8 | NGC 1052 ........ | 1.03 [ 0.00 , 1.45] | 14.35 [ 2.23 , 18.07] | 41.7 [ 0.0 , 41.8] | 41.2 [ 40.4 , 41.3] |
| 11 | NGC 2681 ........ | 0.24 [ 0.07 , 0.26] | 1.85 [ 0.17 , 14.06] | 38.6 [ 38.1 , 38.6] | 39.3 [ 38.3 , 40.2] |
| 13 | UGC 4881 ...... (*) | 0.06 [ 0.01 , 0.10] | 0.09 [ 0.01 , 0.15] | 40.3 [ 39.7 , 40.5] | 40.4 [ 39.8 , 40.7] |
| 14 | 3C 218 ........... | 0.62 [ 0.41 , 0.73] | 1.46 [ 0.40 , 1.82] | 42.2 [ 42.0 , 42.2] | 42.1 [ 41.5 , 42.2] |
| 15 | NGC 2787 ........ | 0.45 [ 0.12 , 0.57] | 0.75 [ 0.00 ,13899.00] | 38.9 [ 38.3 , 39.0] | 38.8 [ 0.0 , 43.2] |
| 16 | NGC 2841 ...... (*) | 0.14 [ 0.10 , 0.18] | 0.21 [ 0.15 , 0.27] | 38.4 [ 38.2 , 38.5] | 38.6 [ 38.4 , 38.7] |
| 17 | UGC 05101 ....... | 0.12 [ 0.05 , 0.16] | 0.81 [ 0.33 , 1.17] | 41.8 [ 41.3 , 41.9] | 42.1 [ 41.7 , 42.2] |
| 20 | NGC 3245 ...... (*) | 0.12 [ 0.05 , 0.18] | 0.18 [ 0.09 , 0.28] | 38.8 [ 38.5 , 39.0] | 39.0 [ 38.6 , 39.2] |
| 21 | NGC 3379 ...... (*) | 0.06 [ 0.04 , 0.09] | 0.10 [ 0.06 , 0.14] | 38.0 [ 37.8 , 38.1] | 38.1 [ 37.9 , 38.3] |
| 22 | NGC 3414 ........ | 0.95 [ 0.61 , 1.17] | 1.63 [ 1.24 , 2.22] | 39.9 [ 39.8 , 40.0] | 39.9 [ 39.8 , 40.0] |
| 23 | NGC 3507 ........ | 0.18 [ 0.00 , 0.23] | 0.00 [ 0.00 , 0.00] | 39.1 [ 0.0 , 39.2] | 37.2 [ 0.0 , 37.4] |
| 24 | NGC 3607 ...... (*) | 0.06 [ 0.04 , 0.09] | 0.09 [ 0.05 , 0.14] | 38.6 [ 38.4 , 38.8] | 38.8 [ 38.5 , 38.9] |
| 25 | NGC 3608 ...... (*) | 0.02 [ 0.00 , 0.03] | 0.03 [ 0.00 , 0.05] | 38.0 [ 36.9 , 38.3] | 38.2 [ 37.1 , 38.5] |
| 28 | NGC 3628 ...... (*) | 0.03 [ 0.02 , 0.05] | 0.05 [ 0.03 , 0.07] | 37.4 [ 37.2 , 37.6] | 37.5 [ 37.3 , 37.7] |
| 29 | NGC 3690B ...... | 0.95 [ 0.00 , 1.42] | 1.29 [ 0.14 , 2.03] | 41.8 [ 0.0 , 42.0] | 40.9 [ 40.0 , 41.1] |
| 30 | NGC 3898 ........ | 0.19 [ 0.01 , 0.27] | 0.20 [ 0.00 ,10468.50] | 42.7 [ 41.5 , 42.9] | 38.8 [ 0.0 , 43.5] |
| 31 | NGC 3945 ........ | 0.79 [ 0.44 , 0.98] | 0.38 [ 0.27 , 0.49] | 39.8 [ 39.5 , 39.9] | 39.1 [ 39.0 , 39.2] |
| 32 | NGC 3998 ........ | 31.27 [ 25.73 , 34.34] | 82.46 [ 60.30 , 91.31] | 41.2 [ 41.1 , 41.2] | 41.3 [ 41.2 , 41.4] |
| 33 | NGC 4036 ...... (*) | 0.12 [ 0.07 , 0.17] | 0.18 [ 0.10 , 0.27] | 39.0 [ 38.7 , 39.1] | 39.1 [ 38.9 , 39.3] |
| 34 | NGC 4111 ........ | 0.59 [ 0.46 , 0.70] | 3.39 [ 0.02 , 104.04] | 40.9 [ 40.8 , 40.9] | 40.4 [ 0.0 , 41.9] |
| 35 | NGC 4125 ........ | 0.07 [ 0.00 , 0.09] | 0.07 [ 0.00 , 0.12] | 39.9 [ 0.0 , 40.0] | 38.7 [ 0.0 , 39.0] |
| 37 | NGC 4261 ........ | 1.88 [ 1.50 , 2.07] | 4.32 [ 0.08 , 5.24] | 41.3 [ 41.2 , 41.3] | 41.1 [ 39.4 , 41.2] |
| 38 | NGC 4278 ........ | 2.10 [ 1.52 , 2.38] | 1.71 [ 1.19 , 2.07] | 39.6 [ 39.4 , 39.6] | 39.2 [ 39.1 , 39.3] |
| 39 | NGC 4314 ...... (*) | 0.11 [ 0.06 , 0.16] | 0.17 [ 0.09 , 0.24] | 38.1 [ 37.8 , 38.3] | 38.3 [ 38.0 , 38.4] |
| 40 | NGC 4321 ........ | 0.55 [ 0.07 , 0.64] | 5.86 [ 3.48 , 45.58] | 40.4 [ 39.6 , 40.5] | 40.5 [ 40.3 , 41.4] |
| 41 | NGC 4374 ........ | 0.53 [ 0.07 , 0.61] | 0.81 [ 0.41 , 44.31] | 39.5 [ 38.6 , 39.5] | 39.5 [ 39.2 , 41.3] |
| 42 | NGC 4410A ...... | 0.75 [ 0.34 , 0.81] | 1.18 [ 0.82 , 1.43] | 41.2 [ 40.9 , 41.2] | 41.2 [ 41.0 , 41.3] |
| 43 | NGC 4438 ........ | 1.11 [ 0.31 , 1.26] | 0.32 [ 0.00 , 990.00] | 40.1 [ 39.4 , 40.1] | 39.0 [ 0.0 , 42.5] |
| 44 | NGC 4457 ........ | 0.37 [ 0.01 , 0.44] | 0.17 [ 0.12 , 0.24] | 39.6 [ 39.4 , 39.6] | 38.8 [ 38.6 , 38.9] |
| 45 | NGC 4459 ...... (*) | 0.05 [ 0.00 , 0.09] | 0.08 [ 0.01 , 0.14] | 38.2 [ 37.4 , 38.5] | 38.4 [ 37.7 , 38.6] |
| 46 | NGC 4486 ........ | 5.15 [ 4.59 , 5.52] | 16.43 [ 11.87 , 18.33] | 40.9 [ 40.9 , 41.0] | 40.8 [ 40.7 , 40.9] |
| 47 | NGC 4494 ........ | 0.42 [ 0.12 , 0.53] | 0.16 [ 0.04 , 0.30] | 39.8 [ 39.3 , 39.8] | 38.8 [ 38.1 , 39.0] |
| 48 | NGC 4552 ........ | 0.72 [ 0.42 , 0.81] | 0.63 [ 0.19 , 0.85] | 39.5 [ 39.3 , 39.6] | 39.2 [ 38.8 , 39.4] |
| 49 | NGC 4589 ...... (*) | 0.09 [ 0.05 , 0.14] | 0.14 [ 0.07 , 0.22] | 38.8 [ 38.5 , 38.9] | 38.9 [ 38.6 , 39.1] |
| 50 | NGC 4579 ........ | 12.38 [ 10.65 , 12.38] | 43.88 [ 41.56 , 45.85] | 40.9 [ 40.8 , 40.9] | 41.2 [ 41.1 , 41.2] |
| 51 | NGC 4596 ...... (*) | 0.06 [ 0.01 , 0.10] | 0.09 [ 0.02 , 0.15] | 38.3 [ 37.6 , 38.5] | 38.5 [ 37.8 , 38.7] |
| 52 | NGC 4594 ........ | 2.25 [ 1.90 , 2.53] | 8.02 [ 7.22 , 8.91] | 39.6 [ 39.6 , 39.7] | 40.0 [ 39.9 , 40.0] |
| 53 | NGC 4636 ........ | 0.41 [ 0.36 , 0.47] | 0.63 [ 0.55 , 0.72] | 39.1 [ 39.0 , 39.1] | 39.2 [ 39.2 , 39.3] |
| 54 | NGC 4676A ..... (*) | 0.05 [ 0.03 , 0.07] | 0.07 [ 0.04 , 0.11] | 39.7 [ 39.4 , 39.9] | 39.9 [ 39.6 , 40.0] |
| 55 | NGC 4676B ..... (*) | 0.09 [ 0.06 , 0.12] | 0.14 [ 0.09 , 0.19] | 40.0 [ 39.8 , 40.1] | 40.1 [ 40.0 , 40.3] |
| 56 | NGC 4698 ...... (*) | 0.11 [ 0.07 , 0.14] | 0.16 [ 0.11 , 0.22] | 38.6 [ 38.4 , 38.7] | 38.7 [ 38.6 , 38.9] |
| 57 | NGC 4696 ........ | 2.44 [ 0.22 , 3.07] | 0.41 [ 0.10 , 0.63] | 41.6 [ 40.6 , 41.7] | 40.0 [ 39.3 , 40.2] |
| 58 | NGC 4736 ........ | 1.17 [ 0.83 , 1.29] | 1.32 [ 0.95 , 1.54] | 38.8 [ 38.6 , 38.8] | 38.6 [ 38.5 , 38.7] |
| 60 | NGC 5055 ........ | 0.18 [ 0.04 , 0.24] | 0.09 [ 0.04 , 0.13] | 38.6 [ 38.1 , 38.7] | 37.8 [ 37.4 , 37.9] |
| 61 | MRK 266NE ...... | 0.25 [ 0.11 , 0.33] | 2.15 [ 0.24 , 3.48] | 41.0 [ 40.7 , 41.1] | 41.7 [ 40.8 , 41.9] |
| 62 | UGC 08696 ....... | 0.25 [ 0.00 , 0.34] | 5.37 [ 0.29 , 6.20] | 43.2 [ 41.9 , 43.3] | 43.0 [ 41.7 , 43.1] |
| 63 | CGCG 162-010 .. | 0.87 [ 0.46 , 1.12] | 0.27 [ 0.18 , 0.36] | 42.0 [ 41.8 , 42.0] | 41.4 [ 41.3 , 41.5] |
| 66 | IRAS 14348-1447(*) | 0.05 [ 0.01 , 0.08] | 0.08 [ 0.03 , 0.14] | 40.9 [ 40.4 , 41.1] | 41.1 [ 40.5 , 41.3] |
| 67 | NGC 5746 ........ | 0.18 [ 0.04 , 0.24] | 1.53 [ 0.50 , 2.01] | 39.7 [ 39.0 , 39.8] | 40.2 [ 39.8 , 40.3] |
| 68 | NGC 5813 ........ | 0.61 [ 0.20 , 0.72] | 0.04 [ 0.01 , 0.07] | 40.4 [ 39.9 , 40.4] | 38.8 [ 38.3 , 39.0] |
| 69 | NGC 5838 ...... (*) | 0.10 [ 0.05 , 0.15] | 0.16 [ 0.08 , 0.24] | 39.0 [ 38.7 , 39.2] | 39.2 [ 38.9 , 39.4] |

**Table 5.** Continuation

| Num | Name | Flux(0.5-2.0keV) ($10^{-13}$ erg s$^{-1}$ cm$^{-2}$) | Flux(2.0-10.0keV) ($10^{-13}$ erg s$^{-1}$ cm$^{-2}$) | Log(Lum(0.5-2.0keV)) | Log(Lum(2.0-10.0keV)) |
|---|---|---|---|---|---|
| (1) | (2) | (3) | (4) | (5) | (6) |
| 70 | NGC 5846......... | 0.77 [ 0.05 , 0.90] | 9.44 [ 0.00 , 2306.31] | 40.2 [ 39.4 , 40.3] | 40.8 [ 0.0 , 43.2] |
| 71 | NGC 5866......(*) | 0.05 [ 0.03 , 0.07] | 0.07 [ 0.04 , 0.10] | 38.1 [ 37.9 , 38.3] | 38.3 [ 38.0 , 38.5] |
| 72 | MRK 0848......(*) | 0.16 [ 0.10 , 0.22] | 0.25 [ 0.16 , 0.34] | 40.8 [ 40.6 , 40.9] | 40.9 [ 40.7 , 41.1] |
| 73 | NGC 6251......... | 2.17 [ 0.89 , 2.14] | 3.62 [ 2.70 , 4.27] | 41.5 [ 41.0 , 41.5] | 41.6 [ 41.5 , 41.7] |
| 74 | NGC 6240......... | 1.45 [ 1.04 , 1.72] | 10.40 [ 7.11 , 12.03] | 42.2 [ 42.1 , 42.3] | 42.4 [ 42.2 , 42.5] |
| 75 | IRAS 17208-0014.. | 0.07 [ 0.01 , 0.10] | 0.40 [ 0.21 , 0.53] | 40.8 [ 40.3 , 40.9] | 41.2 [ 40.9 , 41.3] |
| 76 | NGC 6482......... | 0.68 [ 0.41 , 0.83] | 0.05 [ 0.03 , 0.07] | 40.7 [ 40.5 , 40.8] | 39.3 [ 39.1 , 39.5] |
| 77 | NGC 7130......... | 1.35 [ 0.70 , 1.55] | 0.93 [ 0.46 , 1.10] | 41.8 [ 41.5 , 41.8] | 40.8 [ 40.5 , 40.9] |
| 79 | NGC 7331......(*) | 0.08 [ 0.05 , 0.10] | 0.14 [ 0.09 , 0.19] | 38.3 [ 38.1 , 38.4] | 38.5 [ 38.2 , 38.6] |
| 80 | IC 1459.......... | 2.92 [ 2.07 , 4.83] | 5.21 [ 2.48 , 6.57] | 40.6 [ 40.4 , 40.8] | 40.5 [ 40.1 , 40.6] |
| 81 | NPM1G -12.0625.. | 0.41 [ 0.05 , 0.48] | 0.18 [ 0.00 , 46.65] | 42.6 [ 41.1 , 42.7] | 41.5 [ 0.0 , 43.9] |

(*) Fluxes and luminosities assuming a power-law model with an index of 1.8 and Galactic absorption.

**Table 6.** Observed fluxes and absorption corrected luminosities with *XMM-Newton* data.

| Num | Name | Flux(0.5-2.0keV) ($10^{-13}$ erg s$^{-1}$ cm$^{-2}$) | Flux(2.0-10.0keV) ($10^{-13}$ erg s$^{-1}$ cm$^{-2}$) | Log(Lum(0.5-2.0keV)) | Log(Lum(2.0-10.0keV)) |
|---|---|---|---|---|---|
| (1) | (2) | (3) | (4) | (5) | (6) |
| 1 | NGC 0315......... | 4.05[ 2.97, 9.40] | 7.11[ 6.45, 15.84] | 41.6[ 41.5, 42.0] | 41.6[ 41.5, 41.9] |
| 2 | NGC 0410......... | 5.29[ 4.43, 5.67] | 0.80[ 0.38, 1.10] | 41.7[ 41.6, 41.8] | 40.7[ 40.4, 40.8] |
| 3 | NGC 0474......... | 0.00[ 0.00, 0.15] | 0.00[ 0.00, 0.26] | 37.6[ 0.0, 39.3] | 37.8[ 0.0, 39.5] |
| 4 | IIIZw 035......... | 0.02[ 0.00, 0.11] | 0.04[ 0.00, 0.20] | 39.5[ 0.0, 40.3] | 39.7[ 0.0, 40.5] |
| 7 | NGC 0835......... | 0.70[ 0.44, 0.80] | 1.22[ 0.20, 1.66] | 42.0[ 41.8, 42.0] | 41.5[ 40.7, 41.7] |
| 8 | NGC 1052......... | 2.67[ 2.39, 2.77] | 43.49[ 35.17, 48.05] | 41.0[ 40.9, 41.0] | 41.4[ 41.3, 41.5] |
| 9 | NGC 2639......... | 1.09[ 0.00, 1.19] | 0.01[ 0.00, 0.01] | 42.2[ 0.0, 42.3] | 38.3[ 0.0, 38.4] |
| 10 | NGC 2655......... | 2.13[ 1.33, 2.43] | 10.18[ 0.21, 11.40] | 41.5[ 41.3, 41.6] | 41.2[ 39.5, 41.3] |
| 12 | NGC 2685......... | 0.20[ 0.11, 0.29] | 0.35[ 0.19, 0.51] | 38.8[ 38.6, 39.0] | 39.0[ 38.8, 39.2] |
| 13 | UGC 04881....... | 0.17[ 0.00, 0.22] | 0.00[ 0.00, 0.00] | 42.2[ 0.0, 42.3] | 38.4[ 0.0, 38.5] |
| 14 | 3C 218 .......... | 55.47[ 54.08, 56.92] | 64.11[ 62.55, 66.45] | 43.7[ 43.6, 43.7] | 43.6[ 43.6, 43.6] |
| 15 | NGC 2787......... | 0.87[ 0.37, 2.29] | 1.46[ 0.28, 2.88] | 39.3[ 38.9, 39.6] | 39.2[ 38.7, 39.5] |
| 16 | NGC 2841......... | 1.31[ 0.53, 1.39] | 1.62[ 0.32, 2.19] | 39.4[ 39.0, 39.4] | 39.2[ 38.2, 39.3] |
| 17 | UGC 05101....... | 0.23[ 0.10, 0.29] | 1.86[ 0.57, 2.62] | 42.0[ 41.7, 42.1] | 42.5[ 42.1, 42.6] |
| 18 | NGC 3185......... | 0.26[ 0.17, 0.35] | 0.43[ 0.29, 0.59] | 39.2[ 39.0, 39.3] | 39.4[ 39.2, 39.5] |
| 19 | NGC 3226......... | 1.82[ 1.23, 2.22] | 8.63[ 6.25, 9.89] | 40.7[ 40.5, 40.8] | 40.8[ 40.7, 40.9] |
| 26 | NGC 3623......... | 0.79[ 0.58, 0.94] | 1.73[ 0.00, 48.34] | 39.1[ 39.0, 39.2] | 39.4[ 0.0, 40.8] |
| 27 | NGC 3627......... | 1.22[ 1.02, 1.42] | 2.05[ 1.73, 2.39] | 39.2[ 39.1, 39.3] | 39.4[ 39.3, 39.5] |
| 28 | NGC 3628......... | 1.11[ 0.94, 1.24] | 5.57[ 5.03, 6.10] | 39.6[ 39.6, 39.7] | 39.9[ 39.9, 40.0] |
| 29 | NGC 3690B....... | 6.80[ 5.73, 7.24] | 8.28[ 7.24, 9.05] | 41.4[ 41.3, 41.4] | 41.2[ 41.2, 41.3] |
| 32 | NGC 3998......... | 65.51[ 64.28, 66.62] | 103.01[ 101.35, 105.20] | 41.2[ 41.2, 41.2] | 41.4[ 41.4, 41.4] |
| 35 | NGC 4125......... | 2.04[ 0.79, 2.02] | 0.50[ 0.11, 0.94] | 39.9[ 39.4, 39.9] | 39.3[ 38.7, 39.6] |
| 36 | IRAS 12112+0305 | 0.09[ 0.06, 0.12] | 0.15[ 0.10, 0.20] | 41.0[ 40.8, 41.2] | 41.2[ 41.0, 41.3] |
| 37 | NGC 4261......... | 5.04[ 4.61, 5.34] | 6.48[ 1.29, 7.57] | 41.4[ 41.4, 41.5] | 41.2[ 40.5, 41.2] |
| 38 | NGC 4278(MEPL) | 15.38[ 14.46, 15.94] | 19.23[ 18.41, 20.31] | 40.2[ 40.2, 40.2] | 40.2[ 40.2, 40.3] |
|    | NGC 4278...(2PL) | 15.38[ 14.69, 16.07] | 18.93[ 16.47, 21.12] | 40.4[ 40.4, 40.4] | 40.3[ 40.2, 40.3] |
| 39 | NGC 4314......... | 0.60[ 0.08, 0.68] | 0.63[ 0.00, 343.88] | 39.6[ 38.6, 39.6] | 39.1[ 0.0, 41.8] |
| 40 | NGC 4321......... | 2.53[ 1.50, 2.79] | 1.70[ 1.28, 1.99] | 40.6[ 40.4, 40.6] | 40.0[ 39.8, 40.0] |
| 46 | NGC 4486......... | 41.90[ 241.35, 242.36] | 189.55[ 188.48, 190.57] | 42.3[ 42.3, 42.3] | 41.8[ 41.8, 41.8] |
| 47 | NGC 4494......... | 0.63[ 0.40, 0.67] | 1.04[ 0.00, 154.49] | 39.4[ 39.2, 39.4] | 39.6[ 0.0, 41.8] |
| 48 | NGC 4552......... | 7.32[ 5.80, 7.63] | 3.76[ 3.14, 4.42] | 39.3[ 39.2, 39.3] | 39.0[ 38.9, 39.0] |
| 50 | NGC 4579......... | 25.16[ 23.63, 25.44] | 38.26[ 34.37, 40.84] | 41.3[ 41.3, 41.3] | 41.4[ 41.4, 41.4] |
| 52 | NGC 4594......... | 6.76[ 5.46, 13.56] | 13.59[ 11.55, 22.45] | 40.5[ 40.4, 40.8] | 40.5[ 40.4, 40.7] |
| 53 | NGC 4636......... | 18.06[ 15.34, 18.33] | 0.42[ 0.39, 0.44] | 40.9[ 40.8, 40.9] | 39.0[ 0.0, 44.4] |
| 56 | NGC 4698......... | 0.28[ 0.19, 0.38] | 0.46[ 0.31, 0.61] | 39.0[ 38.8, 39.1] | 39.2[ 39.0, 39.3] |
| 57 | NGC 4696......... | 30.73[ 30.43, 31.02] | 59.94[ 59.37, 59.00] | 60.5[ 41.7, 41.7] | 41.8[ 41.9, 41.9] |
| 58 | NGC 4736......... | 10.98[ 8.11, 11.08] | 15.73[ 14.75, 16.91] | 39.4[ 39.2, 39.4] | 39.5[ 39.5, 39.5] |
| 59 | NGC 5005......... | 3.33[ 1.36, 3.51] | 3.75[ 0.00, 259.58] | 40.7[ 40.2, 40.7] | 39.9[ 0.0, 41.7] |
| 61 | MRK 0266NE..... | 2.03[ 1.28, 2.09] | 2.90[ 0.41, 3.62] | 42.5[ 42.3, 42.5] | 42.1[ 41.3, 42.2] |
| 62 | UGC 08696....... | 0.93[ 0.58, 1.01] | 3.89[ 0.85, 5.07] | 42.4[ 42.2, 42.5] | 42.6[ 42.0, 42.7] |
| 63 | CGC G162-010... | 72.19[ 71.02, 73.54] | 91.87[ 89.97, 93.82] | 43.9[ 43.8, 43.9] | 43.9[ 43.9, 43.9] |
| 64 | NGC 5363......... | 1.27[ 0.65, 1.35] | 1.88[ 0.19, 2.59] | 39.9[ 39.6, 39.9] | 39.8[ 38.7, 39.9] |
| 65 | IC 4395.......... | 0.40[ 0.06, 0.38] | 0.23[ 0.00, 168.79] | 41.1[ 40.3, 41.0] | 40.8[ 0.0, 43.7] |
| 66 | IRAS 14348-1447 | 0.23[ 0.14, 0.32] | 0.35[ 0.11, 0.56] | 41.6[ 41.4, 41.7] | 41.7[ 41.2, 42.0] |
| 68 | NGC 5813......... | 13.65[ 11.49, 13.80] | 0.69[ 0.56, 2.26] | 41.3[ 41.3, 41.3] | 39.9[ 39.8, 40.4] |
| 70 | NGC 5846......... | 11.65[ 10.72, 12.03] | 1.34[ 0.00, 93.41] | 41.2[ 41.2, 41.2] | 40.3[ 0.0, 42.1] |
| 72 | MRK 0848......... | 0.35[ 0.09, 0.46] | 0.42[ 0.34, 1.07] | 41.2[ 40.5, 41.3] | 41.1[ 41.0, 41.6] |
| 73 | NGC 6251......... | 20.72[ 20.16, 21.25] | 36.75[ 33.84, 38.87] | 42.7[ 42.7, 42.7] | 42.8[ 42.7, 42.8] |
| 74 | NGC 6240......... | 4.87[ 4.41, 5.24] | 25.21[ 11.77, 27.17] | 43.9[ 43.9, 44.0] | 44.0[ 43.7, 44.1] |
| 75 | IRAS 17208-0014 | 0.20[ 0.00, 0.27] | 0.04[ 0.00, 0.06] | 42.0[ 0.0, 42.1] | 40.3[ 0.0, 40.5] |
| 76 | NGC 6482......... | 8.65[ 7.61, 9.48] | 0.41[ 0.34, 0.46] | 41.7[ 41.6, 41.7] | 40.2[ 40.1, 40.2] |
| 78 | NGC 7285(MEPL) | 0.80[ 0.38, 0.94] | 4.86[ 2.74, 5.69] | 41.8[ 41.5, 41.9] | 41.3[ 41.1, 41.4] |
|    | NGC 7285...(2PL) | 0.80[ 0.24, 1.03] | 4.91[ 1.44, 5.92] | 41.0[ 40.6, 41.1] | 41.3[ 40.9, 41.4] |
| 79 | NGC 7331......... | 0.85[ 0.22, 1.47] | 1.42[ 0.45, 2.46] | 39.3[ 38.7, 39.5] | 39.5[ 38.9, 39.7] |
| 80 | IC 1459.......... | 5.01[ 4.05, 5.32] | 7.33[ 6.85, 7.87] | 40.7[ 40.6, 40.8] | 40.7[ 40.7, 40.7] |
| 81 | NPM1G -12.0625. | 45.19[ 44.06, 46.13] | 45.33[ 43.03, 47.34] | 44.0[ 44.0, 44.0] | 43.9[ 43.8, 43.9] |
| 82 | NGC 7743......... | 0.53[ 0.00, 0.60] | 0.44[ 0.00, 162.68] | 40.5[ 0.0, 40.6] | 39.5[ 0.0, 42.1] |

**Table 7.** Final compilation of bestfit models for the LINER sample.

| Num | Name | Type | Instrument | Bestfit | $N_{H1}$ ($10^{22}$ cm$^{-2}$) | $N_{H2}$ ($10^{22}$ cm$^{-2}$) | $\Gamma$ | kT (keV) | $\chi^2$ | d.o.f | $\chi^2_\nu$ |
|---|---|---|---|---|---|---|---|---|---|---|---|
| (1) | (2) | (3) | (4) | (5) | (6) | (7) | (8) | (9) | (10) | (11) | (12) |
| 1 | NGC 0315 | AGN | Chandra | ME2PL | $15.21^{21.33}_{10.67}$ | $1.06^{1.40}_{0.87}$ | $2.37^{2.51}_{2.08}$ | $0.46^{0.51}_{0.43}$ | 374.45 | 362 | 1.03 |
| 2 | NGC 0410 | Non-AGN* | XMM-Newton | MEPL | $0.12^{0.17}_{0.09}$ | $0.01^{0.20}_{0.01}$ | $2.43^{3.17}_{1.92}$ | $0.69^{0.71}_{0.67}$ | 218.34 | 178 | 1.23 |
| 3 | NGC 0474 | Non-AGN | Chandra | – | ... | ... | ... | ... | ... | ... | ... |
| 4 | IIIZW 035 | Non-AGN | Chandra | – | ... | ... | ... | ... | ... | ... | ... |
| 5 | NGC 0524 | Non-AGN | Chandra | – | ... | ... | ... | ... | ... | ... | ... |
| 6 | NGC 0833 | AGN | Chandra | ME2PL | $3.51^{10.79}_{1.34}$ | $26.88^{41.78}_{-1.50}$ | $2.15^{5.43}_{-1.50}$ | $0.67^{0.87}_{0.49}$ | 31.63 | 54 | 0.59 |
| 7 | NGC 0835 | AGN | Chandra | ME2PL | $0.16^{3.13}_{0.01}$ | $40.35^{69.46}_{24.69}$ | $2.35^{4.80}_{1.87}$ | $0.47^{0.59}_{0.39}$ | 57.38 | 53 | 1.08 |
| 8 | NGC 1052 | AGN | Chandra | ME2PL | $0.36^{1.88}_{0.00}$ | $12.90^{35.47}_{0.00}$ | $0.61^{1.76}_{-0.02}$ | $0.67^{0.83}_{0.46}$ | 44.79 | 57 | 0.79 |
| 9 | NGC 2639 | Non-AGN* | XMM-Newton | ME | $0.80^{0.92}_{}$ | ... | ... | $0.18^{0.24}_{0.15}$ | 180.55 | 148 | 1.22 |
| 10 | NGC 2655 | AGN* | XMM-Newton | ME2PL | $0.01^{0.07}_{0.01}$ | $30.20^{39.47}_{24.21}$ | $2.48^{2.83}_{2.05}$ | $0.64^{0.68}_{0.59}$ | 94.67 | 111 | 0.85 |
| 11 | NGC 2681 | AGN | Chandra | MEPL | $0.15^{0.28}_{0.01}$ | $0.01^{0.08}_{0.01}$ | $1.57^{1.89}_{1.29}$ | $0.63^{0.68}_{0.54}$ | 66.87 | 103 | 0.65 |
| 12 | NGC 2685 | AGN* | XMM-Newton | – | ... | ... | ... | ... | ... | ... | ... |
| 13 | UGC 4881 | Non-AGN | XMM-Newton | ME | $0.61^{0.79}_{0.32}$ | ... | ... | $0.19^{0.29}_{0.15}$ | 21.71 | 19 | 1.14 |
| 14 | 3C 218 | AGN | Chandra | ME2PL | $0.07^{0.20}_{0.04}$ | $4.05^{5.96}_{2.88}$ | $2.11^{2.76}_{1.87}$ | $1.71^{2.19}_{1.43}$ | 256.46 | 261 | 0.98 |
| 15 | NGC 2787 | AGN | Chandra | PL | $0.11^{0.22}_{0.03}$ | ... | $2.33^{2.94}_{1.87}$ | ... | 21.01 | 16 | 1.31 |
| 16 | NGC 2841 | AGN | XMM-Newton | ME2PL | $0.01^{0.09}_{0.01}$ | $3.30^{5.60}_{1.96}$ | $2.20^{3.01}_{2.07}$ | $0.58^{0.64}_{0.50}$ | 155.16 | 139 | 1.12 |
| 17 | UGC 05101 | AGN | Chandra | ME2PL | $0.21^{0.66}_{0.04}$ | $135.07^{380.33}_{43.51}$ | $1.34^{1.75}_{1.06}$ | $0.66^{1.05}_{0.47}$ | 81.79 | 91 | 0.90 |
| 18 | NGC 3185 | Non-AGN* | XMM-Newton | – | ... | ... | ... | ... | ... | ... | ... |
| 19 | NGC 3226 | AGN | XMM-Newton | 2PL | $0.21^{0.30}_{0.11}$ | $1.22^{1.76}_{0.94}$ | $1.92^{2.12}_{1.80}$ | ... | 169.21 | 179 | 0.95 |
| 20 | NGC 3245 | AGN | Chandra | – | ... | ... | ... | ... | ... | ... | ... |
| 21 | NGC 3379 | Non-AGN | Chandra | – | ... | ... | ... | ... | ... | ... | ... |
| 22 | NGC 3414 | AGN | Chandra | PL | $0.21^{0.30}_{0.13}$ | ... | $2.02^{2.48}_{1.70}$ | ... | 118.01 | 129 | 0.91 |
| 23 | NGC 3507 | Non-AGN | Chandra | ME | $0.08^{0.39}_{0.01}$ | ... | ... | $0.51^{0.64}_{0.28}$ | 60.92 | 47 | 1.30 |
| 24 | NGC 3607 | Non-AGN | Chandra | – | ... | ... | ... | ... | ... | ... | ... |
| 25 | NGC 3608 | Non-AGN | Chandra | – | ... | ... | ... | ... | ... | ... | ... |
| 26 | NGC 3623 | Non-AGN* | XMM-Newton | – | ... | ... | ... | ... | ... | ... | ... |
| 27 | NGC 3627 | Non-AGN* | XMM-Newton | – | ... | ... | ... | ... | ... | ... | ... |
| 28 | NGC 3628 | Non-AGN | XMM-Newton | PL | $0.46^{0.52}_{0.41}$ | ... | $1.61^{1.68}_{1.53}$ | ... | 264.36 | 247 | 1.07 |
| 29 | NGC 3690B | AGN | Chandra | ME2PL | $0.21^{0.41}_{0.03}$ | $9.49^{12.54}_{7.52}$ | $3.52^{4.89}_{}$ | $0.19^{0.25}_{0.14}$ | 119.66 | 148 | 0.81 |
| 30 | NGC 3898 | Non-AGN | Chandra | MEPL | $1.39^{1.75}_{1.12}$ | $0.01^{0.59}_{0.01}$ | $1.81^{2.81}_{1.57}$ | $0.04^{0.05}_{0.04}$ | 3.98 | 8 | 0.50 |
| 31 | NGC 3945 | AGN | Chandra | PL | $0.04^{0.17}_{0.01}$ | ... | $2.60^{3.52}_{2.32}$ | ... | 56.82 | 83 | 0.68 |
| 32 | NGC 3998 | AGN | Chandra | ME2PL | $0.08^{0.15}_{0.06}$ | $2.30^{3.18}_{1.63}$ | $1.81^{2.03}_{1.65}$ | $0.22^{0.25}_{0.20}$ | 490.35 | 454 | 1.08 |
| 33 | NGC 4036 | AGN | Chandra | – | ... | ... | ... | ... | ... | ... | ... |
| 34 | NGC 4111 | AGN | Chandra | ME2PL | $4.67^{9.65}_{1.20}$ | $37.71^{100.00}_{12.52}$ | $3.04^{8.10}_{0.74}$ | $0.66^{0.70}_{0.61}$ | 75.75 | 65 | 1.17 |
| 35 | NGC 4125 | AGN | Chandra | MEPL | $0.53^{0.88}_{0.01}$ | $0.86^{2.13}_{0.08}$ | $2.32^{2.59}_{1.26}$ | $0.57^{0.59}_{0.55}$ | 35.49 | 58 | 0.61 |
| 36 | IRAS 12112+0305 | Non-AGN* | XMM-Newton | – | ... | ... | ... | ... | ... | ... | ... |
| 37 | NGC 4261 | AGN | Chandra | ME2PL | $0.69^{1.38}_{0.31}$ | $16.45^{21.64}_{13.25}$ | $2.37^{2.80}_{1.87}$ | $0.57^{0.59}_{0.55}$ | 313.17 | 242 | 1.29 |
| 38 | NGC 4278 | AGN | Chandra | ME2PL | $0.09^{0.12}_{0.06}$ | $2.65^{4.32}_{1.48}$ | $2.59^{2.66}_{2.28}$ | $0.53^{0.67}_{0.41}$ | 267.30 | 293 | 0.91 |
| 39 | NGC 4314 | Non-AGN | XMM-Newton | MEPL | $0.27^{0.44}_{0.09}$ | $0.01^{0.52}_{0.01}$ | $1.46^{1.88}_{1.11}$ | $0.24^{0.27}_{0.19}$ | 40.89 | 42 | 0.97 |
| 40 | NGC 4321 | Non-AGN | Chandra | MEPL | $0.59^{0.77}_{0.21}$ | $0.19^{0.49}_{0.08}$ | $2.36^{2.77}_{2.06}$ | $0.20^{0.25}_{0.08}$ | 132.14 | 120 | 1.10 |
| 41 | NGC 4374 | AGN | Chandra | MEPL | $0.07^{0.38}_{0.01}$ | $0.13^{0.38}_{0.08}$ | $1.95^{2.28}_{1.53}$ | $0.72^{0.90}_{0.54}$ | 92.38 | 122 | 0.76 |
| 42 | NGC 4410A | AGN | Chandra | MEPL | $0.51^{1.14}_{0.25}$ | $0.01^{0.05}_{0.01}$ | $1.77^{1.93}_{1.13}$ | $0.30^{0.80}_{0.13}$ | 132.47 | 155 | 0.85 |
| 43 | NGC 4438 | AGN | Chandra | MEPL | $0.37^{0.45}_{0.26}$ | $0.01^{0.21}_{0.01}$ | $1.91^{2.77}_{1.58}$ | $0.52^{0.60}_{0.45}$ | 95.94 | 97 | 0.99 |
| 44 | NGC 4457 | AGN | Chandra | MEPL | $0.37^{0.57}_{0.01}$ | $0.17^{2.60}_{0.01}$ | $1.70^{3.54}_{1.18}$ | $0.31^{0.68}_{0.22}$ | 68.74 | 82 | 0.84 |
| 45 | NGC 4459 | Non-AGN | Chandra | – | ... | ... | ... | ... | ... | ... | ... |
| 46 | NGC 4486 | AGN | Chandra | ME2PL | $0.10^{0.14}_{0.09}$ | $3.96^{4.47}_{3.66}$ | $2.40^{2.52}_{2.31}$ | $0.82^{0.96}_{0.70}$ | 567.31 | 429 | 1.32 |
| 47 | NGC 4494 | AGN | Chandra | MEPL | $0.29^{0.72}_{0.01}$ | $0.03^{0.17}_{0.01}$ | $1.72^{2.31}_{0.90}$ | $0.63^{1.02}_{0.24}$ | 36.50 | 56 | 0.65 |
| 48 | NGC 4552 | AGN | Chandra | MEPL | $0.35^{0.56}_{0.01}$ | $0.01^{0.12}_{0.01}$ | $2.02^{2.37}_{1.87}$ | $0.67^{0.81}_{0.56}$ | 147.04 | 162 | 0.91 |
| 49 | NGC 4589 | Non-AGN | Chandra | – | ... | ... | ... | ... | ... | ... | ... |
| 50 | NGC 4579 | AGN | Chandra | MEPL | $0.48^{0.54}_{0.38}$ | $0.45^{0.56}_{0.27}$ | $1.58^{1.66}_{1.47}$ | $0.20^{0.21}_{0.18}$ | 506.16 | 416 | 1.21 |
| 51 | NGC 4596 | Non-AGN | Chandra | – | ... | ... | ... | ... | ... | ... | ... |
| 52 | NGC 4594 | AGN | Chandra | PL | $0.19^{0.23}_{0.17}$ | ... | $1.56^{1.67}_{1.46}$ | ... | 300.39 | 281 | 1.07 |
| 53 | NGC 4636 | Non-AGN | XMM-Newton | MEPL | $0.01^{0.01}_{0.01}$ | $0.01^{0.17}_{0.01}$ | $2.92^{3.91}_{2.36}$ | $0.54^{0.55}_{0.53}$ | 355.63 | 241 | 1.48 |
| 54 | NGC 4676A | Non-AGN | Chandra | – | ... | ... | ... | ... | ... | ... | ... |
| 55 | NGC 4676B | AGN | Chandra | – | ... | ... | ... | ... | ... | ... | ... |
| 56 | NGC 4698 | AGN | Chandra | – | ... | ... | ... | ... | ... | ... | ... |
| 57 | NGC 4696 | Non-AGN | Chandra | MEPL | $0.01^{0.18}_{0.01}$ | $0.01^{6.93}_{0.01}$ | $3.07^{5.56}_{}$ | $0.67^{0.78}_{0.60}$ | 61.63 | 74 | 0.83 |
| 58 | NGC 4736 | AGN | Chandra | MEPL | $0.31^{0.81}_{0.15}$ | $0.04^{0.09}_{0.02}$ | $2.05^{2.28}_{1.82}$ | $0.54^{0.61}_{0.33}$ | 170.20 | 190 | 0.90 |
| 59 | NGC 5005 | AGN* | XMM-Newton | MEPL | $0.61^{0.80}_{0.52}$ | $0.01^{0.07}_{0.01}$ | $1.54^{1.73}_{1.42}$ | $0.27^{0.30}_{0.21}$ | 70.67 | 72 | 0.98 |
| 60 | NGC 5055 | AGN | Chandra | PL | $0.16^{0.33}_{}$ | ... | $2.30^{3.16}_{1.95}$ | ... | 42.78 | 62 | 0.69 |
| 61 | MRK 266NE | AGN | Chandra | ME2PL | $0.01^{0.28}_{0.01}$ | $9.45^{28.94}_{5.69}$ | $1.34^{1.95}_{0.29}$ | $0.83^{1.04}_{0.63}$ | 61.01 | 77 | 0.79 |
| 62 | UGC 08696 | AGN | Chandra | ME2PL | $0.60^{0.96}_{0.22}$ | $50.91^{55.61}_{43.24}$ | $2.05^{2.55}_{1.72}$ | $0.68^{0.77}_{0.58}$ | 210.79 | 229 | 0.92 |
| 63 | CGCG 162-010 | Non-AGN | Chandra | ME | $0.47^{0.63}_{0.39}$ | ... | ... | $1.05^{1.10}_{0.96}$ | 142.58 | 141 | 1.01 |
| 64 | NGC 5363 | AGN* | XMM-Newton | ME2PL | $0.01^{0.08}_{0.01}$ | $2.66^{4.37}_{1.74}$ | $2.14^{2.69}_{1.76}$ | $0.61^{0.65}_{0.56}$ | 79.70 | 81 | 0.90 |
| 65 | IC 4395 | Non-AGN* | XMM-Newton | MEPL | $0.01^{0.42}_{0.01}$ | $0.01^{0.17}_{0.01}$ | $1.78^{4.00}_{1.24}$ | $0.26^{0.30}_{0.19}$ | 25.54 | 22 | 1.16 |
| 66 | IRAS 14348-1447 | Non-AGN | XMM-Newton | ME | $0.01^{0.04}_{0.01}$ | ... | ... | $3.67^{6.46}_{2.50}$ | 25.23 | 34 | 0.74 |
| 67 | NGC 5746 | AGN | Chandra | PL | $0.60^{0.93}_{0.35}$ | ... | $1.28^{1.67}_{0.93}$ | ... | 87.71 | 112 | 0.78 |
| 68 | NGC 5813 | Non-AGN | Chandra | MEPL | $0.12^{0.31}_{0.01}$ | $0.15^{0.51}_{0.01}$ | $2.94^{6.90}_{}$ | $0.49^{0.60}_{0.29}$ | 25.95 | 30 | 0.86 |
| 69 | NGC 5838 | AGN | Chandra | – | ... | ... | ... | ... | ... | ... | ... |
| 70 | NGC 5846 | Non-AGN | Chandra | MEPL | $0.28^{0.43}_{0.01}$ | $0.03^{0.17}_{0.01}$ | $2.55^{3.25}_{2.12}$ | $0.35^{0.58}_{0.28}$ | 71.90 | 85 | 0.85 |
| 71 | NGC 5866 | Non-AGN | Chandra | – | ... | ... | ... | ... | ... | ... | ... |
| 72 | MRK 0848 | Non-AGN | XMM-Newton | PL | $0.07^{0.19}_{}$ | ... | $2.26^{2.75}_{1.73}$ | ... | 31.18 | 42 | 0.74 |
| 73 | NGC 6251 | AGN | Chandra | MEPL | $0.01^{0.96}_{0.01}$ | $0.01^{0.04}_{0.01}$ | $1.48^{1.61}_{1.39}$ | $0.20^{0.22}_{0.14}$ | 273.33 | 250 | 1.09 |
| 74 | NGC 6240 | AGN | Chandra | ME2PL | $0.72^{0.96}_{0.55}$ | $50.12^{103.79}_{22.34}$ | $1.76^{1.97}_{1.63}$ | $1.07^{1.22}_{0.97}$ | 380.08 | 352 | 1.08 |
| 75 | IRAS 17208-0014 | AGN | Chandra | PL | $0.34^{0.61}_{0.18}$ | ... | $1.63^{1.86}_{1.34}$ | ... | 48.01 | 63 | 0.76 |
| 76 | NGC 6482 | Non-AGN | Chandra | ME | $0.19^{0.33}_{0.10}$ | ... | ... | $0.75^{0.81}_{0.65}$ | 60.72 | 71 | 0.85 |
| 77 | NGC 7130 | AGN | Chandra | ME2PL | $0.07^{0.11}_{0.01}$ | $86.01^{160.48}_{60.51}$ | $2.66^{2.82}_{}$ | $0.76^{0.81}_{0.68}$ | 155.74 | 156 | 1.00 |
| 78 | NGC 7285 | AGN* | XMM-Newton | MEPL | $0.68^{0.94}_{0.01}$ | $0.87^{1.13}_{0.59}$ | $1.66^{1.76}_{1.43}$ | $0.13^{0.19}_{0.07}$ | 59.28 | 58 | 1.02 |
| 79 | NGC 7331 | Non-AGN | Chandra | – | ... | ... | ... | ... | ... | ... | ... |

**Table 7.** Continuation

| Num | Name | Type | Instrument | Bestfit | $N_{H1}$ ($10^{22}$ cm$^{-2}$) | $N_{H2}$ ($10^{22}$ cm$^{-2}$) | $\Gamma$ | kT (keV) | $\chi^2$ | d.o.f | $\chi^2_\nu$ |
|---|---|---|---|---|---|---|---|---|---|---|---|
| (1) | (2) | (3) | (4) | (5) | (6) | (7) | (8) | (9) | (10) | (11) | (12) |
| 80 | IC 1459......... | AGN | *Chandra* | ME2PL | $0.20^{0.28}_{0.09}$ | $1.26^{3.17}_{0.66}$ | $2.17^{2.57}_{2.12}$ | $0.61^{0.67}_{0.49}$ | 294.41 | 334 | 0.88 |
| 81 | NPM1G-12.0625. | Non-AGN | *Chandra* | MEPL | $0.71^{0.85}_{0.33}$ | $0.15^{0.32}_{0.06}$ | $2.67^{3.31}_{2.34}$ | $0.31^{0.54}_{0.23}$ | 122.36 | 138 | 0.89 |
| 82 | NGC 7743....... | Non-AGN* | *XMM-Newton* | MEPL | $0.35^{0.58}_{0.18}$ | $1.68^{3.63}_{0.81}$ | $3.16^{6.47}_{1.49}$ | $0.23^{0.28}_{0.18}$ | 43.95 | 25 | 1.76 |

**Table 8.** Final compilation of observed fluxes and absorption corrected luminosities for the LINER sample.

| Num | Name | Type | Instrument | Flux(0.5-2.0keV) ($10^{-13}$ erg s$^{-1}$) | Flux(2.0-10.0keV) ($10^{-13}$ erg s$^{-1}$) | Log(L(0.5-2.0keV)) | Log(L(2.0-10.0keV)) | MEKAL(*) | Power-law(*) | Power-law(*) |
|---|---|---|---|---|---|---|---|---|---|---|
| (1) | (2) | (3) | (4) | (5) | (6) | (7) | (8) | (9) | (10) | (11) |
| 1 | NGC 0315 | AGN | *Chandra* | 1.52[ 0.46, 1.80] | 6.50[ 0.23, 7.76] | 42.0[ 41.4, 42.0] | 41.8[ 40.9, 41.8] | 0/ 36% | 28/ 3% | 72/ 61% |
| 2 | NGC 0410 | Non-AGN* | *XMM-Newton* | 5.29[ 4.43, 5.67] | 0.80[ 0.38, 1.10] | 41.7[ 41.6, 41.8] | 40.7[ 40.4, 40.8] | 90/ 40% | 10/ 60% | 0/ 0% |
| 3 | NGC 0474 | Non-AGN | *Chandra* | 0.01[ 0.00, 0.04] | 0.02[ 0.00, 0.06] | 38.3[ 0.0, 38.7] | 38.5[ 0.0, 38.9] | | | |
| 4 | IIIZW 035 | Non-AGN | *Chandra* | 0.04[ 0.00, 0.08] | 0.07[ 0.02, 0.13] | 39.9[ 39.2, 40.1] | 40.0[ 39.4, 40.3] | | | |
| 5 | NGC 0524 | Non-AGN | *Chandra* | 0.03[ 0.00, 0.06] | 0.05[ 0.00, 0.11] | 38.4[ 36.1, 38.7] | 38.6[ 37.1, 38.9] | | | |
| 6 | NGC 0833 | AGN | *Chandra* | 0.11[ 0.05, 0.15] | 5.52[ 0.42, 6.42] | 41.7[ 41.3, 41.8] | 41.7[ 40.6, 41.8] | 18/ 19% | 82/ 12% | 0/ 69% |
| 7 | NGC 0835 | AGN | *Chandra* | 0.41[ 0.20, 0.52] | 1.37[ 1.21, 17.58] | 41.7[ 41.3, 41.8] | 41.4[ 41.4, 42.5] | 36/ 19% | 64/ 11% | 0/ 70% |
| 8 | NGC 1052 | AGN | *Chandra* | 1.03[ 0.00, 1.45] | 14.35[ 2.23, 18.07] | 41.7[ 0.0, 41.8] | 41.2[ 40.4, 41.3] | 58/ 22% | 42/ 7% | 0/ 71% |
| 9 | NGC 2639 | Non-AGN* | *XMM-Newton* | 1.09[ 0.00, 1.19] | 0.01[ 0.00, 0.01] | 42.2[ 0.0, 42.3] | 38.3[ 0.0, 38.4] | 100/ 100% | 0/ 0% | 0/ 0% |
| 10 | NGC 2655 | AGN* | *XMM-Newton* | 2.13[ 1.33, 2.43] | 10.18[ 0.21, 11.40] | 41.5[ 41.3, 41.6] | 41.2[ 39.5, 41.3] | 45/ 7% | 55/ 1% | 0/ 92% |
| 11 | NGC 2681 | AGN | *Chandra* | 0.24[ 0.07, 0.26] | 1.85[ 0.17, 14.06] | 38.6[ 38.1, 38.6] | 39.3[ 38.3, 40.2] | 65/ 13% | 35/ 87% | 0/ 0% |
| 12 | NGC 2685 | AGN* | *XMM-Newton* | 0.20[ 0.11, 0.29] | 0.35[ 0.19, 0.51] | 38.8[ 38.6, 38.9] | 39.0[ 38.8, 39.2] | | | |
| 13 | UGC 4881 | Non-AGN | *XMM-Newton* | 0.17[ 0.00, 0.22] | 0.00[ 0.00, 0.00] | 42.2[ 0.0, 42.3] | 38.4[ 0.0, 38.5] | 100/ 100% | 0/ 0% | 0/ 0% |
| 14 | 3C 218 | AGN | *Chandra* | 0.62[ 0.41, 0.73] | 1.46[ 0.40, 1.82] | 42.2[ 42.0, 42.2] | 42.1[ 41.5, 42.2] | 56/ 31% | 38/ 8% | 6/ 61% |
| 15 | NGC 2787 | AGN | *Chandra* | 0.45[ 0.12, 0.57] | 0.75[ 0.00,13899.00] | 38.9[ 38.3, 39.0] | 38.8[ 0.0, 43.2] | 0/ 0% | 100/ 100% | 0/ 0% |
| 16 | NGC 2841 | AGN | *XMM-Newton* | 1.31[ 0.53, 1.39] | 1.62[ 0.32, 2.19] | 39.4[ 39.0, 39.4] | 39.2[ 38.2, 39.3] | 74/ 50% | 23/ 0% | 3/ 50% |
| 17 | UGC 05101 | AGN | *Chandra* | 0.12[ 0.05, 0.16] | 0.81[ 0.33, 1.17] | 41.8[ 41.3, 41.9] | 42.1[ 41.7, 42.2] | 77/ 41% | 23/ 6% | 0/ 53% |
| 18 | NGC 3185 | Non-AGN* | *XMM-Newton* | 0.26[ 0.17, 0.35] | 0.43[ 0.29, 0.59] | 39.2[ 39.0, 39.3] | 39.4[ 39.2, 39.5] | | | |
| 19 | NGC 3226 | AGN | *XMM-Newton* | 1.82[ 1.23, 2.22] | 8.63[ 6.25, 9.89] | 40.7[ 40.5, 40.8] | 40.8[ 40.7, 40.9] | 0/ 0% | 50/ 25% | 50/ 75% |
| 20 | NGC 3245 | AGN | *Chandra* | 0.12[ 0.05, 0.18] | 0.18[ 0.09, 0.28] | 38.8[ 38.5, 39.0] | 39.0[ 38.6, 39.2] | | | |
| 21 | NGC 3379 | Non-AGN | *Chandra* | 0.06[ 0.04, 0.09] | 0.10[ 0.06, 0.14] | 38.0[ 37.8, 38.1] | 38.1[ 37.9, 38.3] | | | |
| 22 | NGC 3414 | AGN | *Chandra* | 0.95[ 0.61, 1.17] | 1.63[ 1.24, 2.22] | 39.9[ 39.8, 40.0] | 39.9[ 39.8, 40.0] | 0/ 0% | 100/ 100% | 0/ 0% |
| 23 | NGC 3507 | Non-AGN | *Chandra* | 0.18[ 0.00, 0.23] | 0.00[ 0.00, 0.00] | 39.1[ 0.0, 39.2] | 37.2[ 0.0, 37.4] | 100/ 100% | 0/ 0% | 0/ 0% |
| 24 | NGC 3607 | Non-AGN | *Chandra* | 0.06[ 0.04, 0.09] | 0.09[ 0.05, 0.14] | 38.6[ 38.4, 38.8] | 38.8[ 38.5, 38.9] | | | |
| 25 | NGC 3608 | Non-AGN | *Chandra* | 0.02[ 0.00, 0.03] | 0.03[ 0.00, 0.05] | 38.0[ 36.9, 38.3] | 38.2[ 37.1, 38.5] | | | |
| 26 | NGC 3623 | Non-AGN* | *XMM-Newton* | 0.79[ 0.58, 0.94] | 1.73[ 0.00, 48.34] | 39.1[ 39.0, 39.2] | 39.4[ 0.0, 40.8] | | | |
| 27 | NGC 3627 | Non-AGN* | *XMM-Newton* | 1.22[ 1.02, 1.42] | 2.05[ 1.73, 2.39] | 39.2[ 39.1, 39.3] | 39.4[ 39.3, 39.5] | | | |
| 28 | NGC 3628 | Non-AGN | *XMM-Newton* | 1.11[ 0.94, 1.24] | 5.57[ 5.03, 6.10] | 39.6[ 39.6, 39.7] | 39.9[ 39.9, 40.0] | 0/ 0% | 100/ 100% | 0/ 0% |
| 29 | NGC 3690B | AGN | *Chandra* | 0.95[ 0.00, 1.42] | 1.29[ 0.14, 2.03] | 41.8[ 0.0, 42.0] | 40.9[ 40.0, 41.1] | 63/ 16% | 36/ 4% | 1/ 80% |
| 30 | NGC 3898 | Non-AGN | *Chandra* | 0.19[ 0.01, 0.27] | 0.20[ 0.00,10468.50] | 42.7[ 0.0, 42.9] | 38.8[ 0.0, 43.5] | 36/ 0% | 64/ 100% | 0/ 0% |
| 31 | NGC 3945 | AGN | *Chandra* | 0.79[ 0.44, 0.98] | 0.38[ 0.27, 0.49] | 39.8[ 39.5, 39.9] | 39.1[ 39.0, 39.2] | 0/ 0% | 100/ 100% | 0/ 0% |
| 32 | NGC 3998 | AGN | *Chandra* | 31.27[ 25.73, 34.34] | 82.46[ 60.30, 91.31] | 41.2[ 41.1, 41.2] | 41.3[ 41.2, 41.4] | 87/ 60% | 7/ 0% | 7/ 40% |
| 33 | NGC 4036 | AGN | *Chandra* | 0.12[ 0.07, 0.17] | 0.18[ 0.10, 0.27] | 39.0[ 38.7, 39.1] | 39.1[ 38.9, 39.3] | | | |
| 34 | NGC 4111 | AGN | *Chandra* | 0.59[ 0.46, 0.70] | 3.39[ 0.02, 104.04] | 40.9[ 40.8, 40.9] | 40.4[ 0.0, 41.9] | 3/ 28% | 97/ 22% | 0/ 50% |
| 35 | NGC 4125 | AGN | *Chandra* | 0.07[ 0.00, 0.09] | 0.07[ 0.00, 0.12] | 39.9[ 0.0, 40.0] | 38.7[ 0.0, 39.0] | 35/ 1% | 65/ 99% | 0/ 0% |
| 36 | IRAS 12112+0305 | Non-AGN* | *XMM-Newton* | 0.09[ 0.06, 0.12] | 0.15[ 0.10, 0.21] | 41.0[ 40.8, 41.2] | 41.2[ 41.0, 41.4] | | | |
| 37 | NGC 4261 | AGN | *Chandra* | 1.88[ 1.50, 2.07] | 4.32[ 3.46, 5.24] | 41.3[ 41.2, 41.3] | 41.1[ 39.4, 41.2] | 10/ 13% | 90/ 3% | 0/ 84% |
| 38 | NGC 4278 | AGN | *Chandra* | 2.10[ 1.52, 2.38] | 1.71[ 1.19, 2.07] | 39.6[ 39.4, 39.6] | 39.2[ 39.1, 39.3] | 87/ 72% | 11/ 3% | 2/ 25% |
| 39 | NGC 4314 | Non-AGN | *XMM-Newton* | 0.60[ 0.08, 0.68] | 0.63[ 0.00, 343.88] | 39.6[ 38.6, 39.6] | 39.1[ 0.0, 41.8] | 65/ 7% | 35/ 93% | 0/ 0% |
| 40 | NGC 4321 | Non-AGN | *Chandra* | 0.55[ 0.07, 0.64] | 5.86[ 3.48, 45.58] | 40.4[ 39.6, 40.5] | 40.5[ 40.3, 41.4] | 36/ 0% | 64/ 100% | 0/ 0% |
| 41 | NGC 4374 | AGN | *Chandra* | 0.53[ 0.07, 0.61] | 0.81[ 0.41, 44.31] | 39.5[ 38.6, 39.6] | 39.5[ 39.2, 41.3] | 21/ 7% | 79/ 93% | 0/ 0% |
| 42 | NGC 4410A | AGN | *Chandra* | 0.75[ 0.34, 0.81] | 1.18[ 0.82, 1.43] | 41.2[ 40.9, 41.2] | 41.2[ 41.0, 41.3] | 10/ 9% | 90/ 91% | 0/ 0% |
| 43 | NGC 4438 | AGN | *Chandra* | 1.11[ 0.31, 1.26] | 0.32[ 0.00, 990.00] | 40.1[ 39.4, 40.1] | 39.0[ 0.0, 42.5] | 81/ 17% | 19/ 83% | 0/ 0% |
| 44 | NGC 4457 | AGN | *Chandra* | 0.37[ 0.01, 0.44] | 0.17[ 0.12, 0.24] | 39.6[ 39.4, 39.6] | 38.8[ 38.6, 38.9] | 66/ 6% | 34/ 94% | 0/ 0% |
| 45 | NGC 4459 | Non-AGN | *Chandra* | 0.05[ 0.00, 0.09] | 0.08[ 0.01, 0.14] | 38.2[ 37.4, 38.5] | 38.4[ 37.7, 38.6] | | | |
| 46 | NGC 4486 | AGN | *Chandra* | 5.15[ 4.59, 5.52] | 16.43[ 11.87, 18.33] | 40.9[ 40.9, 41.0] | 40.8[ 40.7, 40.9] | 84/ 30% | 7/ 4% | 9/ 67% |
| 47 | NGC 4494 | AGN | *Chandra* | 0.42[ 0.12, 0.53] | 0.16[ 0.04, 0.30] | 39.8[ 39.3, 39.8] | 38.8[ 38.1, 39.0] | 64/ 20% | 31/ 1% | 5/ 79% |
| 48 | NGC 4552 | AGN | *Chandra* | 0.72[ 0.42, 0.81] | 0.63[ 0.19, 0.85] | 39.5[ 39.3, 39.6] | 39.2[ 38.8, 39.4] | 32/ 9% | 68/ 91% | 0/ 0% |
| 49 | NGC 4589 | Non-AGN | *Chandra* | 0.09[ 0.05, 0.14] | 0.14[ 0.07, 0.22] | 38.8[ 38.5, 38.9] | 38.9[ 38.6, 39.1] | | | |
| 50 | NGC 4579 | AGN | *Chandra* | 12.38[ 10.65, 12.38] | 43.88[ 41.56, 45.85] | 40.9[ 40.8, 40.9] | 41.2[ 41.1, 41.2] | 3/ 3% | 97/ 97% | 0/ 0% |
| 51 | NGC 4596 | Non-AGN | *Chandra* | 0.06[ 0.01, 0.10] | 0.09[ 0.02, 0.15] | 38.3[ 37.6, 38.5] | 38.5[ 37.8, 38.7] | | | |
| 52 | NGC 4594 | AGN | *Chandra* | 2.25[ 1.90, 2.53] | 8.02[ 7.22, 8.91] | 39.6[ 39.6, 39.7] | 40.0[ 39.9, 40.0] | 0/ 0% | 100/ 100% | 0/ 0% |
| 53 | NGC 4636 | Non-AGN | *XMM-Newton* | 18.06[ 15.34, 18.33] | 0.42[ 0.39, 0.44] | 40.9[ 40.8, 40.9] | 39.0[ 0.0, 44.4] | 100/ 76% | 0/ 24% | 0/ 0% |
| 54 | NGC 4676A | Non-AGN | *Chandra* | 0.05[ 0.03, 0.07] | 0.07[ 0.04, 0.11] | 39.7[ 39.4, 39.9] | 39.9[ 39.6, 40.0] | | | |
| 55 | NGC 4676B | AGN | *Chandra* | 0.09[ 0.06, 0.12] | 0.14[ 0.09, 0.19] | 40.0[ 39.8, 40.1] | 40.1[ 40.0, 40.3] | | | |
| 56 | NGC 4698 | AGN | *Chandra* | 0.11[ 0.07, 0.14] | 0.16[ 0.11, 0.22] | 38.6[ 38.4, 38.7] | 38.7[ 38.6, 38.9] | | | |
| 57 | NGC 4696 | Non-AGN | *Chandra* | 2.44[ 0.22, 3.07] | 0.41[ 0.10, 0.63] | 41.6[ 40.6, 41.7] | 40.0[ 39.3, 40.2] | 95/ 33% | 5/ 67% | 0/ 0% |
| 58 | NGC 4736 | AGN | *Chandra* | 1.17[ 0.83, 1.29] | 1.32[ 0.95, 1.54] | 38.8[ 38.6, 38.8] | 38.6[ 38.5, 38.7] | 24/ 18% | 76/ 82% | 0/ 0% |

**Table 8.** Continuation

| Num | Name | Type | Instrument | Flux(0.5-2.0keV) ($10^{-13}$ erg s$^{-1}$) | Flux(2.0-10.0keV) ($10^{-13}$ erg s$^{-1}$) | Log(L(0.5-2.0keV)) | Log(L(2.0-10.0keV)) | MEKAL(*) | Power-law(*) | Power-law(*) |
|---|---|---|---|---|---|---|---|---|---|---|
| (1) | (2) | (3) | (4) | (5) | (6) | (7) | (8) | (9) | (10) | (11) |
| 59 | NGC 5005 | AGN* | *XMM-Newton* | 3.33[ 1.36, 3.51] | 3.75[ 0.00, 259.58] | 40.7[ 40.2, 40.7] | 39.9[ 0.0, 41.7] | 52/ 1% | 48/ 99% | 0/ 0% |
| 60 | NGC 5055 | AGN | *Chandra* | 0.18[ 0.04, 0.24] | 0.09[ 0.04, 0.13] | 38.6[ 38.1, 38.7] | 37.8[ 37.4, 37.9] | 0/ 0% | 100/ 100% | 0/ 0% |
| 61 | MRK 266NE | AGN | *Chandra* | 0.25[ 0.11, 0.33] | 2.15[ 0.24, 3.48] | 41.0[ 40.7, 41.1] | 41.7[ 40.8, 41.9] | | | |
| 62 | UGC 08696 | AGN | *Chandra* | 0.25[ 0.00, 0.34] | 5.37[ 0.29, 6.20] | 43.2[ 41.9, 43.3] | 43.0[ 41.7, 43.1] | 73/ 11% | 27/ 5% | 0/ 84% |
| 63 | CGCG 162-010 | Non-AGN | *Chandra* | 0.87[ 0.46, 1.12] | 0.27[ 0.18, 0.36] | 42.0[ 41.8, 42.0] | 41.4[ 41.3, 41.5] | 100/ 100% | 0/ 0% | 0/ 0% |
| 64 | NGC 5363 | AGN* | *XMM-Newton* | 1.27[ 0.65, 1.35] | 1.88[ 0.19, 2.59] | 39.9[ 39.6, 39.9] | 39.8[ 38.7, 39.9] | 54/ 38% | 40/ 6% | 6/ 57% |
| 65 | IC 4395 | Non-AGN* | *XMM-Newton* | 0.40[ 0.06, 0.38] | 0.23[ 0.00, 168.79] | 41.1[ 40.3, 41.0] | 40.8[ 0.0, 43.7] | 51/ 15% | 49/ 85% | 0/ 0% |
| 66 | IRAS 14348-1447 | Non-AGN | *XMM-Newton* | 0.23[ 0.14, 0.32] | 0.35[ 0.11, 0.56] | 41.6[ 41.4, 41.7] | 41.7[ 41.2, 42.0] | 100/ 100% | 0/ 0% | 0/ 0% |
| 67 | NGC 5746 | AGN | *Chandra* | 0.18[ 0.04, 0.24] | 1.53[ 0.50, 2.01] | 39.7[ 39.0, 39.8] | 40.2[ 39.8, 40.3] | 0/ 0% | 100/ 100% | 0/ 0% |
| 68 | NGC 5813 | Non-AGN | *Chandra* | 0.61[ 0.20, 0.72] | 0.04[ 0.01, 0.07] | 40.4[ 39.9, 40.4] | 38.8[ 38.3, 39.0] | 78/ 9% | 22/ 91% | 0/ 0% |
| 69 | NGC 5838 | AGN | *Chandra* | 0.10[ 0.05, 0.15] | 0.16[ 0.08, 0.24] | 39.0[ 38.7, 39.2] | 39.2[ 38.9, 39.4] | | | |
| 70 | NGC 5846 | Non-AGN | *Chandra* | 0.77[ 0.05, 0.90] | 9.44[ 0.00, 2306.31] | 40.2[ 39.4, 40.3] | 40.8[ 0.0, 43.2] | 76/ 37% | 24/ 63% | 0/ 0% |
| 71 | NGC 5866 | Non-AGN | *Chandra* | 0.05[ 0.03, 0.07] | 0.07[ 0.04, 0.10] | 38.1[ 37.9, 38.3] | 38.3[ 38.0, 38.5] | | | |
| 72 | MRK 0848 | Non-AGN | *XMM-Newton* | 0.35[ 0.09, 0.46] | 0.42[ 0.34, 1.07] | 41.2[ 40.5, 41.3] | 41.1[ 41.0, 41.6] | 0/ 0% | 100/ 100% | 0/ 0% |
| 73 | NGC 6251 | AGN | *Chandra* | 2.17[ 0.89, 2.14] | 3.62[ 2.70, 4.27] | 41.5[ 41.0, 41.5] | 41.6[ 41.5, 41.7] | 38/ 3% | 62/ 97% | 0/ 0% |
| 74 | NGC 6240 | AGN | *Chandra* | 1.45[ 1.04, 1.72] | 10.40[ 7.11, 12.03] | 42.2[ 42.1, 42.3] | 42.4[ 42.2, 42.5] | 92/ 50% | 8/ 16% | 0/ 33% |
| 75 | IRAS 17208-0014 | AGN | *Chandra* | 0.07[ 0.01, 0.10] | 0.40[ 0.21, 0.53] | 40.8[ 40.3, 40.9] | 41.2[ 40.9, 41.3] | 0/ 0% | 100/ 100% | 0/ 0% |
| 76 | NGC 6482 | Non-AGN | *Chandra* | 0.68[ 0.41, 0.83] | 0.05[ 0.03, 0.07] | 40.7[ 40.5, 40.8] | 39.3[ 39.1, 39.5] | 100/ 100% | 0/ 0% | 0/ 0% |
| 77 | NGC 7130 | AGN | *Chandra* | 1.35[ 0.70, 1.55] | 0.93[ 0.46, 1.10] | 41.8[ 41.5, 41.8] | 40.8[ 40.5, 40.9] | 60/ 34% | 40/ 14% | 0/ 51% |
| 78 | NGC 7285 | AGN* | *XMM-Newton* | 0.80[ 0.38, 0.94] | 4.86[ 2.74, 5.69] | 41.8[ 41.5, 41.9] | 41.3[ 41.1, 41.4] | 19/ 4% | 81/ 96% | 0/ 0% |
| 79 | NGC 7331 | Non-AGN | *Chandra* | 0.08[ 0.05, 0.10] | 0.14[ 0.09, 0.19] | 38.3[ 38.1, 38.4] | 38.5[ 38.2, 38.6] | | | |
| 80 | IC 1459 | AGN | *Chandra* | 2.92[ 2.07, 4.83] | 5.21[ 2.48, 6.57] | 40.6[ 40.4, 40.8] | 40.5[ 40.1, 40.6] | 75/ 61% | 13/ 3% | 12/ 37% |
| 81 | NPM1G -12.0625 | Non-AGN | *Chandra* | 0.41[ 0.05, 0.48] | 0.18[ 0.00, 46.65] | 42.6[ 41.1, 42.7] | 41.5[ 0.0, 43.9] | 42/ 2% | 58/ 98% | 0/ 0% |
| 82 | NGC 7743 | Non-AGN* | *XMM-Newton* | 0.53[ 0.00, 0.60] | 0.44[ 0.00, 162.68] | 40.5[ 0.0, 40.6] | 39.5[ 0.0, 42.1] | 70/ 10% | 30/ 90% | 0/ 0% |

(*) Fraction if the flux of the various components at (0.5-2 keV)/(2-10 keV) energy ranges.

**Table 9.** Number of objects per spectral model best fit.

|  | ME (1) | PL (2) | MEPL (3) | 2PL (4) | ME2PL (5) |
|---|---|---|---|---|---|
| Chandra(44) | 3 | 7 | 17 | 0 | 17 |
| XMM-Newton(44) | 5 | 4 | 17 | 1 | 17 |
| Total(60) | 6 | 9 | 24 | 1 | 20 |

**Table 10.** Median and standard deviation properties for the final compilation of our LINER sample in X-rays. [*] **Kolmogorov-Smirnov test probability that the two distributions arise from the same parent distribution.**

|  | Chandra (1) | XMM-Newton (2) | Total (3) | AGN (4) | Non-AGN (5) | **K-S prob.**[*] (6) |
|---|---|---|---|---|---|---|
| $\Gamma$ | 2.05±0.42 | 1.99±0.37 | 2.11±0.52 | 2.02±0.48 | 2.43±0.57 | 66% |
| T(keV) | 0.57±0.22 | 0.61±0.38 | 0.54±0.30 | 0.58±0.30 | 0.35±0.27 | 61% |
| Log(NH1) | 21.45±0.14 | 21.00±0.74 | 21.32±0.71 | 21.32±0.71 | 21.43±0.71 | 99% |
| Log(NH2) | 21.93±1.02 | 21.94±1.23 | 21.93±1.36 | 22.36±1.33 | 20.00±0.77 | 2% |
| Log(L(0.5 − 2 keV)) | 39.95±1.19 | 41.1±1.4 | 40.22±1.33 | 40.60±1.19 | 39.88±1.47 | 21% |
| Log(L(2 − 10 keV)) | 39.85±1.17 | 40.3±1.5 | 39.85±1.26 | 40.22±1.24 | 39.33±1.16 | 29% |
| EW(eV) | 164±160 | 144±213 | 211±214 | 144±167 | 296±442 |  |

**Table 11.** Final compilation of EW(FeK$\alpha$) for the LINER sample

| Num (1) | Name (2) | EW(FeK$\alpha$) (eV) (3) | EW(FeXXV) (eV) (4) | EW(FeXXVI) (eV) (5) |
|---|---|---|---|---|
| 1 | NGC 0315 | 81.89 $^{163.69}_{0.09}$ | 137.37 $^{227.26}_{47.46}$ | 27.08 $^{118.67}_{0.00}$ |
| 2 | NGC 0410 | 47.02 $^{413.54}_{0.00}$ | 1360.47 $^{2466.75}_{249.18}$ | 39.11 $^{474.56}_{0.00}$ |
| 3 | NGC 0474 | ... | ... | ... |
| 4 | IIIZW 035 | ... | ... | ... |
| 5 | NGC 0524 | ... | ... | ... |
| 6 | NGC 0833 | 334.40 $^{621.63}_{47.17}$ | 78.79 $^{228.90}_{0.00}$ | 335.18 $^{736.92}_{0.00}$ |
| 7 | NGC 0835 | 774.92 $^{1065.17}_{485.49}$ | 0.00 $^{59.05}_{0.00}$ | 0.00 $^{493.06}_{0.00}$ |
| 8 | NGC 1052 | 144.40 $^{166.96}_{121.84}$ | 21.51 $^{37.23}_{5.79}$ | 43.36 $^{64.94}_{21.77}$ |
| 9 | NGC 2639 | ... | ... | ... |
| 10 | NGC 2655 | 24.69 $^{155.51}_{0.00}$ | 0.00 $^{131.61}_{0.00}$ | 45.99 $^{189.26}_{0.00}$ |
| 11 | NGC 2681 | 929.23 $^{2211.99}_{0.00}$ | 19.65 $^{593.19}_{0.00}$ | 301.86 $^{2219.93}_{0.00}$ |
| 12 | NGC 2685 | ... | ... | ... |
| 13 | UGC 4881 | ... | ... | ... |
| 14 | 3C 218 | 0.00 $^{6.13}_{0.00}$ | 51.67 $^{81.49}_{18.64}$ | 44.36 $^{67.29}_{21.58}$ |
| 15 | NGC 2787 | 0.00 $^{291.35}_{0.00}$ | 0.00 $^{331.56}_{0.00}$ | 98.79 $^{535.73}_{0.00}$ |
| 16 | NGC 2841 | 0.00 $^{240.58}_{0.00}$ | 0.00 $^{405.89}_{0.00}$ | 0.00 $^{370.30}_{0.00}$ |
| 17 | UGC 05101 | 278.37 $^{456.52}_{100.22}$ | 151.84 $^{263.90}_{41.68}$ | 80.61 $^{230.46}_{0.00}$ |
| 18 | NGC 3185 | ... | ... | ... |
| 19 | NGC 3226 | 33.67 $^{111.38}_{0.00}$ | 0.00 $^{30.43}_{0.00}$ | 47.39 $^{136.50}_{0.00}$ |
| 20 | NGC 3245 | ... | ... | ... |
| 21 | NGC 3379 | ... | ... | ... |
| 22 | NGC 3414 | 0.00 $^{588.89}_{0.00}$ | 0.00 $^{442.41}_{0.00}$ | 0.00 $^{416.08}_{0.00}$ |
| 23 | NGC 3507 | ... | ... | ... |
| 24 | NGC 3607 | ... | ... | ... |
| 25 | NGC 3608 | ... | ... | ... |
| 26 | NGC 3623 | ... | ... | ... |
| 27 | NGC 3627 | ... | ... | ... |
| 28 | NGC 3628 | 0.00 $^{77.39}_{0.00}$ | 98.39 $^{209.02}_{0.00}$ | 17.14 $^{120.28}_{0.00}$ |
| 29 | NGC 3690B | 233.42 $^{341.24}_{125.60}$ | 165.42 $^{255.45}_{75.37}$ | 145.73 $^{262.05}_{29.41}$ |
| 30 | NGC 3898 | ... | ... | ... |
| 31 | NGC 3945 | 0.00 $^{109.52}_{0.00}$ | 0.00 $^{178.38}_{0.00}$ | 0.00 $^{248.41}_{0.00}$ |
| 32 | NGC 3998 | 0.00 $^{31.26}_{0.00}$ | 17.29 $^{54.24}_{0.00}$ | 13.31 $^{53.78}_{0.00}$ |
| 33 | NGC 4036 | ... | ... | ... |
| 34 | NGC 4111 | 0.00 $^{179.91}_{0.00}$ | 0.00 $^{237.77}_{0.00}$ | 0.00 $^{414.51}_{0.00}$ |
| 35 | NGC 4125 | 0.00 $^{1559.70}_{0.00}$ | 67.00 $^{2056.16}_{0.00}$ | 330.40 $^{2947.66}_{0.00}$ |
| 36 | IRAS 12112+0305 | ... | ... | ... |
| 37 | NGC 4261 | 26.11 $^{105.52}_{0.00}$ | 21.86 $^{97.77}_{0.00}$ | 70.68 $^{173.30}_{0.00}$ |
| 38 | NGC 4278 | 0.00 $^{48.20}_{0.00}$ | 0.00 $^{29.31}_{0.00}$ | 0.00 $^{29.63}_{0.00}$ |
| 39 | NGC 4314 | 708.79 $^{1590.35}_{0.00}$ | 0.00 $^{492.54}_{0.00}$ | 204.87 $^{1029.44}_{0.00}$ |
| 40 | NGC 4321 | 0.00 $^{170.94}_{0.00}$ | 344.52 $^{705.80}_{0.00}$ | 231.42 $^{584.46}_{0.00}$ |
| 41 | NGC 4374 | 661.31 $^{1595.10}_{0.00}$ | 0.00 $^{461.28}_{0.00}$ | 0.00 $^{1260.98}_{0.00}$ |
| 42 | NGC 4410A | 117.55 $^{432.16}_{0.00}$ | 245.40 $^{583.21}_{0.00}$ | 205.63 $^{594.11}_{0.00}$ |
| 43 | NGC 4438 | 222.32 $^{2462.99}_{0.00}$ | 13.82 $^{2245.79}_{0.00}$ | 2.51 $^{0.00}_{0.00}$ |
| 44 | NGC 4457 | 0.00 $^{61.92}_{0.00}$ | 0.00 $^{98.06}_{0.00}$ | 0.00 $^{192.89}_{0.00}$ |
| 45 | NGC 4459 | ... | ... | ... |
| 46 | NGC 4486 | 84.04 $^{131.24}_{36.85}$ | 134.79 $^{192.02}_{77.57}$ | 139.92 $^{212.67}_{67.17}$ |
| 47 | NGC 4494 | 0.00 $^{374.52}_{0.00}$ | 0.00 $^{392.19}_{0.00}$ | 0.00 $^{444.36}_{0.00}$ |
| 48 | NGC 4552 | 24.06 $^{317.93}_{0.00}$ | 0.00 $^{272.73}_{0.00}$ | 0.00 $^{314.01}_{0.00}$ |
| 49 | NGC 4589 | ... | ... | ... |

**Table 11.** Continuation

| Num | Name | EW(FeK$\alpha$) (eV) | EW(FeXXV) (eV) | EW(FeXXVI) (eV) |
|---|---|---|---|---|
| (1) | (2) | (3) | (4) | (5) |
| 50 | NGC 4579 | 115.05 $^{157.14}_{72.96}$ | 45.49 $^{81.07}_{9.91}$ | 75.31 $^{121.61}_{29.02}$ |
| 51 | NGC 4596 | ... | ... | ... |
| 52 | NGC 4594 | 28.49 $^{106.10}_{0.00}$ | 96.32 $^{178.57}_{0.00}$ | 0.00 $^{41.78}_{0.00}$ |
| 53 | NGC 4636 | 296.01 $^{1753.93}_{0.00}$ | 346.35 $^{1171.77}_{0.00}$ | 550.43 $^{2472.83}_{0.00}$ |
| 54 | NGC 4676A | ... | ... | ... |
| 55 | NGC 4676B | ... | ... | ... |
| 56 | NGC 4698 | ... | ... | ... |
| 57 | NGC 4696 | ... | ... | ... |
| 58 | NGC 4736 | 8.71 $^{98.56}_{0.00}$ | 56.15 $^{133.71}_{0.00}$ | 0.00 $^{95.73}_{0.00}$ |
| 59 | NGC 5005 | ... | ... | ... |
| 60 | NGC 5055 | 0.00 $^{35.22}_{0.00}$ | 0.00 $^{52.90}_{0.00}$ | 0.00 $^{107.18}_{0.00}$ |
| 61 | MRK 266NE | 275.72 $^{481.38}_{71.24}$ | 47.65 $^{178.02}_{0.00}$ | 0.00 $^{146.25}_{0.00}$ |
| 62 | UGC 08696 | 266.11 $^{382.32}_{149.90}$ | 135.94 $^{226.84}_{45.04}$ | 70.24 $^{173.71}_{0.00}$ |
| 63 | CGCG 162-010 | 0.00 $^{4.18}_{0.00}$ | 105.80 $^{124.19}_{86.36}$ | 17.29 $^{28.43}_{6.18}$ |
| 64 | NGC 5363 | 50.45 $^{337.81}_{0.00}$ | 351.88 $^{852.44}_{0.00}$ | 54.84 $^{700.98}_{0.00}$ |
| 65 | IC 4395 | ... | ... | ... |
| 66 | IRAS 14348-1447 | ... | ... | ... |
| 67 | NGC 5746 | 41.46 $^{375.02}_{0.00}$ | 0.00 $^{213.71}_{0.00}$ | 129.20 $^{563.96}_{0.00}$ |
| 68 | NGC 5813 | 0.00 $^{2609.74}_{0.00}$ | 0.00 $^{2206.17}_{0.00}$ | 0.00 $^{2453.49}_{0.00}$ |
| 69 | NGC 5838 | ... | ... | ... |
| 70 | NGC 5846 | 0.00 $^{177.73}_{0.00}$ | 0.00 $^{204.29}_{0.00}$ | 0.00 $^{248.38}_{0.00}$ |
| 71 | NGC 5866 | ... | ... | ... |
| 72 | MRK 0848 | 196.92 $^{2398.18}_{0.00}$ | 4.16 $^{916.54}_{0.00}$ | 1777.01 $^{5307.68}_{0.00}$ |
| 73 | NGC 6251 | 24.30 $^{53.49}_{0.70}$ | 32.90 $^{63.18}_{0.70}$ | 0.00 $^{21.82}_{0.00}$ |
| 74 | NGC 6240 | 378.06 $^{441.86}_{314.27}$ | 115.35 $^{151.57}_{79.14}$ | 105.08 $^{159.76}_{50.40}$ |
| 75 | IRAS 17208-0014 | 217.03 $^{558.42}_{0.00}$ | 427.05 $^{851.02}_{0.00}$ | 462.96 $^{1020.93}_{0.00}$ |
| 76 | NGC 6482 | ... | ... | ... |
| 77 | NGC 7130 | 382.06 $^{633.04}_{131.08}$ | 147.04 $^{333.58}_{0.00}$ | 34.44 $^{307.20}_{0.00}$ |
| 78 | NGC 7285 | 211.68 $^{382.09}_{39.28}$ | 59.20 $^{247.70}_{0.00}$ | 61.49 $^{295.86}_{0.00}$ |
| 79 | NGC 7331 | ... | ... | ... |
| 80 | IC 1459 | 0.00 $^{72.47}_{0.00}$ | 0.00 $^{76.96}_{0.00}$ | 41.38 $^{173.16}_{0.00}$ |
| 81 | NPM1G -12.0625 | 0.00 $^{4.35}_{0.00}$ | 38.34 $^{58.34}_{16.40}$ | 26.48 $^{41.97}_{11.06}$ |
| 82 | NGC 7743 | 1746.22 $^{8337.29}_{0.00}$ | 0.00 $^{3065.32}_{0.00}$ | 0.00 $^{2121.33}_{0.00}$ |

**Table 12.** Multiwavelength properties of LINERs*.

| Num | Name | X-ray Class. | Log($L_X$) (2-10 keV) | HST Class. | Broad H$\alpha$ | Radio Comp. | Stellar Pop. (%) | UV Var. | X-ray Var. | UV Comp. |
|---|---|---|---|---|---|---|---|---|---|---|
| (1) | (2) | (3) | (4) | (5) | (6) | (7) | (8) | (9) | (10) | (11) |
| 1 | NGC 0315 | AGN | 41.768 | C | Yes | Yes J (1,3) | 3.8 | ... | ... | ... |
| 2 | NGC 0410 | Non-AGN* | 40.715 | ... | ... | No (1,5) | 2.7 | ... | ... | ... |
| 3 | NGC 0474 | Non-AGN | <38.478 | ... | ... | No (1) | ... | ... | ... | ... |
| 4 | IIIZW 035 | Non-AGN | 40.048 | C | ... | Yes F (7) | ... | ... | ... | ... |
| 5 | NGC 0524 | Non-AGN | 38.586 | D | ... | No (8) | ... | ... | ... | ... |
| 6 | NGC 0833 | AGN | 41.734 | ... | ... | No (8) | ... | ... | ... | ... |
| 7 | NGC 0835 | AGN | 41.399 | ... | ... | Yes S (8) | ... | ... | ... | ... |
| 8 | NGC 1052 | AGN | 41.240 | C | Yes | Yes (21) | 4.0 | ... | ... | Yes(1) |
| 9 | NGC 2639 | Non-AGN* | <38.277 | C | Yes | Yes F (9) | ... | ... | ... | ... |
| 10 | NGC 2655 | AGN* | 41.234 | C | ... | Yes S (1) | ... | ... | ... | ... |
| 11 | NGC 2681 | AGN | 39.275 | C | Yes | No (1) | 4.2 | ... | ... | Yes(1) |
| 12 | NGC 2685 | AGN* | 39.042 | ... | ... | No (1) | 1.4 | ... | ... | ... |
| 13 | UGC 4881 | Non-AGN | <38.369 | C | ... | ... | ... | ... | ... | ... |
| 14 | 3C 218 | AGN | 42.081 | ... | ... | Yes J (10) | ... | ... | ... | ... |
| 15 | NGC 2787 | AGN | <38.805 | C | Yes | Yes (1) | ... | ... | ... | No(1) |
| 16 | NGC 2841 | AGN | 39.219 | C | ... | ... | ... | ... | ... | ... |
| 17 | UGC 05101 | AGN | 42.066 | C | ... | Yes S (7) | ... | ... | ... | ... |
| 18 | NGC 3185 | Non-AGN* | 39.374 | C | ... | Yes marg.(9) | ... | ... | ... | ... |
| 19 | NGC 3226 | AGN | 40.800 | ... | Yes | Yes F (5) | 2.2 | ... | ... | ... |
| 20 | NGC 3245 | AGN | 38.982 | C | ... | No (1) | 6.2 | ... | ... | ... |
| 21 | NGC 3379 | Non-AGN | 38.131 | C | ... | No (1) | ... | ... | ... | ... |
| 22 | NGC 3414 | AGN | 39.863 | ... | ... | Yes F (1) | ... | ... | ... | ... |
| 23 | NGC 3507 | Non-AGN | <37.204 | C | ... | No (1) | 32.6 | ... | ... | ... |
| 24 | NGC 3607 | Non-AGN | 38.765 | C | ... | Yes (1) | ... | ... | ... | ... |
| 25 | NGC 3608 | Non-AGN | 38.205 | C | ... | No (1) | ... | ... | ... | ... |
| 26 | NGC 3623 | Non-AGN* | <39.383 | C | ... | No (1) | ... | ... | ... | ... |
| 27 | NGC 3627 | Non-AGN* | 39.415 | D | ... | Yes F (2,4) | 5.0(7.3) | ... | ... | ... |
| 28 | NGC 3628 | Non-AGN | 39.942 | U | ... | No Ext. (1,2) | ... | ... | ... | ... |
| 29 | NGC 3690B | AGN | 40.862 | C | ... | Yes S (11) | ... | ... | ... | ... |
| 30 | NGC 3898 | Non-AGN | <38.776 | C | ... | No (1) | ... | ... | ... | ... |
| 31 | NGC 3945 | AGN | 39.121 | C | ... | Yes (1) | ... | ... | ... | ... |
| 32 | NGC 3998 | AGN | 41.319 | C | Yes | Yes (1) | 54.4 | Yes | ... | Yes(2) |
| 33 | NGC 4036 | AGN | 39.126 | C | Yes | No (1) | ... | ... | ... | ... |
| 34 | NGC 4111 | AGN | <40.364 | C? | ... | No (1) | ... | ... | ... | No(1) |
| 35 | NGC 4125 | AGN | <38.730 | C | ... | No (1) | ... | ... | ... | ... |
| 36 | IRAS 12112+0305 | Non-AGN* | 41.226 | ... | ... | Yes F (7) | ... | ... | ... | ... |
| 37 | NGC 4261 | AGN | 41.069 | U | ... | Yes J (1) | 2.2 | ... | ... | ... |
| 38 | NGC 4278 | AGN | 39.226 | C | Yes | Yes J (1) | ... | ... | ... | ... |
| 39 | NGC 4314 | Non-AGN | <39.095 | C | ... | No (1) | 3.8 | ... | ... | No(1) |

**Table 12.** Continuation

| Num | Name | X-ray Class. | Log($L_X$) (2-10 keV) | HST Class. | Broad H$\alpha$ | Radio Comp. | Stellar Pop. (%) | UV Var. | X-ray Var. | UV Comp. |
|---|---|---|---|---|---|---|---|---|---|---|
| (1) | (2) | (3) | (4) | (5) | (6) | (7) | (8) | (9) | (10) | (11) |
| 40 | NGC 4321 | Non-AGN | 40.491 | C | ..... | No Ext. S (5) | 24.7 | ... | ..... | ..... |
| 41 | NGC 4374 | AGN | 39.533 | C | ..... | Yes J (1) | ..... | ... | ..... | ..... |
| 42 | NGC 4410A | AGN | 41.176 | C | Yes | Yes (5) | ..... | ... | ..... | ..... |
| 43 | NGC 4438 | AGN | <39.047 | D | Yes | No (1) | 2.2 | ... | ..... | No(1) |
| 44 | NGC 4457 | AGN | 38.814 | C | ..... | No (1) | ..... | ... | ..... | ..... |
| 45 | NGC 4459 | Non-AGN | 38.372 | C | ..... | No (1) | 8.7 | ... | ..... | ..... |
| 46 | NGC 4486 | AGN | 40.823 | C | Yes | Yes (1) | ..... | Yes | Yes | Yes(2) |
| 47 | NGC 4494 | AGN | 38.777 | C | ..... | No (1) | ..... | ... | ..... | ..... |
| 48 | NGC 4552 | AGN | 39.249 | C | ..... | Yes (1) | ..... | Yes | Yes | Yes(2) |
| 49 | NGC 4589 | Non-AGN | 38.917 | C | ..... | Yes (1,5) | ..... | ... | ..... | ..... |
| 50 | NGC 4579 | AGN | 41.175 | C | Yes | Yes (1) | ..... | Yes | ..... | Yes(2) |
| 51 | NGC 4596 | Non-AGN | 38.467 | C | ..... | No (1) | 4.1 | ... | ..... | ..... |
| 52 | NGC 4594 | AGN | 39.973 | C | ..... | Yes (13) | ..... | Yes | ..... | Yes(2) |
| 53 | NGC 4636 | Non-AGN | <39.026 | D | Yes | Yes (1) | ..... | ... | ..... | No(1) |
| 54 | NGC 4676A | Non-AGN | 39.855 | D | ..... | ..... | ..... | ... | ..... | ..... |
| 55 | NGC 4676B | AGN | 40.135 | C | ..... | ..... | ..... | ... | ..... | ..... |
| 56 | NGC 4698 | AGN | 38.744 | C | ..... | No (1) | ..... | ... | ..... | ..... |
| 57 | NGC 4696 | Non-AGN | 39.978 | C | ..... | Yes F (14) | ..... | ... | ..... | ..... |
| 58 | NGC 4736 | AGN | 38.599 | C | ..... | Yes (1) | 5.3 | Yes | ..... | Yes(1) |
| 59 | NGC 5005 | AGN* | <39.855 | C? | Yes | No (1) | 3.6 | ... | Yes | ..... |
| 60 | NGC 5055 | AGN | 37.778 | C | ..... | No (1) | 5.7(14.0) | ... | ..... | Yes(2) |
| 61 | MRK 266NE | AGN | 41.656 | C | ..... | ..... | ..... | ... | ..... | ..... |
| 62 | UGC 08696 | AGN | 42.993 | C | ..... | Yes F (7) | ..... | ... | ..... | ..... |
| 63 | CGCG 162-010 | Non-AGN | 41.426 | C | ..... | Yes F (15) | ..... | ... | ..... | ..... |
| 64 | NGC 5363 | AGN* | 39.782 | ..... | ..... | Yes J (1) | ..... | ... | ..... | ..... |
| 65 | IC 4395 | Non-AGN* | <40.797 | ..... | ..... | ..... | ..... | ... | ..... | ..... |
| 66 | IRAS 14348-1447 | Non-AGN | 41.737 | C | ..... | Yes F (11) | ..... | ... | ..... | ..... |
| 67 | NGC 5746 | AGN | 40.223 | C? | ..... | No (1) | ..... | ... | ..... | ..... |
| 68 | NGC 5813 | Non-AGN | 38.770 | C | ..... | Yes (1) | ..... | ... | ..... | ..... |
| 69 | NGC 5838 | AGN | 39.206 | C | ..... | Yes F (2) | ..... | ... | ..... | ..... |
| 70 | NGC 5846 | Non-AGN | <40.814 | C | ..... | Yes J (1,2,4) | ..... | ... | ..... | ..... |
| 71 | NGC 5866 | Non-AGN | 38.296 | ..... | ..... | Yes F (1,2,4) | ..... | ... | ..... | ..... |
| 72 | MRK 0848 | Non-AGN | 41.145 | C | ..... | Yes F (16) | ..... | ... | ..... | ..... |
| 73 | NGC 6251 | AGN | 41.583 | C | ..... | Yes J (17,18) | ..... | ... | ..... | ..... |
| 74 | NGC 6240 | AGN | 42.408 | C | ..... | Yes (19) | ..... | ... | ..... | ..... |
| 75 | IRAS 17208-0014 | AGN | 41.188 | C | ..... | Yes F (7) | ..... | ... | ..... | ..... |
| 76 | NGC 6482 | Non-AGN | 39.335 | ..... | ..... | No (1) | 1.7 | ... | ..... | ..... |
| 77 | NGC 7130 | AGN | 40.795 | C | ..... | Yes S (8) | ..... | ... | ..... | ..... |
| 78 | NGC 7285 | AGN* | 41.323 | ..... | ..... | ..... | ..... | ... | ..... | ..... |
| 79 | NGC 7331 | Non-AGN | 38.456 | C | ..... | Yes F (3) | 1.9(2.1) | ... | ..... | ..... |
| 80 | IC 1459 | AGN | 40.511 | C | ..... | Yes F (14) | ..... | ... | ..... | ..... |
| 81 | NPM1G -12.0625 | Non-AGN | <41.463 | D | ..... | Yes S (20) | ..... | ... | ..... | ..... |
| 82 | NGC 7743 | Non-AGN* | <39.548 | C | ..... | Yes F (9) | ..... | ... | ..... | ..... |

C-T = Compton-Thick candidates. (1) Chiaberge et al. (2005); (2) Maoz et al. (2005). Col. (7) Stellar populations younger than $10^9$ yr taken from Cid-Fernandes et al. (2004) and Gonzalez-Delgado et al. (2004) (within parentesis)
Col. () Broad H$\alpha$ reported by Ho et al. (1997)
Col. (6) References: (1) Nagar et al. (2005); (2) Filho et al. (2000); (3) Filho et al. (2002); (4) Filho et al. (2004); (5) Filho et al. (2006); (6) Anderson and Ulvestad (2005); (7) Baan and Klockner (2006); (8) Corbett et al. (2002); (9) Ho and Ulvestad (2001); (10) Taylor et al. (1990); (11) Condon and Broderick (1991); (12) Smith (2000); (13) Bajaja et al. (1988); (14) Slee et al. (1994); (15) van Breugel et al. (1984); (16) Clemens et al. (2008); (17) Urry and Padovani (1995); (18) Jones et al. (1986); (19) Carral et al. (1990); (20) Sarazin et al. (1995); (21) Vermeulen et al. (2003) Col. (13) Interacting types: 0 = Isolate, 1 = Merger, 1.5 = Close Interacting Pair, 2 = Pair, 2.5 = Wide Pair, 3 = Triplet, 4 = Compact Group, 4.5 = 1st group, 5 = Group, 10 = Cluster Center and, 20 = Cluster Member

**Table 13.** Bestfit model applied to the diffuse emission extracted from *Chandra* data.

| Name | Bestfit | $N_{H1}$ | $N_{H2}$ | $\Gamma$ | kT (keV) | $\chi^2$ | d.o.f | $\chi^2_R$ |
|---|---|---|---|---|---|---|---|---|
| (1) | (2) | (3) | (4) | (5) | (6) | (7) | (8) | (9) |
| NGC 0315 | ME | 0.04 $^{0.13}_{0.00}$ | ... | ... | 0.58 $^{0.62}_{0.52}$ | 90.91 | 79 | 1.15 |
| 3C 218 | ME | 0.00 $^{0.00}_{0.00}$ | ... | ... | 3.00 $^{3.09}_{2.95}$ | 563.84 | 413 | 1.37 |
| NGC 4111 | ME | 0.00 $^{0.07}_{0.00}$ | ... | ... | 0.53 $^{0.60}_{0.49}$ | 23.18 | 35 | 0.66 |
| NGC 4125 | ME | 0.04 $^{0.37}_{0.00}$ | ... | ... | 0.44 $^{0.58}_{0.26}$ | 57.63 | 61 | 0.94 |
| NGC 4261 | ME | 0.00 $^{0.03}_{0.00}$ | ... | ... | 0.60 $^{0.64}_{0.57}$ | 71.68 | 82 | 0.87 |
| NGC 4278 | ME | 0.51 $^{0.74}_{0.29}$ | ... | ... | 0.54 $^{0.66}_{0.00}$ | 78.08 | 67 | 1.17 |
| NGC 4321 | ME | 0.06 $^{0.21}_{0.00}$ | ... | ... | 0.50 $^{0.63}_{0.23}$ | 62.10 | 48 | 1.29 |
| NGC 4374 | MEPL | 0.00 $^{0.13}_{0.00}$ | 0.00 $^{0.47}_{0.00}$ | 1.93 $^{2.96}_{1.29}$ | 0.52 $^{0.61}_{0.47}$ | 60.56 | 68 | 0.89 |
| NGC 4486 | MEPL | 0.10 $^{0.13}_{0.07}$ | 0.13 $^{0.16}_{0.10}$ | 2.70 $^{2.83}_{2.56}$ | 1.08 $^{1.09}_{1.07}$ | 697.62 | 375 | 1.86 |
| NGC 4552 | ME | 0.00 $^{0.03}_{0.00}$ | ... | ... | 0.56 $^{0.59}_{0.53}$ | 154.84 | 103 | 1.50 |
| NGC 4579 | MEPL | 0.00 $^{0.07}_{0.00}$ | 0.00 $^{0.10}_{0.00}$ | 1.98 $^{3.22}_{1.41}$ | 0.31 $^{0.41}_{0.23}$ | 118.32 | 116 | 1.02 |
| NGC 4696 | MEPL | 0.00 $^{0.11}_{0.00}$ | 0.00 $^{0.13}_{0.00}$ | 3.51 $^{4.30}_{2.65}$ | 0.64 $^{0.66}_{0.61}$ | 198.32 | 198 | 1.00 |
| UGC 08696 | ME | 0.44 $^{0.57}_{0.28}$ | ... | ... | 0.62 $^{0.69}_{0.55}$ | 67.39 | 62 | 1.09 |
| CGCG 162-010 | ME | 0.00 $^{0.08}_{0.00}$ | ... | ... | 4.09 $^{4.23}_{3.93}$ | 522.95 | 394 | 1.33 |
| NGC 5813 | ME | 0.10 $^{0.14}_{0.62}$ | ... | ... | 0.61 $^{0.63}_{0.58}$ | 150.65 | 127 | 1.19 |
| NGC 6240 | MEPL | 0.73 $^{0.80}_{0.62}$ | 0.08 $^{0.18}_{0.00}$ | 2.26 $^{2.61}_{1.88}$ | 0.61 $^{0.66}_{0.58}$ | 174.16 | 175 | 1.00 |
| NGC 6482 | ME | 0.09 $^{0.19}_{0.01}$ | ... | ... | 0.76 $^{0.82}_{0.70}$ | 63.62 | 76 | 0.84 |
| IC 1459 | ME | 0.00 $^{0.10}_{0.00}$ | ... | ... | 0.58 $^{0.62}_{0.00}$ | 123.01 | 74 | 1.66 |
| NPM1G -12.0625 | MEPL | 1.00 $^{1.15}_{0.75}$ | 0.23 $^{0.29}_{0.21}$ | 2.61 $^{2.79}_{2.53}$ | 0.37 $^{0.52}_{0.28}$ | 389.61 | 307 | 1.27 |

NOTES: Column density expressed in units of $10^{22}$ cm$^{-2}$.

**Table 14.** Bestfit model applied to *Chandra* data with the *XMM-Newton* extraction region (25 arcsecs).

| Num | Name | Bestfit | $N_{H1}$ | $N_{H2}$ | $\Gamma$ | kT (keV) | $\chi^2$ | d.o.f | $\chi^2_\nu$ |
|---|---|---|---|---|---|---|---|---|---|
| (1) | (2) | (3) | (4) | (5) | (6) | (7) | (8) | (9) | |
| 1 | NGC 0315 | ME2PL | 0.89 $^{1.31}_{0.65}$ | 13.56 $^{18.62}_{10.30}$ | 2.48 $^{3.24}_{2.26}$ | 0.53 $^{0.55}_{0.52}$ | 408.66 | 369 | 1.11 |
| 7 | NGC 0835 | ME2PL | 0.04 $^{0.39}_{0.00}$ | 48.03 $^{90.62}_{21.29}$ | 3.36 $^{4.58}_{2.36}$ | 0.44 $^{0.54}_{0.36}$ | 61.92 | 65 | 0.95 |
| 8 | NGC 1052 | 2PL | 0.00 $^{59.83}_{0.00}$ | 8.31 $^{0.00}_{0.20}$ | 0.47 $^{3.22}_{-0.44}$ | 0.55 $^{0.66}_{0.37}$ | 57.87 | 78 | 0.74 |
| 14 | 3C 218 | MEPL | 0.00 $^{0.04}_{0.00}$ | 0.07 $^{0.09}_{0.02}$ | 2.21 $^{2.26}_{2.07}$ | 0.47 $^{0.83}_{0.44}$ | 484.73 | 427 | 1.14 |
| 15 | NGC 2787 | PL | 0.03 $^{0.08}_{0.00}$ | ... | 1.77 $^{2.01}_{1.60}$ | ... | 136.91 | 167 | 0.82 |
| 17 | UGC 05101 | 2PL | 0.19 $^{0.27}_{0.10}$ | 6.54 $^{13.46}_{4.08}$ | 2.86 $^{3.30}_{1.84}$ | ... | 114.29 | 129 | 0.89 |
| 29 | NGC 3690B | MEPL | 0.00 $^{0.02}_{0.00}$ | 0.00 $^{0.02}_{0.00}$ | 2.07 $^{2.33}_{1.96}$ | 0.19 $^{0.19}_{0.18}$ | 298.07 | 236 | 1.26 |
| 32 | NGC 3998 | MEPL | 0.00 $^{0.16}_{0.00}$ | 0.08 $^{0.11}_{0.03}$ | 1.45 $^{1.98}_{1.39}$ | 0.23 $^{0.25}_{0.19}$ | 530.89 | 450 | 1.18 |
| 35 | NGC 4125 | ME2PL | 0.00 $^{0.03}_{0.00}$ | 55.77 $^{179.87}_{20.79}$ | 2.35 $^{2.76}_{2.18}$ | 0.54 $^{0.57}_{0.51}$ | 192.56 | 194 | 0.99 |
| 37 | NGC 4261 | ME2PL | 0.37 $^{1.58}_{0.21}$ | 18.25 $^{26.04}_{14.73}$ | 2.80 $^{4.87}_{2.27}$ | 0.59 $^{0.61}_{0.58}$ | 346.87 | 296 | 1.17 |
| 38 | NGC 4278 | MEPL | 0.48 $^{0.64}_{0.29}$ | 0.03 $^{0.06}_{0.01}$ | 1.98 $^{2.06}_{1.89}$ | 0.23 $^{0.31}_{0.18}$ | 327.99 | 355 | 0.92 |
| 40 | NGC 4321 | MEPL | 0.00 $^{0.17}_{0.00}$ | 0.02 $^{0.09}_{0.00}$ | 2.32 $^{2.71}_{2.12}$ | 0.59 $^{0.64}_{0.51}$ | 152.67 | 159 | 0.96 |
| 47 | NGC 4494 | MEPL | 0.00 $^{2.42}_{0.00}$ | 0.07 $^{0.23}_{0.00}$ | 1.73 $^{2.34}_{1.28}$ | 0.70 $^{0.86}_{0.47}$ | 84.60 | 92 | 0.92 |
| 48 | NGC 4552 | ME2PL | 0.10 $^{0.16}_{0.05}$ | 15.47 $^{29.23}_{10.04}$ | 2.60 $^{2.94}_{2.45}$ | 0.54 $^{0.56}_{0.52}$ | 269.34 | 253 | 1.06 |
| 50 | NGC 4579 | 2PL | 0.09 $^{0.11}_{0.06}$ | 3.84 $^{4.10}_{3.47}$ | 2.48 $^{2.59}_{2.25}$ | ... | 524.21 | 429 | 1.22 |
| 52 | NGC 4594 | 2PL | 0.19 $^{0.23}_{0.12}$ | 4.37 $^{5.30}_{3.52}$ | 2.69 $^{2.95}_{2.08}$ | ... | 283.02 | 308 | 0.92 |
| 58 | NGC 4736 | MEPL | 0.00 $^{0.02}_{0.00}$ | 0.00 $^{0.01}_{0.00}$ | 1.55 $^{1.98}_{1.52}$ | 0.50 $^{0.51}_{0.47}$ | 386.70 | 384 | 1.01 |
| 61 | MRK 266NE | ME2PL | 0.00 $^{0.10}_{0.00}$ | 11.68 $^{24.59}_{5.23}$ | 2.54 $^{3.35}_{2.33}$ | 0.76 $^{0.84}_{0.64}$ | 141.49 | 132 | 1.07 |
| 62 | UGC 08696 | ME2PL | 0.05 $^{0.11}_{0.00}$ | 48.66 $^{54.13}_{40.06}$ | 2.22 $^{2.52}_{1.80}$ | 0.74 $^{0.82}_{0.66}$ | 243.66 | 265 | 0.92 |
| 63 | CGCG 162-010 | MEPL | 0.06 $^{0.09}_{0.04}$ | 0.01 $^{0.02}_{0.00}$ | 2.01 $^{2.11}_{1.92}$ | 3.25 $^{3.43}_{2.84}$ | 466.00 | 398 | 1.17 |
| 68 | NGC 5813 | MEPL | 0.01 $^{0.05}_{0.00}$ | 0.54 $^{1.17}_{0.29}$ | 3.85 $^{5.27}_{3.12}$ | 0.59 $^{0.60}_{0.57}$ | 243.38 | 192 | 1.27 |
| 70 | NGC 5846 | MEPL | 0.00 $^{0.03}_{0.00}$ | 0.09 $^{0.27}_{0.00}$ | 2.40 $^{3.12}_{1.84}$ | 0.57 $^{0.59}_{0.55}$ | 235.55 | 147 | 1.60 |
| 73 | NGC 6251 | MEPL | 0.00 $^{0.23}_{0.00}$ | 0.00 $^{0.08}_{0.00}$ | 1.53 $^{1.69}_{1.44}$ | 0.21 $^{0.22}_{0.16}$ | 265.07 | 260 | 1.02 |
| 74 | NGC 6240 | ME2PL | 0.22 $^{0.26}_{0.19}$ | 85.75 $^{143.81}_{51.18}$ | 2.06 $^{2.16}_{1.95}$ | 0.83 $^{0.86}_{0.79}$ | 531.24 | 384 | 1.38 |
| 75 | IRAS 17208-0014 | MEPL | 0.83 $^{1.08}_{0.00}$ | 0.05 $^{0.84}_{0.00}$ | 1.41 $^{2.78}_{0.97}$ | 0.22 $^{0.35}_{0.08}$ | 115.00 | 131 | 0.88 |
| 76 | NGC 6482 | ME | 0.20 $^{0.24}_{0.16}$ | ... | ... | 0.67 $^{0.69}_{0.66}$ | 157.58 | 141 | 1.12 |
| 80 | IC 1459 | MEPL | 0.00 $^{0.16}_{0.00}$ | 0.10 $^{0.13}_{0.08}$ | 1.79 $^{1.88}_{1.73}$ | 0.52 $^{0.57}_{0.47}$ | 323.81 | 364 | 0.89 |
| 81 | NPM1G -12.0625 | ME2PL | 0.12 $^{0.14}_{0.11}$ | 0.00 $^{0.00}_{0.00}$ | 2.24 $^{2.31}_{2.19}$ | 2.72 $^{2.81}_{2.60}$ | 419.17 | 417 | 1.01 |

NOTES: Column density expressed in units of $10^{22}$ cm$^{-2}$

**Table 15.** X-ray luminosity for *Chandra* data with the *XMM-Newton* extraction region (25").

| Num | Name | Log(Lx(0.5-2.0 keV)) | Log(Lx(2-10 keV)) |
|---|---|---|---|
| (1) | (2) | (3) | (4) |
| 1 | NGC 0315 | 41.624 [41.482,41.650] | 41.692 [41.595,41.739] |
| 7 | NGC 0835 | 41.711 [41.270,41.816] | 41.472 [40.525,41.712] |
| 8 | NGC 1052 | 40.211 [39.624,40.360] | 41.100 [39.818,41.282] |
| 14 | 3C 218 | 43.438 [43.386,43.444] | 43.414 [43.400,43.447] |
| 15 | NGC 2787 | 38.787 [38.631,38.861] | 39.029 [38.949,39.133] |
| 17 | UGC 05101 | 40.833 [40.557,40.894] | 41.443 [41.313,41.523] |
| 29 | NGC 3690B | 41.284 [40.601,41.295] | 40.814 [40.677,40.900] |
| 32 | NGC 3998 | 40.935 [40.908,40.943] | 41.356 [41.347,41.371] |
| 35 | NGC 4125 | 40.062 [39.706,40.064] | 39.720 [39.546,39.826] |
| 37 | NGC 4261 | 41.752 [41.680,41.779] | 41.250 [40.199,41.330] |
| 38 | NGC 4278 | 40.445 [40.344,40.472] | 40.033 [39.999,40.067] |
| 40 | NGC 4321 | 39.731 [39.399,39.746] | 39.385 [39.184,39.602] |
| 47 | NGC 4494 | 39.381 [39.278,39.462] | 39.543 [39.441,39.625] |
| 48 | NGC 4552 | 40.224 [40.065,40.228] | 39.852 [39.751,39.901] |
| 50 | NGC 4579 | 43.428 [43.333,43.418] | 41.211 [41.188,41.227] |
| 52 | NGC 4594 | 41.248 [40.808,41.634] | 41.292 [40.582,41.599] |
| 58 | NGC 4736 | 39.470 [39.364,39.473] | 39.615 [39.588,39.641] |
| 61 | MRK 266NE | 42.068 [41.819,42.087] | 41.800 [41.054,41.938] |
| 62 | UGC 08696 | 42.947 [42.785,43.000] | 42.929 [42.141,43.019] |
| 63 | CGCG 162-010 | 43.554 [43.517,43.589] | 43.597 [43.579,43.676] |
| 68 | NGC 5813 | 41.214 [41.131,41.238] | 39.804 [39.491,40.055] |
| 70 | NGC 5846 | 40.197 [40.158,40.231] | 40.361 [40.324,40.398] |
| 73 | NGC 6251 | 41.588 [41.129,41.596] | 41.587 [41.451,41.671] |
| 74 | NGC 6240 | 42.634 [42.594,42.663] | 42.642 [42.525,42.693] |
| 75 | IRAS 17208-0014 | 41.088 [40.710,41.200] | 41.273 [41.119,41.419] |
| 76 | NGC 6482 | 41.700 [41.648,41.739] | 40.193 [40.128,40.244] |
| 80 | IC 1459 | 40.750 [40.655,40.761] | 40.844 [40.808,40.884] |
| 81 | NPM1G -12.0625 | 43.833 [43.823,43.842] | 43.751 [43.742,43.764] |

## Appendix A: *XMM-Newton* versus *Chandra* results

We have added *XMM-Newton* data in an attempt to achieve three main objectives: (1) to enlarge the sample with spectral fits; (2) to obtain more accurate spectral analysis since *XMM-Newton* data have higher sensitivity than *Chandra* data; and (3) to get information about the iron FeK$\alpha$ line due to the superb sensitivity of the *XMM-Newton*/EPIC camera at such energies.

However, the high spatial resolution of *Chandra* data have demonstrated that LINERs present complex morphologies with point-like sources close to the nucleus and diffuse emission contaminating the nuclear extraction apertures of *XMM-Newton* data. Thus a careful analysis of the limitation of *XMM-Newton* data must be done to fully understand in which cases *XMM-Newton* data are still valid for our purposes. We have followed three approaches: (1) Comparison with *Chandra* data for the objects in common, (2) Re-analysis of *Chandra* data for these objects in common with the same aperture (25 arcsecs) and (3) statistical comparison between the 68 objects observed with *Chandra* and the 55 objects observed with *XMM-Newton*. These three steps will be discussed in the following subsections.

### A.1. Comparison with Chandra data

There are 40 objects with *Chandra* and *XMM-Newton* data and 28 with spectral fits. It has to be noticed that about half of them (14 cases) shows the same spectral fit. Seven objects have a more complex model in *XMM-Newton* data than in *Chandra* data (NGC 2787, NGC 4579, NGC 4594, CGCG 162-010, NGC 5846, NGC 6251 and NPM1G -12.0625), five have a more complex model with *Chandra* data (3C 218, NGC 3690B, NGC 4278, NGC 4494 and IC 1459) and IRAS 17208-0014 appears as a ME with *XMM-Newton* data and as PL with *Chandra* data. Statistically speaking, the reduced $\chi^2$ shows median value and standard deviation $<\chi_r^2> = 0.91 \pm 0.18$ and $<\chi_r^2> = 1.07 \pm 0.21$ for the objects in common between *Chandra* and *XMM-Newton* samples, respectively. *XMM-Newton* median value is larger than in *Chandra* data most probably because of its higher sensitivity. However, the standard deviation is similar for both datasets.

Table A.1 show the results on the linear fit between parameters and luminosities in *Chandra* and *XMM-Newton* data in common. Three objects have much higher temperature with *XMM-Newton* data than *Chandra* data (3C 218, CGCG 162-010 and NPM1G -12.0625). They are the central galaxies of the clusters Abell 780, Abell 1795 and Abell 2597 and thus a strong thermal emission from the cluster might be included in the *XMM-Newton* aperture. For *XMM-Newton* data the aperture is 25 arcsec in comparison with the *Chandra* apertures of 2.2, 3.9 and 1.8 arcsecs, respectively. The resulting *XMM-Newton* and *Chandra* temperatures have larger departures when different best-fits are selected. Objects with low temperatures at *Chandra* data ($\leq 0.2$ keV) tend to be located at higher temperatures (kT $\sim$ 0.6 keV) with *XMM-Newton* data. It has to be noticed that three objects depart from the quoted correlation, NGC 3690B, NGC 4579 and NGC 6251, all with different fits. Only NGC 3690B and UGC 05101 show departures on the resulting spectral index. NH1 column density shows a poor correlation although all the objects show nevertheless compatible NH1 column densities. Four objects show departures in the NH2 column density correlation, namely NGC 3690B, NGC 4579, IC 1459 and NPM1G -12.0625, all of them with different fits. Finally, L(0.5-2keV) and L(2-10keV) are the best correlated, however, both *XMM-Newton* luminosities tend to be larger than those calculated with *Chandra* data, a median factor around of 6.0 and 2.4 dex in the energy ranges (0.5-2.0 keV) and (2-10 keV), respectively. When only results from the same best fit are selected, temperature and NH2 column density tend to be similar while the other parameters does not seem to be affected.

### A.2. Aperture effects. Chandra 25" aperture data versus XMM-Newton data

The finding of higher luminosities, both soft and hard, and the discrepancies in temperatures point to the idea that these effects might be due to the difference in the window extraction. To test such a possibility we have performed a new spectral analysis on *Chandra* data using a fixed 25" radius for the extraction region (*Chandra 25"* hereinafter), same that used in *XMM-Newton* data, for the 28 objects in common. The process has been the same than in the previous analysis and the best fit and flux and luminosity results are shown in Tables 14 and 15. The values of new correlations are given in Table A.1.

Sixteen out of the 28 objects show the same spectral model when it is compared to *XMM-Newton* data. It has to be noticed that among them, in ten cases the small aperture of *Chandra* data also show the same spectral model indicating the dominance of the nuclear source over the surrounding medium. For the other six galaxies (3C 218, NGC 3690B, NGC 4278, CGCG 162-010, IC 1459 and NPM1G -12.0625), the best-fit model changes from ME2PL with *Chandra* small aperture to MEPL with *Chandra* 25" and *XMM-Newton* data in four objects. Thus, it seems that the large aperture tends to dilute the final best-fit to a simplified model. For the remaining galaxies, the comparison is rather confusing. Six of them (NGC 1052, NGC 2787, UGC 05101, NGC 4494, NGC 5846 and NGC 6251) show the same spectral model for *Chandra* small and large apertures, but it is different to that from *XMM-Newton* data analysis. The discrepancy with *XMM-Newton* spectral fit has to be found in the poor quality data for these *Chandra* observations. Another two galaxies (NGC 4125 and NGC 4552) are fitted with the same best-fit in *Chandra* small aperture and *XMM-Newton* data (MEPL) while *Chandra* 25" data reports a different model (ME2PL). Finally, there are four cases (NGC 3998, NGC 4579, NGC 4594 and IRAS 17208-0014) showing different best-fits for the three datasets.

CGCG 162-010 and NPM1G -12.0625 recover the *XMM-Newton* temperature with the same fit when we use *Chandra* 25" data, whereas 3C 218 results in a lower temperature and a different model with the larger aperture. It then appears that the differences between *Chandra* and *XMM-Newton* temperatures do not seem to be due to aperture effects in this object. Objects with rather low temperatures (around 0.2 keV) show again a lower temperature with the large aperture (NGC 3690B, NGC 4278, NGC 6251, IRAS 17208-0014). We think that the only explanation for this effect has to be related to the lower sensitivity of *XMM-Newton* data at low energies. NH1 column densities show a much better correlation (r=0.80) than with *Chandra* small aperture (r=0.38). Thus, NH1 column density discrepancies seem to be related to aperture effects. Higher discrepancies when using different models are found in spectral index and NH2 column density, suggesting that the differences from *Chandra* and *XMM-Newton* data may related to the selection of the best-fit model rather than to aperture effects.

Regarding luminosities, both L(0.5-2keV) and L(2-10keV) are well correlated, where offsets have disappeared. Therefore, the difference in luminosities can be entirely attributed to aperture effects. However, the effect is smaller for the harder lumi-nosity, since there is only a factor of 2.4 dex between *Chandra* small aperture and *XMM-Newton* data in L(2-10 keV).

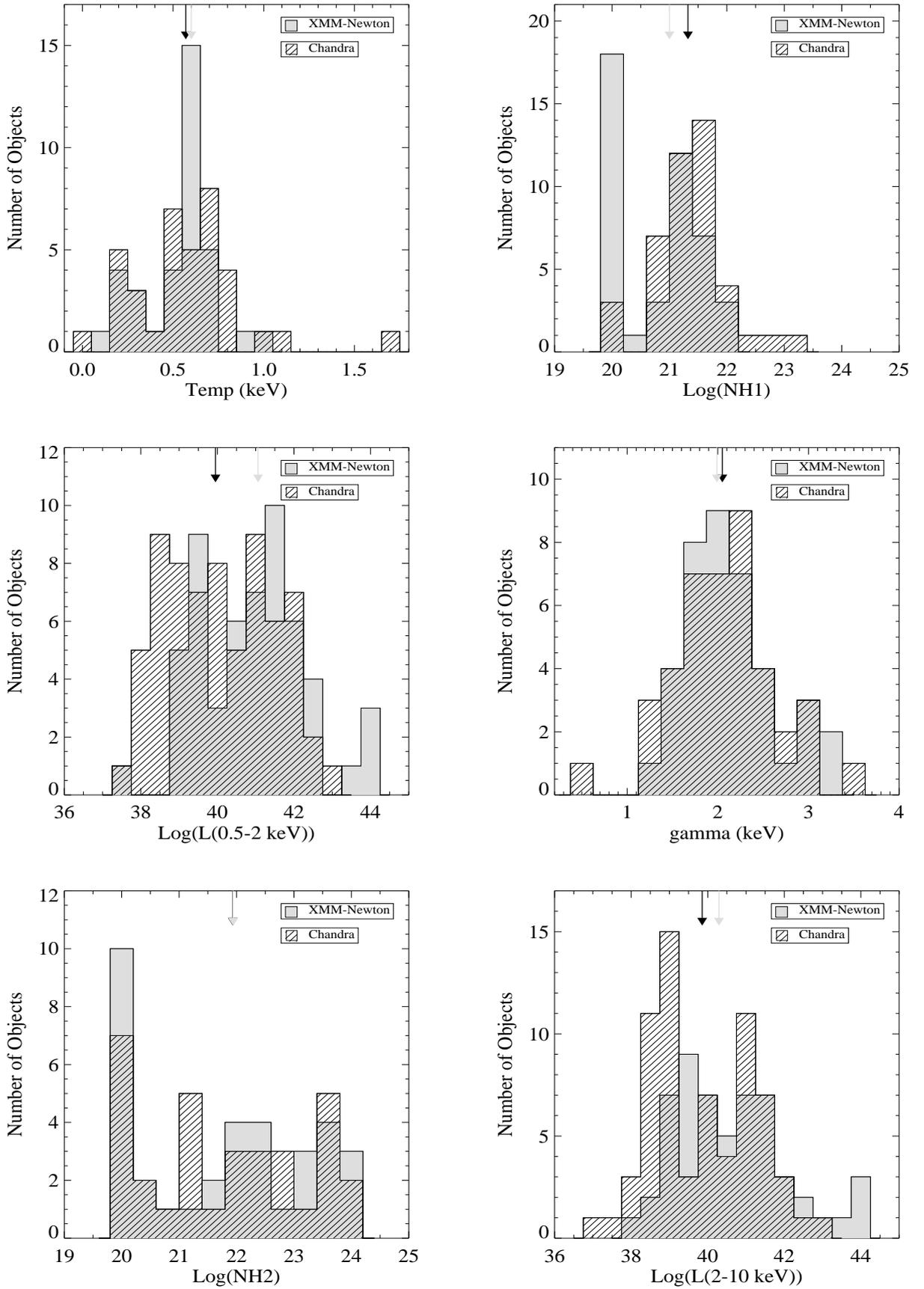

**Fig. A.1.** Temperature (*Top-Left*), spectral index (*Bottom-Left*), NH1 (*Top-Centre*), NH2 (*Bottom-Centre*), soft (0.5-2 keV) luminosity (*Top-Right*) and hard (2-10 keV) luminosity (*Bottom-Right*) histograms. The *Chandra* data is shown as dashed histogram while the grey histogram plots the *XMM-Newton* data. The median value for *Chandra* and *XMM-Newton* data are marked as black and grey arrows, respectively. Four objects (3C 218, CGCG 162-010, IRAS 14348-1447 and NPM1G -12.0625) have been excluded from the *XMM-Newton* temperature histograms because they show a temperature above 2 keV.

**Table A.1.** Correlation between X-ray parameters and luminosities.

| | Diff. Apertures | | | Same Apertures (25") | | | K-S test |
|---|---|---|---|---|---|---|---|
| | Slope | Zero point | Corr. Coeff. | Slope | Zero point | Corr. Coeff. | |
| Γ .................. | 0.63± 0.08 | 0.73± 0.16 | 0.43 | 0.89± 0.06 | 0.28± 0.12 | 0.59 | 83.2% |
| Temp............... | 0.31± 0.08 | 0.45± 0.05 | 0.54 | 0.30± 0.08 | 0.44± 0.05 | 0.21 | 33.3% |
| Log(NH1).......... | 1.20± 0.02 | -4.55± 0.35 | 0.38 | 0.83± 0.01 | 3.64± 0.32 | 0.80 | 0.4% |
| Log(NH2).......... | 1.60± 0.01 | -14.24± 0.17 | 0.49 | 1.06± 0.00 | -1.62± 0.09 | 0.74 | 25.8% |
| Log(Lum(0.5-2 keV)) | 1.34± 0.01 | -12.72± 0.44 | 0.85 | 1.06± 0.00 | -2.34± 0.13 | 0.87 | 1.2% |
| Log(Lum(2-10 keV)) | 1.15± 0.01 | -5.63± 0.49 | 0.87 | 1.06± 0.00 | -2.58± 0.16 | 0.94 | 1.0% |

### A.3. Statistical comparison between Chandra *small aperture and* XMM-Newton *samples*

A reliable comparison between the fitting parameters from the two sets of data is difficult because of the small number of objects in common. The statistical comparison of the two samples is important to ensure the final conclusions about the utility of the *XMM-Newton* sample.

Fig. A.1 shows the histograms of parameters and luminosities and the Kolmogorov-Smirnov (K-S) test probability is included in Table A.1, Only the spectral index shows a high probability that the two distributions are the same. Then the intrinsic spectral index seems to be equally derived from both sets of data.

Temperatures show median values compatible within one standard deviation (see Table 10). Bimodal distribution is also found in both samples although four objects, all located at the centre of a galaxy cluster, show values larger than 2 keV (3C 218, CGCG 162-010, IRAS 14348-1447 and NPM1G -12.0625), and the peak at 0.60 keV is much stronger in *XMM-Newton* data When the K-S test is performed taking out all the sources with temperatures larger than 2 keV, a higher probability is found (44%). These differences is easily explained because of the large aperture used for *XMM-Newton* spectra. Cluster centres show a strong thermal contribution that can be described by a thermal model with temperatures around 2-3 keV (Kaastra et al., 2008). The large aperture of *XMM-Newton* spectra includes a substantial amount of cluster contribution which is easily removed in *Chandra* data. In fact, *Chandra 25"* data recovers the *XMM-Newton* temperature in two of the three galaxies (CGCG 162-010 and NPM1G 12.0625) common in both samples. The soft, diffuse observed emission is a thermal component at around 0.6-0.7 keV, which is the case for most of the galaxies (see figures in Appendix C). Regarding the differences found at low energies, temperatures around 0.1-0.2 keV seem to be most probably related with differences in sensitivity between both datasets at low energies, with *XMM-Newton* data worse suited to trace it.

The K-S test shows that neither NH1 nor NH2 column densities are drawn from the same distribution. In NH1 column density, both samples show a peak around $3\times10^{21}$ cm$^{-2}$ but *XMM-Newton* data show an additional peak compatible with the value of Galactic absorption. Aperture effects and lower sensitivity at soft energies seems to be the responsible. However, the median values, also for NH2 column density, are not too far (see Table 10).

The largest discrepancies between both sets of data are found in the luminosities (see median values in Table 10). *XMM-Newton* data result in soft (hard) luminosity around 10 (2.5) times brighter than *Chandra* data. From previous section we now know that luminosities, specially the soft band, are contaminated by the surrounding medium. When we make the offset proposed in the previous section (offsets 1.13 and 0.44 dex, respectively) the K-S test shows that (0.5-2.0 keV) and (2-10 keV) luminosities have 50% and 53% probabilities that these two distributions are the same.

In summary, differences in luminosities, both soft and hard are a factor of ∼5-10 and ∼2.5 times the *Chandra* luminosity, respectively. NH1 column density and temperature discrepancies might be due to different sensitivity between the two instruments. NH1 column density discrepancies also seem to be related to aperture effects. NH2 column density is strongly linked to the best-fit model selected and independent on the selected aperture or sensitivity. Finally, the spectral index has shown to be the most robust parameter while temperature is strongly dependent on the aperture effect and different sensitivity of the instruments at lower energies.

## Appendix B: Notes on individual sources

Some of the notes given here was already published in paper GM+06. It is duplicated for the easier access of the reader to the information. Here we present an actualization of the previous reported notes and the addition of the new LINERs included in this paper.

**NGC 315 (UGC 597, B2 0055+30).** NGC 315 is a giant cD radiogalaxy located in the Zwicky cluster 0107.5+3212 (Zwicky et al., 1961, see Appendix F, Fig. F.1), with two-sided well resolved radiojets shown both with the *VLA* and VLBI observations (Venturi et al., 1993; Cotton et al., 1999). Nagar et al. (2005) reported also an unresolved core in addition to the radio jet at VLBI resolutions. The high spatial resolution provided by *Chandra* imaging allowed the detection of X-ray jets, the most striking one being the one along ∼10" to the NW (see Fig. C.1 and Donato et al., 2004; Worrall et al., 2003, 2007) In both studies, based on a short 4.67 ks exposure, they found that the central source is of quite high (2-10 keV) luminosity $10^{41}$erg s$^{-1}$ given its AGN character.

With a more recent, longer exposure (≈ 52 ksec) dataset, Satyapal et al. (2005) obtained very similar results (kT=0.54 keV, $N_H$=0.8 × 10$^{22}$cm$^{-2}$, Γ=1.60, and a hard X-ray luminosity of 4.67 × 10$^{41}$erg s$^{-1}$). Both results do agree with the ones reported in GM+06. Here we present a reanalysis of the longest, previous *Chandra* data (≈ 52 ksec) and analize the 43.6 ksec *XMM-Newton* data finding that both datasets fit with ME2PL with not very discrepant parameters, which leads to a similar (2-10 keV) X-ray luminosity, 5.8 × 10$^{41}$erg s$^{-1}$ and 4.2 × 10$^{41}$erg s$^{-1}$ respectively. A clear FeKα has been detected with equivalent width of 82 eV. The AGN nature of this galaxy was noted by Ho et al. (1997), who reported a broad Hα component.

**NGC 410 (UGC 735)**. NGC 410 is the central galaxy of the NORAS cluster RXC 0112.2+3302 (Böhringer et al., 2000), see Appendix F, Fig. F.1. After the *ROSAT* observations new data on the galaxy have not been reported. Here, we analyse the archival 19.6 ksec *XMM-Newton* observations. The bestfit model obtained is MEPL ($\Gamma$=2.4, kT=0.69 keV, NH1 = $1.2 \times 10^{21}$ cm$^{-2}$ and NH2 < $2. \times 10^{21}$ cm$^{-2}$).

**NGC 474 (UGC 864)**. This lenticular galaxy is a member of the wide pair Arp 227, together with NGC 470 (see Appendix F, Fig. F.1), which is the dominant member of a loose group. Rampazzo et al. (2006) reported *XMM-Newton* EPIC data for the group. They calculated X-ray luminosities of $L_X$(0.5-2 keV)= $1.12 \times 10^{39}$ erg s$^{-1}$ and $L_X$(2-10 keV) = $6.2 \times 10^{39}$ erg s$^{-1}$ assuming $\Gamma \sim 1.7$ and galactic column density. The low count rate of these data does not allow us to perform any spectral fitting. We have computed upper limits in the hard X-ray luminosity of $3.0^{+8.38}_{-0.001} \times 10^{38}$ erg s$^{-1}$ for *Chandra* observations and $6.04^{+326}_{-0.01} \times 10^{37}$ erg s$^{-1}$ for *XMM-Newton*, both consistent with an upper limit of $10^{39}$ erg s$^{-1}$, which is roughly consistent with the value quoted by Rampazzo et al. (2006) ($2.5 \times 10^{39}$ erg s$^{-1}$).

**IIIZw 035 (CGPG 0141.8+1650)**. This galaxy belongs to a close interacting galaxy pair (see Appendix F, Fig. F.1). There are not previous X-ray data reported in the literature. We have analysed both the 14.5 ksec *Chandra* and 16.3 ksec *XMM-Newton* data. Neither have a high enough count-rate for the spectral analysis to be performed. The X-ray morphology shows two diffuse peaks both seen at hard energies, whith the northern one coincident with the LINER nuclear position, see Fig. C.4). Hattori et al. (2004) indicate that the morphological appearance in radio continuum emission suggests that star-forming activity dominates the energetics of the northern galaxy (Pihlstrim et al., 2001); the classification as a LINER may be due to a contribution from shock heating, possibly driven by superwind activity (Taniguchi et al., 1999; Lutz et al., 1999). However Baan and Klockner (2006) by means of observations at 1.4 and 8.4 GHz suggest an AGN nature of the radio continuum emission based in the compactness and flat index spectral nature of the nuclear source (see Tab. 12).

**NGC 524 (UGC 968)**. This massive S0 galaxy dominates a small group (see Appendix F, Fig. F.1). The 15.4 ksec *Chandra* ACIS-S data show a morphology with diffuse emission at low energies (< 2 keV) and no X-ray detection in the hard band (2-10 keV), with a total X-ray luminosity amounting to $6.48 \times 10^{38}$ erg s$^{-1}$. Filho et al. (2002) reported a slightly resolved 2 mJy core in the center of the optical galaxy from their *VLA* data at 5 GHz and 1" resolution. They compare their data with those by Wrobel and Heeschen (1991), who detected a 1.4 mJy core at 5" resolution, concluding that the source has to be compact and have a flattish spectrum. Dwarakanath and Nath (2006) on the contrary, based on the analysis at radiofrequencies, conclude that this group do not show any evidence of either current or past AGN activity.

**NGC 835 and NGC 833 (Arp 318A and B ; HCG 16 A and B)**. Both members of an unusually active compact group, they have been optically classified as LINER and Seyfert 2, respectively (Coziol et al., 2004), and are strongly interacting with each other (see Appendix F, Figs. F.1 and F.2). Only NGC 835 was detected at radiofrequencies (Corbett et al., 2002) as a nuclear compact source. At X-ray frequencies no point sources were detected in neither hard band images (4.5-8* keV or 6-7 keV), but the X-ray soft emission is extended (see Fig. C.6). Turner et al. (2001) analyzed the 40 ksec EPIC *XMM-Newton* first-light observations and confirmed the presence of an AGN in both galaxies A and B. They fitted three components to the EPIC X-ray spectrum of **NGC 833** (Arp 318B): (1) a power-law for the obscured AGN, with $\Gamma$=1.8 and $N_H$=$2.4 \times 10^{23}$cm$^{-2}$, (2) an unabsorbed power-law for the radiation scattered into our line of sight by thin, hot plasma directly illuminated by the AGN, and (3) an optically-thin thermal plasma with kT=0.47 keV; the luminosity of the AGN component of $6.2 \times 10^{41}$ erg s$^{-1}$ turns out to be 100 times brighter than the thermal X-ray emission. The core of **NGC 835** (Arp 318A) shows a very similar spectrum, with absorbed and scattered power laws indicating a heavily obscured AGN ($N_H$=$4.6 \times 10^{23}$cm$^{-2}$ and $\Gamma$=2.25) of $5.3 \times 10^{41}$erg$^{-1}$ (0.5-10 keV) and a soft thermal component with kT=0.51 keV contributing to 2% of the total luminosity. Due to a missidentification, the sources used in GM+06 do not correspond to the nuclear sources we analyse here. FeK$\alpha$ line has been detected in both galaxies with equivalent width of 334 and 774 eV respectively.

**NGC 1052**. NGC 1052 is the brightest member of a small group (see Appendix F, Fig. F.2) which together with the NGC 1069 group makes up the Cetus I cluster (Wiklind et al., 1995). The X-ray morphology clearly indicates the presence of an unresolved nuclear source in the hard bands (Fig. C.8), in agreement with the classification by Satyapal et al. (2004), that made use of the same dataset. Evidences for the AGN nature of this object have already been given with the detection of (a) broad lines in spectropolarimetric measurements by Barth et al. (1999) and a broad underlying component in H$\alpha$ reported in Ho et al. (1997); (b) a variable radio core (Vermeulen et al., 2003); (c) H$_2$0 megamaser emission (Claussen et al., 1998); (d) highly probable UV variability (Maoz et al., 2005). Guainazzi et al. (2000) confirm that its X-ray spectrum may therefore resemble that of Seyfert galaxies with the analysis of its *BeppoSAX* spectrum (0.1-100 keV). They obtain a very good fit with a two-component model for the spectrum, constituted by an absorbed ($N_H$=$2.0 \times 10^{23}$cm$^{-2}$) and rather flat ($\Gamma \approx$1.4) power-law plus a "soft excess" below 2 keV. The corresponding flux in the 2-10 keV energy range is $4.0 \times 10^{-12}$ erg cm$^{-2}$ s$^{-1}$. We have best modelled the *Chandra* data with two power-laws with a flat index ($\Gamma$ = 1.15) and column density compatible with that reported by Guainazzi et al. (2000), although we are using a different instrument. Kadler et al. (2004) obtained a much flatter spectral index with another set of *Chandra* data (2.3 ksec) that they atributed to piled up effects. In addition to this, we also analise the 47.2 ksec *XMM-Newton* observation. We find the soft excess reported by Guainazzi et al. (2000), with the best-fit model resulting to be ME2PL ($\Gamma$ = 1.34), steeper than the *Chandra* spectrum, and kT=0.60 keV. The FeK$\alpha$ line has been detected with an equivalent width of 144 eV.

**NGC 2639 (UGC 4544)**. This a rather isolated object (Marquez et al., 2003), see Appendix F, Fig. F.1). *XMM-Newton*/pn images show a very faint source with almost no emission detection at hard energies, what argues in favour of classifying it as a Non-AGN like object. The spectral analysis at soft energies results in MEKAL as the best fit (kT=0.18 keV) with $N_H$ = $8 \times 10^{21}$ cm$^{-2}$. However this galaxy is a well known type 1.9 LINER with broad H$\alpha$ emission (Ho et al., 1997) and it is one of the best known detections of a water megamaser (Wilson et al., 1995). At radio frequencies Ho and Ulvestad (2001) reported an inverted Spectrum in the compact nuclear source detected. Then it appears that this galaxy can be a very good candidate to a heavily obscured AGN. This could explain the fit at soft energies with a kT value typical of what it is found in some Seyfert 1 like objects (Teng et al., 2005) due to the presence of a warm absorber. Terashima et al. (2002) comment that the large EW (> 1.1 keV) of the FeK$\alpha$) emission line detected in the *ASCA* data suggests that the nucleus is highly obscured.

**NGC 2655 (UGC 4637, Arp 225).** Arp 225 is an Sa galaxy which shows traces of a strong interaction or merger event (Mllenhoff and Heidt, 2001): faint outer stellar loops (see Appendix F, Fig. F.2) and extended HI-envelope (Huchtmeier and Richter, 1982). The X-ray *XMM-Newton* morphology of this galaxy shows a clear diffuse extended emission in the softer and harder bands that almost disappears in the medium band (1.0-2.0 keV) (see Fig. C.10). The bestfit model is ME2PL ($\Gamma = 2.48$, kT=0.64 keV, no absorption for the soft component and NH2 = $3.0 \times 10^{23}$ cm$^{-2}$). Terashima et al. (2002) comment on the lack of a strong FeK$\alpha$) fluorescent line in *ASCA* data, ruling out the possibility of "cold" reflection as the origin of the observed flat spectral slope. Ho et al. (1997) report a questionable detection of a broad band H$\alpha$ component in its optical spectrum. Moreover, core radio emission has been detected at a flux density of 1.1mJy and indications of polarized light in the nucleus pointing to their AGN nature (Nagar et al., 2000).

**NGC 2681 (UGC 4645).** A small companion at the same redshift, namely MCG09-15-039, appears in Appendix F, Fig. F.2. An unresolved nuclear source was clearly detected at hard X-ray energies (Fig. C.11). Satyapal et al. (2005), who made use of the same archival *Chandra* ACIS observations of this galaxy, classed them as an AGN-LINER, and derived $\Gamma$=1.57 and kT=0.73 keV for an apec plus power-law fit to the nuclear spectrum. These values are in perfect agreement, within the errors, with the parameters we derive for our best model (MEPL) ($\Gamma$=1.57 and kT=0.63 keV. Ho et al. (1997) report several arguments for the reality of the broad H$\alpha$ component derived from the profile fitting, whereas they indicate that the actual parameters of the broad H$\alpha$ component are not well constrained.

**NGC 2685 (UGC 4666, Arp 336).** This is a rather isolated object (see Appendix F, Fig. F.2). We report a snapshot (0.86ksec) *XMM-Newton* observation. No nuclear point-like source is detected at hard X-ray, but only a very faint source at high energies (> 2 keV). A (2-10 keV) luminosity of $2.63 \times 10^{39}$ erg s$^{-1}$ has been estimated assuming a power law ($\Gamma$=1.8) and galactic column density, a factor of three lower than the value reported by Cappi et al. (2006), who analysed the same dataset. But we would like to notice that, in addition to the low count rate in these data, these authors determine a highly unrealistic value for $\Gamma$= 0.5, what makes us confident in our results.

**UGC 4881 (Arp 55).** UGC 4881 is a member of an strongly interacting galaxy pair (see Appendix F, Fig. F.3). We present the analysis of 19 ksec *XMM-Newton* and 14.6 ksec *Chandra* observations. The X-ray morphology (see Fig. C.13) shows two peaks at soft X-rays with no point-like source at high energies (> 2 keV). The best fit for the *XMM-Newton* spectrum is obtained with a MEKAL model (kT=0.19 keV) with a resulting luminosity L$_X$(2 − 10 keV) $\simeq 10^{41}$ erg s$^{-1}$. These values could be uncertain since the two members of the pair are included in the *XMM-Newton* aperture (see Appendix F, Fig. F.3). Although the low quality of the *Chandra* data does not allow any spectral fitting, we have estimated a very high luminosity of $2.7 \times 10^{40}$erg$^{-1}$.

**3C 218 (Hydra A).** 3C 218 is one of the most luminous radio sources in the local (z < 0.1) Universe, only surpassed by Cygnus A. It has been optically identified with the cD2 galaxy Hydra A (Simkin, 1979), which dominates the poor cluster Abell 780 (see Appendix F, Fig. F.3). Its X-ray morphology shows strong thermal emission (showing caves and bubbles) at the whole X-ray energy range, and a point-like source at the nuclear position on the hard band (> 2 keV). Sambruna et al. (2000) discovered with *Chandra* a LLAGN in the LINER harboured by this nearby cD galaxy. They reported the existence of a compact source at energies larger than 2 keV. Their best fit *Chandra* spectrum was found to be an obscured (N$_H$ = $2.8 \times 10^{22}$ cm$^{-2}$) power law with $\Gamma$=1.75 plus an unabsorbed Raymond-Smith with kT=1.05 with abundance 0.1 solar. Here we present *Chandra*/ACIS (80 ksec) and *XMM-Newton*/pn (24 ksec) data. The spectrum for *Chandra* data is fitted by a ME2PL model ($\Gamma$ = 2.1 and kT = 1.71 keV) and the *XMM-Newton* data by a MEPL with $\Gamma$ = 2.2 and kT = 2.77. It has to be noticed that we obtain a difference in luminosities of two orders of magnitude. This might be due to the large contribution of the diffuse thermal emission at hard X-rays from the cluster thermal emission. Previous studies conclude that both cooling flow and radio jet emission are important (Lane et al., 2004) and also thermal emission form supercavities/bubbles (Wise et al., 2007, based on 227 ksec *Chandra* observations).

**NGC 2787 (UGC 4914).** This a rather isolated object (see Appendix F, Fig. F.3). Snapshot *Chandra* data of this galaxy were reported by Ho et al. (2001), who estimated an X-ray luminosity of $2 \times 10^{41}$ erg s$^{-1}$ by assuming a power-law model with $\Gamma$ = 1.8 and galactic column density. We report the analysis of 29.4 ksec *Chandra* and 33 ksec *XMM-Newton* observations. The *Chandra* data have been fitted to a PL model ($\Gamma$ = 2.3) while *XMM-Newton* data is better modeled by a moderately obscured (NH2=$1.0 \times 10^{22}$cm$^{-2}$) ME2PL model. The *Chandra* X-ray morphology shows a point-like source coincident with the nucleus and an extranuclear source to its SE. Evidences on its AGN nature have been reported at other frequencies. Nagar et al. (2000) confirm its AGN nature at radiofrequencies, based on its flat radio spectrum between 20 and 6 cm. Ho et al. (1997) reported a fairly prominent broad H$\alpha$ component classifying it as a type 1.9 LINER. Tremaine et al. (2002) have estimated a mass for the central black hole of $4.1 \times 10^7 M_\odot$.

**NGC 2841 (UGC 4966).** This is a rather isolated object (see Appendix F, Fig. F.3). The *Chandra* X-ray morphology shows a number of point sources with diffuse emission, one of these coinciding with the nuclear position. We report here the analysis of both *Chandra* and *XMM-Newton* data. Snapshot *Chandra* data of this galaxy was reported previously by Ho et al. (2001). They estimated an X-ray luminosity of $1.81 \times 10^{38}$ erg s$^{-1}$ ergs/s by assuming a power law with $\Gamma$ = 1.8 and galactic column density. Our estimated value for the *Chandra* data is a factor of 2 larger than this value. For the *XMM-Newton* spectrum we have obtained the best fit with a ME2PL model with $\Gamma$ = 2.20 and kT= 0.58 keV. Ho et al. (1997) estated that broad H$\alpha$ emission is absent in this galaxy.

**UGC 05101 (IRAS 09320+6134).** A clearly perturbed morphology characterises this ultra-luminous infrared galaxy, with a long tail extending to the West (see Appendix F, Fig. F.3). In addition to the hard-band point-like nuclear source, extended emission is seen in both (4.5-8.0* keV) and (6-7 keV) bands in the image obtained from *Chandra* data (Fig. C.17). The evidence of a heavily obscured active galactic nucleus in this galaxy has been provided by Imanishi et al. (2001) and Imanishi et al. (2003), based in near-IR spectroscopy and on its *XMM-Newton* EPIC spectrum, respectively. They fit the spectrum with an absorbed power-law ($\Gamma$=1.8 fixed), a narrow Gaussian for the 6.4 keV Fe

K$\alpha$ emission line, which is clearly seen in their spectrum, and a 0.7 keV thermal component; they derive $N_H$=14 × $10^{23}$cm$^{-2}$ and EW(Fe K$\alpha$)=0.41 keV. The resulting (2-10 keV) luminosity (~5×$10^{42}$erg s$^{-1}$) is within a factor of 2-3 of the values we obtain both from *Chandra* and *XMM-Newton* data. Our best fit model for the spectra is that of a double power-law for *Chandra* data and a ME2PL for *XMM-Newton* data, with consistent spectral slopes for both sets of data. The use of a fixed power law slope of 1.8 to estimate the luminosity can explain the difference between the value we calculate from the spectral fitting and that estimated by Satyapal et al. (2004) for the same *Chandra* dataset. The type 1 AGN nature of the nucleus was already reported by Sanders et al. (1988). Farrah et al. (2003) found that this system is a composite object containing both starburst and AGN contributions, consistent with the result for its optical spectrum (Goncalves et al., 1999). New evidences come from radiofrequencies where it is found a compact nucleus with flat spectral shape (Baan and Klockner, 2006). The Fe-K emission is marginally detected in the analysis of *Chandra* data by Ptak et al. (2003). We have measured an equivalent width of 278 for the FeK$\alpha$ line.

**NGC 3185 (UGC 5554, HCG 44c).** It makes up the compact group HCG 44, together with NGC 3193, NGC 3190 and NGC 3187 (see Appendix F, Fig. F.3). Cappi et al. (2006) report their analysis of the *XMM-Newton* available data for this galaxy, finding a very weak nuclear emission with an X-ray spectrum consistent with a power-law with $\Gamma$ =2.1, which results in a luminosity of $10^{39}$ ergs/s, in very good agreement with the one we estimate here (see Table 8). Ho and Ulvestad (2001) reported this galaxy as marginally detected at 6 cm but not at 20 cm, using *VLA* 1" resolution data.

**NGC 3226 (UGC 5617, Arp 94a).** Strongly interacting with NGC 3227 (see Appendix F, Fig. F.4), several point sources have been detected at the (4-8) keV band image of this galaxy, with Fe emission unambiguously present in the nucleus. The analysis of HETGS *Chandra* data by George et al. (2001), whose properties strongly suggested that this galaxy hosted a central AGN, resulted in an adequate fit with a photon index $\Gamma$=1.94 and $N_H$=4.8 × $10^{21}$cm$^{-2}$, with the resulting luminosity L(2-10 keV)≈3.2×$10^{40}$ erg s$^{-1}$. The *XMM-Newton* observations of this dwarf elliptical galaxy most probably indicate the presence of a sub-Eddington, super-massive black hole in a radiatively inefficient stage (Gondoin et al., 2004). They conclude that, since the best fit is provided by a bremsstrahlung model absorbed by neutral material, the X-ray emission may therefore be reminiscent of advection-dominated accretion flows. Nevertheless, an acceptable fit is also obtained by including a power-law model ($\Gamma$=1.96) absorbed by neutral ($N_H$=4.1× $10^{21}$cm$^{-2}$) and ionized material. The resulting (2-10 keV) luminosity, calculated for the distance we use, is 1.8 × $10^{40}$ erg s$^{-1}$, a factor of 4 higher than the one we estimate.

Terashima and Wilson (2003) fit the 22 ksec *Chandra* ACIS nuclear spectrum with a power-law with $\Gamma$=2.21 (from 1.62 to 2.76) and $N_H$=0.93 × $10^{22}$cm$^{-2}$. Notice that substantial absorption is also derived from the position of this galaxy in the color-color diagrams in GM+06, whereas the power-law index is somewhat steeper. We analyze 31 ksec *XMM-Newton* observations finding that the best fit is a combination of two power-laws ($\Gamma$ = 1.92, NH1 = 2.1 × $10^{21}$ cm$^{-2}$ and NH2 = 1.2 × $10^{22}$ cm$^{-2}$). This is a more complicated model than that proposed by Gondoin et al. (2004). A flat compact radio source has been detected at the nuclear region (Condon et al., 2002; Filho et al., 2006). Ho et al. (1997) succeeded in extracting a moderately strong broad H$\alpha$ component from a complicated, three narrow-line component blend.

**NGC 3245 (UGC 5663).** It forms a wide pair together with NGC 3245A (see Appendix F, Fig. F.4), with cz=1322 km/s. A nuclear source is detected in the (4.5-8.0*  keV) band image from *Chandra* data. This agrees with the analysis by Filho et al. (2004) who already noticed a hard nuclear X-ray source coincident with the optical nucleus. The luminosity they calculated with a fixed $\Gamma$=1.7 is in excellent agreement with our estimation. Filho et al. (2002) concluded that at radio frequencies the source could be consistent with a flat and compact spectrum. Wrobel and Heeschen (1991) found an unresolved 3.3 mJy core at 5 GHz, 5" resolution.

**NGC 3379 (UGC 5902, M 105),** is the dominant elliptical galaxy in the nearby Leo Group (see Appendix F, Fig. F.4). David et al. (2005) published their study of its X-ray emission as traced by ACIS-S *Chandra* observations. That work is mainly devoted to the analysis of extra-nuclear X-ray sources and diffuse emission, and they derive a power-law index for the diffuse emission of 1.6-1.7, in agreement with the value reported by Georgantopoulos et al. (2002). David et al. (2005) do not fit the spectrum of the nuclear source (their source 1) due to the too low net counts in the S3 chip data for this object. This is also the reason for having neither a fit nor an estimation of the spectral parameters by GM+06. The X-ray image is used for the morphological classification (SB or Non-AGN) and for estimating the (2-10 keV) luminosity (1.3 × $10^{38}$erg s$^{-1}$).

**NGC 3414 (UGC 5959, Arp 162).** UGC 5959 is a peculiar galaxy, with two companions at very similar redshifts (NGC 3418 and UGC 5958) within 250 kpc (see Appendix F, Fig. F.4). The *Chandra* X-ray morphology shows a point-like source coincident with the optical nucleus. The best spectral fit to the *Chandra* data is a PL with $N_H$ = 2.1 × $10^{21}$ cm$^{-2}$ and $\Gamma$ = 2.0, which provides a luminosity of Lx = 9.7 × $10^{40}$ erg s$^{-1}$. To our knowledge no previous X-ray data have been reported in the literature. Condon et al. (2002) suggested that the radio source is powered by an AGN. This has been later confirmed with the data reported by Nagar et al. (2005).

**NGC 3507 (UGC 6123, KPG 263b).** It makes up a wide isolated physical pair (number 263 in Karachentsev's catalogue of isolated pairs) together with NGC 3501 (see Appendix F, Fig. F.4). No hard nuclear point source has been detected in the *Chandra* images of this galaxy (Fig. C.9). The only previously published X-ray study is based on observations obtained with *ASCA*. Terashima et al. (2002) obtained a power-law to be the best model to fit the data, with $\Gamma$=1.71 and $N_H$<7.2 × $10^{20}$cm$^{-2}$. However, our best fit is MEKAL with kT=0.5 keV and absorption consistent with the galactic value. Our estimated luminosity amounts to Lx = 1.6 × $10^{37}$ erg s$^{-1}$, which seems to be much lower than the value reported by Terashima et al. (2002).

**NGC 3607 (UGC 6297),** is the brightest member of the Leo II group, which NGC 3608 and NGC 3605 also belong to (see Appendix F, Fig. F.4). No hard nuclear point source has been detected in the *Chandra* images of this galaxy (Fig. C.10). No spectral fitting can be made with the data. Based on observations obtained with *ASCA*, Terashima et al. (2002) find no clear evidence for the presence of an AGN in this LINER, in agreement with our classification as a Non-AGN candidate.

**NGC 3608 (UGC 6299),** member of the Leo II group, it forms a non-interacting pair with NGC 3607 (see Appendix F, Fig. F.5). No hard nuclear point source has been detected in the *Chandra* images of this galaxy (Fig. C.11). The previous X-ray study of this galaxy is that by O'Sullivan et al. (2001), who present a catalogue of X-ray bolometric luminosities for 401 early-type galaxies obtained with *ROSAT* PSPC pointed observations. Adjusted to our adopted distance, this luminosity result to be $1.37 \times 10^{40}$ erg s$^{-1}$, about 2 orders of magnitudes brighter than our estimation. However our results are in good agreement with the analysis reported by Flohic et al. (2006), maybe suggesting that O'Sullivan data correspond to extranuclear sources (see Fig. C.11).

**NGC 3623 (Arp 317a, M 65).** It makes up the Leo Triplet together with NGC 3628 and NGC 3627, with which it forms a non-interacting pair (see Appendix F, Fig. F.5). We report the results on the unique X-ray data provided by *XMM-Newton* observations, althought their quality does not allow a spectral analysis. We have estimated an X-ray luminosity of $2.4 \times 10^{39}$ erg s$^{-1}$ assuming a power law with spectral index $\Gamma=1.8$ and galactic column density. Satyapal et al. (2004) use 1.76 ksec *Chandra* observations to get a luminosity which is in good agreement with our determination. A black hole mass of $10^{7}$ erg s$^{-1}$ has been estimated by Dong and Robertis (2006).

**NGC 3627 (UGC 6346, Arp 16, M 66, Arp 317b).** It makes up the Leo Triplet together with NGC 3628 and NGC 3623, with which it forms a non-interacting pair (see Appendix F, Fig. F.5). Soft X-ray emission extends over about 2' in a Northwest-Southeast direction, similar to the extent and orientation of the triple-radio source (Filho et al., 2004). No obvious nuclear X-ray source - hard or soft - was found on the available snapshot *Chandra* image (see Fig. C.13 and Ho et al., 2001). Panessa et al. (2006) noted that another source at 10" is present in this image with similar flux than that of the nucleus, which probably contaminates other data with worse spatial resolution (see also Georgantopoulos et al., 2002). They derive an upper limit on the *Chandra* (2-10) keV luminosity of $7.6 \times 10^{37}$ erg s$^{-1}$ with a fixed $\Gamma=1.8$. Previous analyses of *ROSAT*, *ASCA* and *BeppoSAX* data resulted in a moderately absorbed power-law component with $\Gamma \approx 2$-2.5 required to fit the spectra (Roberts and Warwick, 2000; Georgantopoulos et al., 2002; Dadina, 2007). The *XMM-Newton* data do not have enough count-rate for a spectral analysis to be perfomed. Our estimated luminosity assuming a power-law model with a fixed $\Gamma = 1.8$ and galactic absorption is $2.4 \times 10^{39}$ erg s$^{-1}$, much larger than the value by Panessa et al. (2006). Its AGN nature was assessed by Filho et al. (2000) based in the compact nuclear source which appears to have a variable flat radio-spectrum.

**NGC 3628 (UGC 6350, Arp 317c).** It makes up the Leo Triplet together with NGC 3623 and NGC 3627 (see Appendix F, Fig. F.5). The hard X-ray morphology provided by *Chandra* data shows an unresolved nuclear component that also appears in the Fe image (Fig. C.14). *Chandra* X-ray and ground-based optical H$\alpha$, arc-second resolution imaging is studied by Strickland et al. (2004), with the main aim of determining both spectral and spatial properties of the diffuse X-ray emission. They also show the total counts for the nuclear region (an extraction of 1 kpc radius around the dynamical center that, for this galaxy, corresponds to the central 20"), but no spectral fitting was attempted. Our morphological classification does not agree with that of Dudik et al. (2005), who have classified this galaxy as an object displaying no nuclear source according to its morphology in previous *Chandra* ACIS snapshot (1.8 ksec) data; this galaxy is taken as a LINER/transition object and an upper limit of $2.7 \times 10^{37}$ erg s$^{-1}$ (corrected to our adopted distance) is given for its (2-10) keV nuclear luminosity, which is consistent with our result. Note that high absorption is derived from the position of this galaxy in the color-color diagrams by GM+06. Here we report 41.6 ksec observation of *XMM-Newton*/pn data. The best fit model is a single power-law ($\Gamma = 1.6$) with $N_H = 4.6 \times 10^{21}$ cm$^{-2}$. The X-ray luminosity with *XMM-Newton* data is two orders of magnitude brighter than that from *Chandra* data. This can be explained because there are three point-like sources close to the diffuse nucleus, which may be contributing to the *XMM-Newton* aperture. Another explanation could be the high level of variability detected in this galaxy (Roberts et al., 2001).

**NGC 3690B (Arp 299, Mrk 171).** This galaxy is strongly interacting, in a probable merger, with IC 694 (see Appendix F, Fig. F.5). X-ray emission from *Chandra* data has plenty of features, with a hard unresolved source clearly detected in the nuclear position, which is also seen in the 6-7 keV band (Fig. C.15). The EPIC-pn *XMM-Newton* spatially resolved data have clearly demonstrated the existence of an AGN in NGC 3690, for which a strong 6.4 keV line is detected, and suggested that the nucleus of its companion IC 694 might also host an AGN,[10] since a strong 6.7 keV Fe-K$\alpha$ line is present (Ballo et al., 2004). *Chandra* and *XMM-Newton* data have been fitted to ME2PL and MEPL, respectively, with ($\Gamma = 3.5$) and kT=0.19 for *Chandra* and ($\Gamma = 1.84$) and kT=0.63 for *XMM-Newton*. The *XMM-Newton* X-ray luminosity results to be almost one order of magnitude brighter than that from *Chandra*. This can be understood due to the inclusion of the companion galaxy NGC 3690A in the *XMM-Newton* extraction. Condon and Broderick (1991) suggested an AGN nature of this source because of its compact flat radio-spectrum.

**NGC 3898 (UGC 6787).** This object is a member of the cluster of galaxies Abell 1377 (see Appendix F, Fig. F.5). There are not previous reported results on X-ray data for this object in the literature. Here we analyze the available *Chandra* data and find MEPL to be the best fit model ($\Gamma = 1.8$, kT=0.04 keV, NH1 = $1.4 \times 10^{22}$ cm$^{-2}$ and NH2 $< 6 \times 10^{21}$ cm$^{-2}$) to describe the X-ray spectral energy distribution. The X-ray morphology shows a point-like source only at soft energies, questioning its AGN nature. Thus, we have classified this object as Non-AGN like object. Ho et al. (1997) adopted the conservative assumption that broad H$\alpha$ is not present due to the ambiguity of its detection, but they claim that it would be highly desirable to verify this with data of higher S/N.

**NGC 3945 (UGC 6860).** This is a rather isolated object (see Appendix F, Fig. F.6). We have detected a nuclear unresolved source at hard energies with *Chandra* data (Fig. C.17), leading to an AGN-like classification. At softer energies (0.4-1 keV) it shows a ring like or arm like structure of diffuse extended emission. The spectral analysis of the nucleus gives as best fit model a single PL with $\Gamma = 2.6$ and column density consistent with the galactic value. There are not previous reported X-ray data in the literature for this object. Nagar et al. (2005) detect a compact continuum radiosource.

**NGC 3998 (UGC 6946).** Five galaxies (NGC 3990, NGC 3977, NGC 3972, NGC 3982 and UGC 6919) are seen within 250 kpc (see Appendix F, Fig. F.6), all but one at cz compatible with sharing the same physical association. The AGN nature of this galaxy was assed by Ho et al. (1997) based on the clear detection of a broad H$\alpha$ line, the detection of a variable radio core (Filho et al., 2002) and a 20% UV flux variation reported by Maoz et al. (2005). Ptak et al. (2004) published the analysis of the same 10 ks *XMM-Newton* data on this LINER galaxy. They fitted the X-ray spectrum with a simple-absorber power-law with $\Gamma=1.88$, $N_H = 3.3\times 10^{20}$ cm$^{-2}$ and obtained an observed flux F(2-10 keV) = $1.1\times 10^{-11}$ cm$^{-2}$ s$^{-1}$, already in agreement with previously published data from *BeppoSAX* (Pellegrini et al., 2000; Georgantopoulos et al., 2002; Dadina, 2007) and *ASCA* (Terashima et al., 2000, 2002). Our spectral fitting for *XMM-Newton* data results in a single PL with spectral index $\Gamma=1.87$ at $N_H = 2.0\times 10^{20}$ cm$^{-2}$. However, the fitting to the *Chandra* spectrum is improved when using a model with two power-laws and a thermal contribution. Both sets of data gives the same value for the X-ray luminosity. A *BeppoSAX* observation (Pellegrini et al., 2000) showed that the Fe K$\alpha$ line was not detected to an EW upper limit of 40 eV. Nevertheless, in the *ASCA* spectrum by Terashima et al. (2002) an Fe K$\alpha$ line is marginally detected at 6.4 keV.

---

[10] However, both galaxies are found in the comparative sample of starburst galaxies in Satyapal et al. (2004)

**NGC 4036 (UGC 7005).** It forms a wide pair (see Appendix F, Fig. F.6) together with NGC 4041 (cz=1234 km/s). The *Chandra* images show a point-like source within diffuse emission extending less than 5" (Fig. C.19). Also several knotty regions are present within 20" radius. We estimate $L_x(2-10$ keV$) = 1\times 10^{39}$ erg s$^{-1}$ by assuming a single PL model with ($\Gamma = 1.8$ fixed) and galactic absorption. Its AGN nature is also confirmed by the optical data, since Ho et al. (1997) reported a faint, broad H$\alpha$ line.

**NGC 4111 (UGC 7103).** At least four galaxies are found within 250 kpc (see Appendix F, Fig. F.6) with redshifts from 600 to 900 km/s. A hard nuclear point source has been detected for this galaxy (Fig. C.20). A previous X-ray spectral analysis was based in *ASCA* data by Terashima et al. (2000) (see also Terashima et al., 2002) who could not fit the spectrum with a single-component model, but instead they required a combination of a power-law together with a Raymond-Smith plasma, with $\Gamma=0.9$, kT=0.65 keV and $N_H = 1.4\times 10^{20}$ cm$^{-2}$. These parameters agree with those estimated from its position in the color-color diagrams by GM+06. We have fitted the *Chandra* spectrum with ME2PL ($\Gamma = 3.0$, kT=0.66 keV, NH1 = $4.7\times 10^{21}$ cm$^{-2}$ and NH2 = $3.8\times 10^{22}$ cm$^{-2}$) which leads to an estimated hard luminosity of $2.5\times 10^{40}$ erg s$^{-1}$, a factor of 3 brighter than Terashima et al.'s estimation. The agreement is remarkably good taking into account the different instruments used and the different models assumed for the spectral fitting.

**NGC 4125 (UGC 7118).** It forms a pair with NGC 4121, at 3.6 arcmin (see Appendix F, Fig. F.6) and less than 60 km/s in cz. Figure C.21 shows the presence of a nuclear hard point-like source. The best fit that Georgantopoulos et al. (2002) obtained for the central $2'$ *BeppoSAX* spectrum is provided by an absorbed power-law with $\Gamma=2.52$ and $N_H=3\times 10^{22}$cm$^{-2}$, that resulted in L(2-10 keV)=$0.68\times 10^{40}$ erg s$^{-1}$. Based on the same *Chandra* ACIS dataset, Satyapal et al. (2004) class this galaxy among those revealing a hard nuclear source embedded in soft diffuse emission. They estimate the luminosity by assuming an intrinsic power-law slope of 1.8, which results in L(2-10 keV)=$7.3\times 10^{38}$ erg s$^{-1}$, in very good agreement with the value estimated by GM+06. We provide here a new fit to the *Chandra* data, with a best-fit MEPL ($\Gamma = 2.32$ and kT=0.57 keV). The 35.3 ksec *XMM-Newton* data are reproduced by an unabsorbed MEPL ($\Gamma = 2.36$ and kT=0.54 keV). The much larger luminosity for *Chandra* data are difficult to understand but it may be attributed to the difference in the column density between both sets of data. In principle a larger luminosity should be expected for *XMM-Newton* data since in addition to the nucleus, an ULX to the NE is included in this extraction aperture.

**IRAS 12112+0305.** This merging system (see Appendix F, Fig. F.6) contains two separate nuclei with a pair of tidal tails (Scoville et al., 2000). Very faint extended X-ray emission has been detected in this Ultraluminous Infrared Galaxy with no evidence of unresolved hard nuclear emission. Franceschini et al. (2003) presented *XMM-Newton* first results for this galaxy and made a formal fitting to the spectrum with MEPL in spite of the low count-rate of the data. We have not tried any spectral fitting and estimated an X-ray luminosity L(2-10 keV)=$1.5\times 10^{41}$ erg s$^{-1}$, which is in very well agreement with Franceschini et al.'s value. Condon and Broderick (1991) detect a compact radio core with a flat continuum.

**NGC 4261 (UGC 7360, 3C 270).** NGC 4261 is the main galaxy in a group of 33 galaxies (Nolthenius, 1993) in the Virgo West cloud (see Appendix F, Fig. F.7). Noel-Storr et al. (2003) pointed out this galaxy to host a nuclear dust disk together with a twin radio jet morphology in the *VLA* and VLBI images. The nuclear hard band emission of this galaxy is clearly unresolved both in the (4.5-8.0* keV) and 6-7 keV bands (Fig. C.23). The various features seen at soft energies (Fig. C.23) were already shown by Donato et al. (2004), who analysed its *Chandra* ACIS data for the core component (core radius of 0.98"); they fit it with a PL+apec model with $\Gamma=1.09$, kT=0.60 keV, and a high column density $N_H=7.0\times 10^{22}$cm$^{-2}$, reported to be the largest intrinsic column density among the 25 radio galaxies in their study. These parameters agree with those obtained by Rinn et al. (2005) and Satyapal et al. (2005) for the same data. Zezas et al. (2005) published the analysis of the same 35ks *Chandra* ACIS-S observations we use here. They reported a point-like emission above 4.0 keV and the evidence for an X-ray jet component down to arc-second scales from the nucleus (barely visible in our Fig. C.23). A three-component model was given as the best fit for the X-ray spectrum of the nuclear 2": a heavily obscured flat power-law ($\Gamma=1.54$ and $N_H=8.4\times 10^{22}$cm$^{-2}$), a less absorbed steeper power-law ($\Gamma=2.25$ and $N_H<3.7\times 10^{20}$cm$^{-2}$), and a thermal component (kT=0.50 keV), which resulted in L(2-10 keV)=$10.8\times 10^{40}$ erg s$^{-1}$, in agreement with our results. They reported an equally good fit with a single power-law ($\Gamma=1.37$) seen through a partially covering absorber ($N_H=7.7\times 10^{22}$cm$^{-2}$, $f_{cov}$=0.92) plus a thermal component. GM+06 did not include this object in the subsample with spectral fits due to its complexity which gave as unexpected parameters with any of the five models we tested. We provide here an acceptable fit by using ME2PL.

Sambruna et al. (2003) published its nuclear EPIC-pn *XMM-Newton* spectrum (the central 10"), which was best-fitted with a two-component model with a power law ($\Gamma=1.4$) absorbed by a column density of $N_H\approx 4\times 10^{22}$cm$^{-2}$ plus a thermal component with kT$\approx$0.7 keV (in agreement with *Chandra* spectral results by Gliozzi et al. (2003) and Chiaberge et al. (2003)); an unresolved FeK emission line with EW $\approx$ 0.28 keV was detected at $\sim$7 keV. They also reported short-term flux variability from the nucleus (timescale of 3-5 ks), which they argued as being originated in the inner jet. We analyse 27 ksec *XMM-Newton* data finding that the bestfit is ME2PL ($\Gamma=2.38$, kT=0.62 keV, NH1 = $2.2\times 10^{21}$ cm$^{-2}$ and NH2 = $1.4\times 10^{23}$ cm$^{-2}$), which is a more complicated model than Sambruna et al. (2003) proposed, with a temperature compatible with theirs but a higher value for the spectral index. Our best fit to the *XMM-Newton* spectrum agrees with that obtained for *Chandra* data with consistent parameters. The reported luminosity is also consistent (Lx(2 − 10 keV) = 1.3 and $1.5 \times 10^{41}$ erg s$^{-1}$, respectively). No obvious signs of broad H$\alpha$ have been reported in optical spectroscopic data either from the ground (Ho et al., 1997) or from small-aperture (0.1") HST spectra (Ferrarese et al., 1996).

**NGC 4278 (M 98).** It is a member of a group together with NGC 4314, with 3 of its members visible in Appendix F, Fig. F.7. Ho et al. (2001) used snapshot (1.43 ksec) *Chandra* data to class the X-ray morphology of this galaxy as type I, i.e. dominated by a nuclear source (see also Dudik et al., 2005). The same dataset is used by Terashima and Wilson (2003) who, in addition to the unresolved nucleus, report the presence of a faint elongated structure about 50" long along PA ≈ 70 in the (0.5-2 keV) band. Their best spectral fitting corresponds to an unabsorbed power-law with Γ= 1.64, after the correction of a slight pile-up effect. They also reported flux variability from the comparison between *Chandra* and previous *ASCA* data. We fitted the same ≈ 100 ksec dataset with ME2PL (Γ = 2.37, kT=0.57 keV). The best fit model we derive for the 30.5 ksec *XMM-Newton* data 2PL (Γ = 2.3, NH1 = $6 \times 10^{20}$ cm$^{-2}$, NH2 = $7.2 \times 10^{22}$ cm$^{-2}$). An equally good fit is obtained with MEPL with Γ = 1.99, kT=0.65 keV, NH1 = $3.9 \times 10^{21}$ cm$^{-2}$, NH2 = $1.0 \times 10^{20}$ cm$^{-2}$ The estimated *XMM-Newton* hard X-ray luminosity is a factor of 10 brighter than that from *Chandra* data This difference can be in part explained as due to the contamination by a number of point-like sources around the AGN that can be seen on *Chandra* X-ray image (Fig. C.23).The AGN nature of this source was well known since earlies 80's when Jones (1984) observed this object using VLBI and found that the core flux density is 180 and 190 mJy at 18 and 6 cm, on size scales less than 5 and less than 1 mas, respectively. Later on, Ho et al. (1997) classified it as a type 1.9 LINER based on the broad component detected in the H$\alpha$ line. Nagar et al. (2005) have detected a radiojet at 2cm.

**NGC 4314 (UGC 7443).** It is a member of a group together with NGC 4278, wich can be seen in the SW corner in Appendix F, Fig. F.7. No nuclear source has been detected in the hard X-ray band images from *Chandra* data (Fig. C.25). Satyapal et al. (2004) used the same *Chandra* ACIS dataset to classify this galaxy among those exhibiting multiple, hard off-nuclear point sources of comparable brightness to the nuclear source. With an assumed power-law index of 1.8, the corresponding luminosity, corrected to our adopted distance, results in L(2-10 keV)=$8\times 10^{37}$ erg s$^{-1}$, in excellent agreement with the one that was estimated by GM+06. We report here the analysis of the 22 ksec *XMM-Newton* data. We have found that the X-ray spectral distribution can be best fitted by MEPL (Γ = 1.5, kT=0.24 keV, N$_H$ = $2.7 \times 10^{21}$ cm$^{-2}$).

**NGC 4321 (UGC 7450, M 100).** NGC 4321 is a well-studied, grand design spiral galaxy, located in the Virgo Cluster (see Appendix F, Fig. F.7). Based on snapshot imaging data, Ho and Ulvestad (2001) classed its *Chandra* X-ray morphology as type I, i.e., with a dominant nuclear source at variance with our results as an Non-AGN candidate. Roberts et al. (2001) analysed previous *ASCA* data and derived a best model for the spectral fitting with an absorbed power-law (N$_H$=$0.2\times 10^{22}$cm$^{-2}$ and Γ=1.9) and a MEKAL (kT=0.67 keV) components; these parameters are far from the range we have derived from the analysis of the available *XMM-Newton* and *Chandra* data, which agree perfectly well. No evidence for an AGN is found at radio frequencies. Filho et al. (2000, 2006) found a resolved extended source at 6cm.

**NGC 4374 (UGC 7494, 3C 272.1, M 84).** NGC 4374 is one of the brightest giant elliptical galaxies in the center of the Virgo cluster (see Appendix F, Fig. F.7). It shows strong radio emission and a two-sided jet emerging from its compact core (Xu et al., 2000). Bower et al. (1998) find a black hole mass of (0.9-2.6) $\times 10^9$ M$_\odot$ from velocities measured in the central emission-gas disk. An unresolved nuclear source is detected both in (4.5-8.0* keV) and 6-7 keV band images from *Chandra* data (Fig. C.27). Satyapal et al. (2004) have already described the X-ray morphology traced by the same *Chandra* ACIS dataset of this galaxy as revealing a hard nuclear source embedded in soft diffuse emission. The *Chandra* ACIS-S data are also analysed by Finoguenov and Jones (2001)[11]; they report a remarkable interaction of the radio lobes and the diffuse X-ray emission, and provide the parameters for a fit with an absorbed (N$_H$=$2.7\times 10^{21}$cm$^{-2}$) power law (Γ=2.3) and the corresponding L(0.5-10 keV)=$4.7\times 10^{39}$ erg s$^{-1}$, all in very good agreement with the ones we give in GM+06. These values somewhat differ from those obtained from the *ASCA* spectrum (Terashima et al., 2002), most probably due to the different spatial resolutions.

**NGC 4410A (UGC 7535, Mrk 1325).** NGC 4410A is a member of a compact group of galaxies (see Appendix F, Fig. F.7), being NGC 4410A associated with the *VLA* radio source (Hummel et al., 1986). Both (4.5-8* keV) and 6-7 keV band images show the unresolved nature of the nuclear source at these energies (Fig. C.28). The same ACIS-S *Chandra* observations we use for the NGC 4410 group are presented in Smith et al. (2003), who obtained an adequate fit for the spectrum of the inner 1" with a power law with Γ≈2 and a fixed N$_H$=$5\times 10^{20}$cm$^{-2}$, in agreement with a previous analysis of *ROSAT* X-ray observations (Tschke et al., 1999). The best fit model by GM+06 only needs the inclusion of a power law with Γ=1.75 (consistent with theirs within the errors). The reanalysis of *Chandra* data results in a best fit model MEPL with Γ = 1.77, kT=0.30 keV and N$_H$1=$5.1\times 10^{21}$cm$^{-2}$ and no absorption for the hard component. This fit agrees with the reported by GM+06. AGN nature of this object was obtained throught the detection of a rather broad H$\alpha$ component by Donahue et al. (2002).

**NGC 4438 (UGC 7574, Arp 120b).** This galaxy is in a pair with NGC 4435 (see Appendix F, Fig. F.8) in the Virgo cluster (Rauscher, 1995). Ho et al. (1997) reported this galaxy as that with the weakest broad H$\alpha$ nucleus. The results from 25 ks *Chandra* ACIS-S observations of this galaxy are also presented in Machacek et al. (2004), who suggest the presence of an AGN, based on the steep spectral index and the location of the hard emission at the center of the galaxy, in contrast to our morphological classification. The spectrum of the central 5" is claimed to be best-fitted by a combination of an absorbed power law (with N$_H$=$2.9\times 10^{22}$cm$^{-2}$ and a fixed Γ=2.0) and a MEKAL with kT=0.58 keV thermal component, providing L(2-10 keV)=$2.5\times 10^{39}$ erg s$^{-1}$. Nevertheless, Satyapal et al. (2005) class this galaxy as a non-AGN LINER based on the same ACIS *Chandra* dataset, in agreement with GM+06 classification. The new fit for *Chandra* data results in MEPL with kT=0.52 keV, Γ=1.91 and N$_H$=$3.7\times 10^{21}$cm$^{-2}$, wich is consistent, within the errors, with that provided by GM+06.

**NGC 4457 (UGC 7609).** This is a rather isolated object in the Virgo Cluster , with an unphysical companion within 250 kpc (UGC 7644 at cz=4222 km/s). Unresolved hard X-ray emission is seen in the nucleus on this galaxy (Fig. C.30). The spectral analysis of the same ACIS *Chandra* data by Satyapal et al. (2005) gives Γ=1.57, kT=0.69 keV, and no additional absorption, in very good agreement with GM+06 results. The best fit model reported here is also MEPL, with compatible parameters but with a larger column density, NH1 = $3.7 \times 10^{21}$ cm$^{-2}$.

---

[11] See also Kataoka and Stawarz (2005) for the analysis of the two extra-nuclear knots.

**NGC 4459 (UGC 7614).** With NGC 4468 at 8.5 arcmin and NGC 4474 at 13.5 arcmin to the E-NE, NGC 4477 and NGC 4479 are the two similar-sized galaxies to the SE within 250kpc (see Appendix F, Fig. F.8) close in redshift. GM+06 morphologically classified this galaxy as a Non-AGN candidate (see Fig. C.31), in agreement with Satyapal et al. (2005) who, also based on these ACIS *Chandra* data, gave no additional X-ray information on this object. A mass of $M_{BH} = 7 \times 10^7 M_\odot$ has been reported for its nuclear black hole (Tremaine et al., 2002, based on Space Telescope Imaging Spectrograph (STIS) measurements of ionized-gas disks by Sarzi et al. (2001)) .

**NGC 4486 (UGC 7654, Virgo A, Arp 152, 3C 274, M 87).** This well known giant elliptical galaxy located at the center of the Virgo cluster (see Appendix F, Fig. F.8), hosts a very strong central radio source and a synchrotron jet which is visible from radio to X-ray wavelengths (Marshall et al., 2002). Both the unresolved nuclear emission and the jet-like feature extending ∼15" to the W-NW, in the direction of the optical jet, are seen in Fig. C.32. Deep 500 ksec *Chandra* observations of this galaxy are shown in Forman et al. (2005), where the same salient features present in our Fig. C.32 can be seen, with X-ray jets clearly detected, but unfortunately no spectral analysis is made. Donato et al. (2004) analyse both *Chandra* and *XMM-Newton* data providing a radius for the core of 0.22". Dudik et al. (2005), based on 38 ksec *Chandra* observations, classed it among objects exhibiting a dominant hard nuclear point source and estimated its luminosity as L(2-10 keV)=$3.3 \times 10^{40}$ erg s$^{-1}$ with a fixed $\Gamma=1.8$ power law, in good agreement with the one estimated by GM+06 with the same dataset. Here we use a much longer exposure time dataset ($\approx$ 100 ksec) and improve the fitting by ME2PL ($\Gamma = 2.4$, kT= 0.82, NH1 = $1.0 \times 10^{21}$ cm$^{-2}$ and NH2 = $3.96 \times 10^{22}$ cm$^{-2}$), what results in a luminosity of $7.2 \times 10^{41}$ erg s$^{-1}$, 20 times lager than the value obtained by Forman et al. (2005). The spectrum extracted from a 19 ksec *XMM-Newton* observation suffers from strong pile-up effects, so the resulting parameters will not be discussed any further. Noel-Storr et al. (2003) found by using optical spectroscopic STIS data that a broad component is needed to fit the spectrum. A black hole of mass $M_{BH} \approx 2.6 \times 10^9$ M$_\odot$ is found by Lauer et al. (1992). The active nucleus in M87 emits a nonstellar continuum, which is found to vary in strength over time in UV (Perlman et al., 2003; Maoz et al., 2005), optical (Tsvetanov et al., 1998) and X-rays (Harris et al., 1997).

**NGC 4494 (UGC 7662).** This elliptical galaxy is located in the Coma I cloud (see Appendix F, Fig. F.8). Hard nuclear emission from *Chandra* data is point-like (Fig. C.33). The *XMM-Newton* EPIC spectrum extracted from a 45" region has been published by O'Sullivan and Ponman (2004). A MEPL model results in their best model for the spectral fitting, for which they get $\Gamma=1.5$ (consistent with GM+06 value) but for hydrogen column density fixed at the Galactic value ($N_H=1.56 \times 10^{20}$cm$^{-2}$) and kT=0.25 keV. Dudik et al. (2005) classed it as a hard nuclear point-dominated source and estimated L(2-10 keV)=$7.2 \times 10^{38}$ erg s$^{-1}$ with a fixed $\Gamma=1.8$ power law, about a factor of 6 fainter than the one GM+06 calculated with the spectral fitting. The new fitted model agrees with that reported in GM+06. We have also analised the 24.5 ksec *XMM-Newton* observation, obtaining a single PL as bestfit model with a spectral index consistent with *Chandra* spectral index and absorption consistent with galactic absorption.

**NGC 4552 (UGC 7760, M 89).** This Virgo elliptical galaxy (see Appendix F, Fig. F.8) has no detected broad band H$\alpha$ component (Cappellari et al., 1999); its nuclear source shows long-term variability at UV wavelengths (Cappellari et al., 1999; Maoz et al., 2005) and a radio jet with VLBI observations (Nagar et al., 2005). This galaxy shows an unresolved source in the hard X-rays band over an extended nebulosity with the peak of emission coincident with the galaxy center determined from 2MASS data (Fig. C.34). Xu et al. (2005) found from *Chandra* ACIS-S data that the central source is the brightest in the field and that it coincides with the optical/IR/radio center of the galaxy within 0.5". The X-ray-identified source is compact and variable on short time scales of 1 h. Their best-fitted model of the source is consistent with an absorbed power-law with spectral index $\Gamma=2.24$, in rather good agreement with the *ASCA* data reported by Colbert and Mushotzky (1999). The inferred luminosity in the 2-10 keV is $4 \times 10^{39}$ erg s$^{-1}$, consistent with our result ($2.6 \times 10^{39}$erg s$^{-1}$). Their main conclusion based on the variability, the spectral analysis, and multi-wavelength data is that the central source is more likely a low-luminosity AGN than contribution from LMXBs (Low Mass X-ray Binaries). GM+06 best-fit parameters are consistent with a model of a power law ($\Gamma=1.81$) plus a thermal RS (kT= 0.83 keV), in much better agreement with the results by Filho et al. (2004) on the analysis of *Chandra* archival data, with $\Gamma=1.51$ and kT=0.95. The new *Chandra* spectral analysis is consistent with those obtained before. We also report 32 ksec *XMM-Newton* observation. The bestfit model and parameters agree with these obtained with *Chandra* data. However the X-ray luminosity is a factor of 10 larger than *Chandra* data (Lx(2 − 10 keV) = $1 \times 10^{40}$ erg s$^{-1}$). This difference can be easily explained by the contribution of point-like sources within the *XMM-Newton* extraction region that can be seen in the hard X-ray image (see Fig. C.34).

**NGC 4589 (UGC 7797).** It is part of a small group (Wiklind et al. 1995), from wich NGC 4648 and NGC 4572 are visible in Appendix F, Fig. F.9. Roberts et al. (1991) already reported it as a faint X-ray source. We present here the new 54 ksec *Chandra* data for this galaxy, which allow to derive a luminosity Lx(2 − 10 keV) = $7.9 \times 10^{38}$ erg s$^{-1}$). We cannot perform a proper spectral fitting due to the low count rate. The X-ray morphology shows a diffuse emission without any point-like source at hard energies (> 2keV). However a radio jet has been reported by Nagar et al. (2005), pointing to its AGN nature.

**NGC 4579 (UGC 7796, M 58).** Another galaxy (NGC 4564) lies in the field within 250 kpc (see Appendix F, Fig. F.9, SW corner), which is close in redshift (cz=1142 km/s). NGC 4579 shows a compact nuclear source sitting in a diffuse halo (Fig. C.36), as already reported by Ho and Ulvestad (2001). Eracleous et al. (2002) fitted the compact unresolved central source detected in *Chandra* X-ray data, coincident with the broad-line region detected in UV by Barth et al. (2002), with a simple power-law spectra with $\Gamma=1.88$, which gives an estimated luminosity of $1.7 \times 10^{41}$erg s$^{-1}$. Dewangan et al. (2004) presented *XMM-Newton* data to search for the presence of an FeK$\alpha$ line. The best-fit spectrum is rather complex: an absorbed power-law with $\Gamma=1.77$ plus a narrow Gaussian at 6.4 keV and a broad Gaussian at 6.79 keV with FWHM ∼ 20.000 km s$^{-1}$. This broad component is interpreted as arising from the inner accretion disk. The estimated luminosity amounts to $1.2 \times 10^{40}$erg s$^{-1}$, lower than both Eracleous's estimation and GM+06 ($1.4 \times 10^{41}$erg s$^{-1}$). We have made a new analysis on both 30 ksec *Chandra* and 19.6 ksec *XMM-Newton* data. The spectrum extracted from *XMM-Newton* observations suffers from strong pile-up effects, so the resulting parameters will not be discussed any further. The fitting to *Chandra* data gives MEPL with $\Gamma = 1.58$, kT=0.20 keV and column densities for the soft and hard frequencies $N_H = 5 \times 10^{21}$ cm$^{-2}$. The derived X-ray luminosity is $\sim 10^{41}$ erg s$^{-1}$, consistent with those by Barth et al. (2002) and GM+06. The AGN nature of this LINER is confirmed by the detection of broad wings in the H$\alpha$ line Stauffer (1982); Keel (1983); Filippenko and Sargent (1985); Ho et al. (1997) along with broad lines in the UV (Maoz et al., 1998), large UV variability (Barth et al., 1996; Maoz et al., 1998, 2005) and a flat-spectrum radio core (Hummel et al., 1987; Nagar et al., 2005).

**NGC 4596 (UGC 7828).** It forms a wide pair (see Appendix F, Fig. F.9) together with NGC 4608 (cz=1864 km/s). This galaxy is very faint at X-ray frequencies, showing diffuse X-ray morphology, from a *Chandra* observation, in all the spectral bands (Fig. C.37). In fact, information on its spectral properties could not be obtained based on the present data due to the lack of sufficient counts in the hard band (4.5-8.0* keV). No previous X-ray data have been reported for this galaxy. A black hole mass of $7.8\times 10^7$ M$_\odot$ is calculated by Tremaine et al. (2002) based on Space Telescope Imaging Spectrograph (STIS) measurements of ionized-gas disks by Sarzi et al. (2001).

**NGC 4594 (M 104, Sombrero Galaxy).** The famous galaxy NGC 4594, with no evidence of a similar size galaxy within 250 kpc (see Appendix F, Fig. F.9), was one of the earliest galaxies to show evidences for the possible presence of a central supermassive (up to $10^9$ M$_\odot$) black hole (Kormendy, 1988). Its nucleus shows large short-term variability in the UV (Maoz et al., 2005) and show a radio compact core Its X-ray morphology shows a compact unresolved nuclear source on top of a diffuse halo (Fig. C.38). Dudik et al. (2005) used the classification by Ho and Ulvestad (2001) based on snapshot *Chandra* observations (< 2 ksec), that classed it with the objects that exhibit a dominant hard nuclear point source. We have made for the *Chandra* spectrum a new analysis and found a best fit model consistent with a single PL ($\Gamma = 1.56$ with $N_H = 1.9 \times 10^{21}$ cm$^{-2}$). This is in close agreement, within the errors, with the spectral fitting values provided by GM+06 ($\Gamma = 1.41$ with $N_H = 2.0 \times 10^{21}$ cm$^{-2}$). We also present here the 15.8 ksec *XMM-Newton* spectrum, which is better modelled with ME2PL ($\Gamma = 2.05$, $N_H = 6.4 \times 10^{21}$ cm$^{-2}$ and kT = 0.69 keV). Pellegrini et al. (2003) presented a spectral analysis based on 40 ksec *XMM-Newton* of the 7" central nuclear source, which they derive to be consistent with an absorbed power law with $\Gamma$=1.88 and a column density of $N_H$=1.8$\times 10^{21}$cm$^{-2}$. The value of our estimated 2-10 keV luminosity, $1.6\times 10^{40}$ erg s$^{-1}$, agrees fairly well with that reported by Pellegrini from *XMM-Newton*. The *XMM-Newton* computed hard X-ray luminosity is a factor of 1.7 higher than that obtained from *Chandra* data. The morphology of hard X-rays and the larger extraction aperture of *XMM-Newton* data cannot explain this difference.

**NGC 4636 (UGC 7878).** It belongs to a galaxy group Mahtessian (number 98j in 1998), but it shows no companion within 250 kpc (see Appendix F, Fig. F.9). This galaxy does not show emission at high energies (Fig. C.39), although a moderately strong broad component at H$\alpha$ was detected by Ho et al. (1997) in the starlight-subtracted optical spectrum. *Chandra* data do not have high enough quality to allow a proper fitting to the spectrum. The difference in our estimation of the X-ray luminosity ($1.77 \times 10^{39}$erg s$^{-1}$) and the value reported by Loewenstein et al. (2001) for the nucleus ($2 \times 10^{38}$erg s$^{-1}$) is due to the different apertures used, 13" and 3", respectively.

Xu et al. (2002) and O'Sullivan et al. (2005) presented *XMM-Newton* data for this source and find that it can be consistent with thermal plasma with a temperature kT between 0.53 and 0.71 keV. We present the 16.4 ksec *XMM-Newton* observation. The extracted nuclear spectrum is better modelled with MEPL ($\Gamma = 2.9$ and kT=0.54 keV) with no additional absorption. The arm-like structure reported by Jones et al. (2002) at soft energies can be produced by shocks driven by symmetric off-center outbursts, preventing the deposition of gas in the center. O'Sullivan et al. (2005) suggest that the X-ray morphology can be the result of a past AGN that is quiescent at the present. There is a two orders of magnitude difference in the luminosities obtained with *XMM-Newton* data compared to *Chandra* data, which can be attributed to the diffuse emission at hard X-rays shown in *Chandra* images (see Fig. C.39).

**NGC 4676A and B (Arp 242, The Mice Galaxy).** These two galaxies are the members of the well known interacting pair named "The Mice" (Arp 242, see Appendix F, Figs. F.9 and F.10). No high energy X-rays emission is detected (Fig. C.40) for component A, but it is present for component B (Fig. C.41). Read (2003) presented the first *Chandra* analysis of the Mice Galaxy and found a compact source in component B with a rather diffuse emission in A. Their spectral fitting in B is both consistent with MEKAL and power law models. We did not perform any fitting due to poor counting statistics. From the color-color diagrams and based on the same dataset, GM+06 concluded that the spectrum for component A is consistent with a power law with a spectral index in the range 0.8-1.2. GM+06 did not make any estimation for component B since the errors in the count-rate for the hardest band is greater than 80%. GM+06 estimated the luminosities for both components which agree remarkably well with the results by Read (2003), who explain the X-ray emission as produced by starbursts in both components. Our new estimation is also in agreement with these previously published values.

**NGC 4698 (UGC 7970).** This seems to be a rather isolated object, since no companion is visible within 250 kpc (see Appendix F, Fig. F.10). This galaxy shows a very faint, high-energy X-ray emission from its central region. The largest extension is found at intermediate energies, between 1 and 4 keV (Fig. C.42). Georgantopoulos and Zezas (2003) made a careful analysis of the *Chandra* data on this source and found that the X-ray nuclear position coincides with the faint radio source reported by Ho and Ulvestad (2001). They find that the best-fit model consists of an absorbed power law with $\Gamma$=2.18 and column density of $N_H$=5 $\times 10^{20}$cm$^{-2}$, which gives a a nuclear luminosity of $10^{39}$erg s$^{-1}$. GM+06 found, from the color-color diagrams obtained from the same *Chandra* data, that they may be well reproduced by a combined model with a power law with $\Gamma$=[1.2-1.6] and a thermal component with kT=[0.7-0.8] keV, what results in a luminosity fainter by a factor of two than that estimated by Georgantopoulos and Zezas (2003). Cappi et al. (2006) fit its *XMM-Newton* spectrum with a single power law model with $\Gamma$=2.0 and get L(2-10 keV)=1.6$\times 10^{39}$erg s$^{-1}$, a factor of 3 brighter than our determination from *Chandra* data. Our *XMM-Newton* (2-10 keV) luminosity is consistent with the result reported by Cappi et al. (2006). The discrepancies between both measurements can be explained because of the off-nuclear point sources located within the *XMM-Newton* extraction region. No trace of broad H$\alpha$ is visible in the relatively high S/N spectrum presented by Ho et al. (1997).

**NGC 4696 (Abell 3526).** NGC 4696 is the brightest member of the rich Centaurus Cluster, Abell 3526 (see Appendix F, Fig.

F.10). This galaxy is rather diffuse at high X-ray energies, having a clear nuclear halo morphology at soft energies (Fig. C.43). In fact, Satyapal et al. (2004) classed it as an object that reveals a hard nuclear point source embedded in soft diffuse emission. Taylor et al. (2006) used 196.6 ksec *Chandra* data, extracted the nuclear source with a 0.9" aperture, and obtained the best-fit model by a MEKAL thermal plasma with kT=0.75 keV and abundance 0.22 solar. Rinn et al. (2005) fit its *XMM-Newton* spectrum with a thermal model with kT=0.7 keV but with a matellicity 1.2 times solar. At variance with them, GM+06 best-fit model was a power law but with a rather high and unrealistic spectral index of 4.26. In spite of this difference, the estimated luminosities are within a factor of 2 ($6 \times 10^{39}$ erg s$^{-1}$ and $1.2 \times 10^{40}$ erg s$^{-1}$ for our analysis and Taylor's, respectively). GM+06 classified this source as a good candidate for a Starburst due to the absence of a nuclear-unresolved source at hard energies (Fig. C.43). Nevertheless, the VLBA data reported by Taylor and collaborators reveal a weak nucleus and a broad, one-sided jet extending over 25 pc, suggesting an AGN nature of this peculiar source. We have reanalysed the same *Chandra* data than in GM+06 and obtained that the best-fit model is MEPL ($\Gamma = 3.1$ and kT=0.67 keV) without additional absorption. The X-ray luminosity is now $1 \times 10^{40}$ erg s$^{-1}$. The available *XMM-Newton* data (40 ksec) produce a spectrum with strong pile-up effects, so it will not be used any further.

**NGC 4736 (UGC 7996, M 94).** This is a rather isolated object in terms of similar size galaxies within 250 kpc projected distance (see Appendix F, Fig. F.10). An unresolved nuclear source was reported for this galaxy using 0.15 arcsec resolution *VLA* data at 2cm (Nagar et al., 2005). Maoz et al. (2005) reported a factor of 2.5 long-term variability at UV. This galaxy shows a large number of unresolved compact sources in the few central arcseconds, which makes the extraction of the true nuclear source rather difficult (Fig. C.44). The high abundance of extranuclear sources can be due to the blue knots and HII regions located in an external ring (Roberts et al., 2001; Maoz et al., 2005)). With the same *Chandra* dataset we have used for this galaxy, Eracleous et al. (2002) identified 3 X-ray sources in the nuclear region, all of them showing hard spectra with power law indices ranging from 1.13, for the brightest one, to 1.8 for X-3, and luminosities in the 2-10 keV band between $4 \times 10^{38}$ erg s$^{-1}$ and $9.1 \times 10^{39}$ erg s$^{-1}$. We assign the source X-2 by Eracleous to be the nucleus of the galaxy since it coincides with the 2MASS near-IR nucleus within 0.82". Eracleous et al. (2002) stressed the complications of defining an AGN or SB character to this source, suggesting that even if the brightest source is associated with an AGN it will only contribute 20% to the energy balance in X-rays. The radio monitoring observations made by Krding et al. (2005) with the VLBI found a double structure, with the radio position N4736-b coinciding with our X-ray nucleus. From this double structure the brightest knot N4736-b also appears to be variable, pointing to an AGN nature for this low luminosity AGN. Our new fit to *Chandra* data agrees with that reported in GM+06 (MEPL). We report also our analysis on 16.8 ksec *XMM-Newton* data. The spectrum is modelized by a ME2PL. An order of magnitude difference is found between the luminosities obtained from *Chandra* and *XMM-Newton* data as expected due to the inclusion of all the X-ray sources mentioned by Eracleous et al. (2002) in the *XMM-Newton* extraction aperture. *ASCA* and *ROSAT* results are given by Roberts et al. (1999) and Roberts et al. (2001), where they reported a marginal detection of an ionized Fe K emission line. Terashima et al. (2002) found a possible hint of Fe K emission in the *ASCA* spectrum. In our *XMM-Newton* spectrum there is only marginal evidence for such an emission line (see Appendices E and E).

**NGC 5005 (UGC 8256).** This is a rather isolated object with no similar size galaxies within 250 kpc projected distance (see Appendix F, Fig. F.10), but two galaxies (NGC 5002 at cz=1091 km/s and NGC 5014 at cz=1126 km/s) are just out of the 250 kpc box. A broad H$\alpha$ component was found in its optical spectrum Terashima et al. (2002); Ho et al. (1997). Its *ASCA* X-ray spectrum (Terashima et al., 2002) was fitted with an absorbed ($N_H=0.1 \times 10^{22}$ cm$^{-2}$) power-law with $\Gamma = 0.97$, and a thermal (Raymond-Smith) component with kT=0.76 keV. They also reported a factor about 2 variability for this source. Dudik et al. (2005) classed its *Chandra* X-ray morphology as type III, i.e., a nuclear source embedded in diffuse emission; their spectral fitting provided $N_H=1.1 \times 10^{20}$ cm$^{-2}$, $\Gamma=1.9$ and kT=0.9 keV. Guainazzi et al. (2005b) analysed both *Chandra* and *XMM-Newton* data, to disclaim the *Compton-thick* nature of this source (Risaliti et al., 1999), deriving $N_H = 3 \times 10^{22}$ cm$^{-2}$ and $\Gamma = 1.6$. Gallo et al. (2006) obtained, with the same *XMM-Newton* dataset, $N_H < 1.4 \times 10^{20}$ cm$^{-2}$ and $\Gamma = 1.58$, and a total (2-10) keV luminosity of $\approx 10^{40}$ erg s$^{-1}$. We report our results on the available *XMM-Newton* observations (0.86 ksec). The best fit model is a MEPL model with ($\Gamma = 1.5$ and kT=0.27 keV) with a column density of $6 \times 10^{21}$ cm$^{-2}$, even lower that in Guainazzi et al. (2005a), and a luminosity of $Lx(2 - 10 \text{ keV}) = 2 \times 10^{40}$ erg s$^{-1}$.

**NGC 5055 (UGC 8334, M 63).** Only a dwarf spiral, namely UGC 8313, at very similar redshift (cz=593 km/s) appears close to this galaxy (see Appendix F, Fig. F.10). Its shows a clearly unresolved nuclear source coincident with the 2MASS position for the nucleus (Fig. C.46). No previous study of *Chandra* data has been reported. The only data available were *ROSAT* PSPC and HRI observations (Read et al., 1997; Roberts and Warwick, 2000) that pointed to the nucleated nature of this source within their low spatial resolution (10" at best). In the course of an investigation of ULXs over a sample of 313 nearby galaxies, Liu and Bregman (2005) found 10 ULX in this galaxy, one of which is close to the nucleus with a luminosity variation from 0.96 and $1.59 \times 10^{39}$ erg s$^{-1}$ in 1.6 days. The new *Chandra* data can be fitted with a PL model ($\Gamma = 2.3$), with a very low luminosity (Lx = $6 \times 10^{37}$ erg s$^{-1}$). This seems to be consistent with the spatially resolved UV source detected by (Maoz et al., 1998, 2005), which they reported as an extended, non-varying source, who suggested a young star origin to the observed emission. The 0.2" resolution observations by Nagar et al. (2000) give an upper limit of 1.1 mJy to any small-scale radio emission at 15 GHz. Ho et al. (1997) classified it as a type 1.9 LINER based in the detection of a broad H$\alpha$ component.

**Mrk 266NE (NGC 5256, UGC 8632, IZw 67).** Mrk 266 is a merging system (see Appendix F, Fig. F.11) with two nuclei separated by 10" (Hutchings and Neff, 1988; Wang et al., 1997): a Seyfert 2 nucleus to the southwest and a LINER nucleus to the northeast. Here we pay attention to the LINER nucleus. Its X-ray morphology shows the double structure of these merging, luminous infrared system with the northeast nucleus brighter than the southwestern one. Also the southwest nucleus shows hard emission being more diffuse (Fig. C.47). Our nuclear morphology agrees with that reported by Satyapal et al. (2004). Here we find the best fit for the *Chandra* spectrum to be ME2PL ($N_H = 2.2 \times 10^{23}$ cm$^{-2}$, $\Gamma = 1.34$ and kT=0.83 keV). The resulting X-ray luminosity is $4.6 \times 10^{41}$ erg s$^{-1}$. We also report *XMM-Newton* data which seem to be consistent with the same model but with a steeper spectral index ($\Gamma = 2.7$) and an X-ray luminosity a bit larger but consistent with *Chandra* data. A clear FeK$\alpha$ has been detected with equivalent width of 276 eV.

**UGC 08696 (Mrk 273, IRAS 13428+5608).** Mrk 273 is one of the prototypical ultra-luminous galaxies, showing a very complex structure at optical frequencies with a double nucleus, with a projected separation of $\approx$ 1", and a long tidal tail, indicative of a merging system (see Appendix F, Fig. F.11). A compact flat spectrum radio source has been detected between 2 and 6 cm (Baan and Klockner, 2006). At high X-ray energies only the northern nucleus is detected (Fig. C.48), which is coincident with the compact radio source shown in VLBI observations (Cole et al., 1999; Carilli and Taylor, 2000). Based on *Chandra* ACIS imaging, Satyapal et al. (2004) classed this galaxy among those revealing a hard nuclear source embedded in soft diffuse emission. X-ray *Chandra* data have been previously analysed by (Xia et al., 2002), who carefully studied both the nucleus and the extended emission. They showed that the compact nucleus is well described by an absorbed power law ($N_H = 4.1 \times 10^{20}$ cm$^{-2}$, $\Gamma$=2.1, L(2-10 keV)=2.9$\times 10^{42}$ erg s$^{-1}$) plus a narrow FeK$\alpha$ line. The most remarkable result of this analysis is that the spectrum of the central 10" is consistent with 1.5$Z_\odot$ metallicity, whereas the extended halo seems to be consistent with a thermal plasma with 0.1 $Z_\odot$ metallicity. The results reported by Ptak et al. (2003) on the same *Chandra* dataset (the only available up to now) point out that most of the observed X-ray emission (95%) comes from the nucleus. GM+06 best-fit model agrees with those data within the errors ($\Gamma$=1.74, kT=0.75 keV, $N_H$=3.9 $\times 10^{20}$cm$^{-2}$ and L(2-10 keV)=1.5 $\times 10^{42}$ erg s$^{-1}$). Our new best fit to the *Chandra* data is consistent with a ME2PL model with a column density 10 times higher than previously reported and almost a factor of ten larger X-ray luminosity. In addition, we report the results for the available 18 ksec *XMM-Newton* observation obtaining the same bestfit model with consistent spectral parameters and luminosity. Using *XMM-Newton* data, Balestra et al. (2005) analysed the FeK$\alpha$ line and concluded that, alike the case of NGC 6240, the line is the result of the superposition of neutral FeK$\alpha$ and a blend of highly ionized lines of FeXXV and FeXXVI. We have measured an equivalent width of 266 eV for the FeK$\alpha$ line.

**CGCG 162-010 (Abell 1795, 4C 26.42).** This galaxy is the central cD galaxy of the cluster A1795 (see Appendix F, Fig. F.11), which hosts the powerful type I radio source 4C26.42. The X-ray morphology shows a rather diffuse emission at high energies and a very clear long filament at soft energies (Fig. C.49). A full description of the nature of this filament was made in Crawford et al. (2005), who attributed the observed structure to a large event of star formation induced by the interaction of the radio jet with the intra-cluster medium. Satyapal et al. (2004) classed this galaxy among those revealing a hard nuclear source embedded in soft diffuse emission, based on *Chandra* ACIS imaging. Nevertheless, Donato et al. (2004), investigating the nature of the X-ray central compact core in a sample of type I radio galaxies, classified this galaxy among sources without a detected compact core, in agreement with GM+06 classification. The X-ray spectroscopic analysis of GM+06 results in this object being one of the five most luminous in the sample, with a value for the luminosity in very good agreement with that estimated by Satyapal et al. (2004) for an intrinsic power-law slope of 1.8 for the same dataset. Here we report a new analysis on these 20 ksec *Chandra* data and 42.3 ksec *XMM-Newton* data. The spectrum extracted from *Chandra* data is better fitted with ME (kT=1.1 keV) whereas that from *XMM-Newton* data is better described by MEPL ($\Gamma$ = 2.1 and kT=3.3 keV). The hard X-ray luminosity estimated from *XMM-Newton* data appears to be three orders of magnitude brighter than the *Chandra* value. These large differences can be attributed to the contribution of an extranuclear hard X-ray component from the cooling flow of the galaxy cluster (Fabian, 1994). In fact, the difference vanishes for the spectrum extracted from *Chandra* data and 25" aperture.

**NGC 5363 (UGC 8847).** It makes a wide pair together with NGC 5364 (cz=1241), in a group with several smaller galaxies, as NGC 5356 and NGC 5360 (see Appendix F, Fig. F.11). We report here for the first time the analysis on archival 19.3 ksec *XMM-Newton* observations. The extracted spectrum is better fitted with ME2PL ($\Gamma$ = 2.14 and kT=0.61 keV). Evidences for its AGN nature can be found in the radio data reported by Nagar et al. (2005).

**IC 4395 (UGC 9141).** This galaxy is disturbed by a neighboring edge-on galaxy, UGC 9141 at cz=1102 km/s) (see Appendix F, Fig. F.11). The only previously published X-ray data on this galaxy correspond to the *XMM-Newton* RGS spectrum by Guainazzi and Bianchi (2007), in which neither of the studied emission lines have been detected. We report the first analysis of archival 18 ksec *XMM-Newton* observation, that results in MEPL ($\Gamma$ = 1.78 and kT=0.26 keV) with no additional absorption as the best fit.

**IRAS 14348-1447.** IRAS 14348-1447 is part of a merging galaxy pair (see Appendix F, Fig. F.11). Our *Chandra* image shows a diffuse morphology in the whole X-ray energy range, but does not allow any kind of spectral analysis. We estimate a (2-10 keV) X-ray luminosity of $1 \times 10^{41}$ erg s$^{-1}$ by assuming a single power-law with fixed spectral index to 1.8 and galactic absorption. Franceschini et al. (2003) analysed the same *XMM-Newton* dataset presented in this paper, deriving a two component model as the best fit, with a thermal component with kT$\approx$0.62 keV, and an absorbed power-law with $\Gamma \approx$ 2.2 and $N_H > 10^{21}$ cm$^{-2}$ accounting for a significant hard X-ray component. They described the (0.2-10) keV X-ray morphology at larger spatial scales (2 $\times$ 2 arcminutes), with a bow-like structure extending about 30" in the NS direction, together with another relatively bright blob at about 20" to the SE which lacks any optical counterpart. The spectrum we have extracted from *XMM-Newton* data is fitted with an unabsorbed ME with kT=3.67 keV; this value for the temperature is a rather extreme, but no other model is able to provide a good fit with physically reasonable parameters. The derived luminosity is 4.9$\times 10^{41}$erg s$^{-1}$, consistent with the value estimated with *Chandra* data.

**NGC 5746 (UGC 9499).** This galaxy is part of a very wide galaxy pair with NGC 5740 (cz=1572 km/s), at $\approx$ 18' (out of the plotted field of view in Appendix F, Fig. F.12). No previous X-ray data analysis have been reported. Here we make use of the archival *Chandra* observations, that indicate an X-ray morphology showing a clearly compact, unresolved nuclear source (Fig. C.53). We obtain that a single power-law model ($\Gamma$ = 1.28) with moderate obscuration ($N_H = 6 \times 10^{21}$ cm$^{-2}$) can explain the observed spectrum. The analysis reported by GM+06 with the same data, both the fitting and the position in the color-color diagrams, provided very similar results. Nagar et al. (2002) detected a compact radio source suggesting the AGN nature of the nuclear source in this galaxy.

**NGC 5813 (UGC 9655).** NGC 5813 belongs to the group of galaxies #50 in the catalog by de Vaucouleurs (1975), whith NGC 5846 being the brightest member of the group; the closest galaxy to NGC 5813 is NGC 5814 at 4.8 arcmin to the S-SE (see Appendix F, Fig. F.12), but it lies too far away to be a physical companion (cz=10581 km/s). The X-ray morphology is extremely diffuse, with very extended emission at softer energies and without any emission at hard energies. We present the analysis on the spectra extracted from 48.4 ksec *Chandra* and 28 ksec *XMM-Newton* observations. Both *Chandra* and *XMM-Newton* data are best fitted with a MEPL model. The fitting parameters are compatible, excepting for the hydrogen column density (consistent with zero for *Chandra* data and $2.6 \times 10^{21}$ cm$^{-2}$ for *XMM-Newton* data). This leads to a high discrepancy between *Chandra* and *XMM-Newton* luminosities ($6.3 \times 10^{38}$ erg s$^{-1}$ and $1 \times 10^{40}$ erg s$^{-1}$, respectivelly). This discrepancy can be explained as due to the inclusion in the *XMM-Newton* spectrum of diffuse emission coming from the core galaxy cluster group. In fact, the higher luminosity is recovered when the aperture used from extracting the spectrum with *Chandra* data is fixed to 25". radiofrequencies Nagar et al. (2005) found a well detected compact radiocore.

**NGC 5838 (UGC 9692, CGCG 020-057).** Also belonging to the NGC 5846 group, a number of small galaxies are seen within 250 kpc (see Appendix F, Fig. F.12). There is not previous reported X-ray data in the literature. We present the analysis on the available 13.6 ksec *Chandra* observations, for which unfortunatelly no spectral fitting can be made due to the low countrate. The estimated hard X-ray luminosity is $1.6 \times 10^{39}$ erg s$^{-1}$ assuming a power-law model with fixed spectral index to 1.8 and galactic absorption. The hard X-ray morphology appears extended and diffuse with a faint nuclear source. Filho et al. (2002) detected a slightly resolved 2.2 mJy source, and confirm its compactness from subarcsecond-resolution 8.4 GHz images. Based on previous data at radio frequencies, they also conclude that the nuclear radio source in NGC 5838 must have a flat radio spectrum.

**NGC 5846 (UGC 9706).** NGC 5846 is the brightest member of the G50 group in the catalog of de Vaucouleurs (1975). In Appendix F, Fig. F.12 the two galaxies closest to it are the two small ones to the W (NGC 5845, at cz=1458 km/s) and NGC 5839, at 1225 km/s); the barred spiral to the E is NGC 5850, at 2556 km/s, and hence it does not conform a close interacting pair. Based on *Chandra* data, Trinchieri and Goudfrooij (2002) revealed a complex X-ray morphology with no clear nuclear identification (see also Fig. C.56). They detected, however, a large amount of individual, compact sources in the luminosity range from 3 to $20 \times 10^{38}$ erg s$^{-1}$. Filho et al. (2004) reanalysed the data already presented in Trinchieri and Goudfrooij (2002) and reported a weak, hard (2-10 keV) nuclear source with $\Gamma$=2.29, which is compatible within the errors with the value we obtain from the spectral fitting. Satyapal et al. (2005) analysed the *Chandra* data of this galaxy that they classed within Non-AGN LINERs, fitting its spectrum with a single thermal model with kT=0.65 keV, exactly the same as in GM+06 for our single RS model. We have reanalysed the spectra extracted from *Chandra* and *XMM-Newton* data, that result to best-fitted by MEPL for *Chandra* and ME2PL for *XMM-Newton*. At radiofrequencies it appears as a clearly compact radio core with a flat continuum (Filho et al., 2000, 2006).

**NGC 5866 (UGC 9723).** With several galaxies in the field of view in Appendix in Fig. F.12, two of them are physically close to it, namely NGC 5666A (cz=585 km/s) and NGC 5826 (cz=823 km/s). It forms a wide physical group with NGC 5879 (at 80 arcmin and cz=929 km/s) and NGC 5907 (at 85 arcmin and cz=779 km/s). The data for this galaxy reveals a rather complex morphology at hard X-ray energies with an identifiable nuclear region and extended emission in the northwest direction (Fig. C.57). Previous X-ray data analysis by Pellegrini (1994) based on *ROSAT* PSPC observations, pointed out a high excess of soft X-ray emission in S0 galaxies. Filho et al. (2004) and Terashima and Wilson (2003) failed to detect any hard nuclear X-ray emission in the *Chandra* image of this galaxy, and Satyapal et al. (2005) classed it as a Non-AGN-LINER, which agrees with GM+06 morphological classification. We estimate a hard X-ray luminosity Lx(2 − 10 keV) = $2 \times 10^{38}$ erg s$^{-1}$. Multifrequency radio observations suggest it harbors a compact, flat-spectrum radio core Hummel (1980); Wrobel and Heeschen (1991); Nagar et al. (2005); Filho et al. (2000, 2004); Falcke et al. (2000).

**IZw 107 (Mkn 848, VV 705).** Mark 848S is a Luminous Infrared Galaxy (Goldader et al., 1997) belonging to a close pair (see Appendix F, Fig. F.12) of interacting galaxies (Armus et al., 1990). The *Chandra* X-ray imaging (see Fig. C.58) shows a diffuse source at the nuclear position and a point-like source to its North. The spectrum extracted from *XMM-Newton* data is better fitted with a single PL ($\Gamma$ = 2.3) without additional absorption. The reported X-ray Luminosity is $1.6 \times 10^{41}$ erg s$^{-1}$.

**NGC 6251 (UGC 10501).** Paired with NGC 6252 at 2.4 arcmin (cz=6428 km/s) (see Appendix F, Fig. F.13), this is a well-known radio galaxy hosting a giant radio jet (Birkinshaw and Worrall, 1993; Urry and Padovani, 1995; Sudou and Taniguchi, 2000). The high-energy X-ray morphology shows a well-defined unresolved nuclear source without any extended halo (Fig. C.59). Guainazzi et al. (2003) reported a full analysis of the nuclear energy source comparing *Chandra*, *BeppoSAX*, and *ASCA* data. They found that the spectrum can be modeled with a combination of a thermal plasma at kT=1.4 keV, plus a power law with $\Gamma$=1.76 and $N_H$=1.6× $10^{21}$cm$^{-2}$, but they do not find evidence for the broad FeK$\alpha$ claimed in previous *ASCA* observations. However, the high sensitivity of *XMM-Newton* leads Gliozzi et al. (2004) to suggest again that such a broad ($\sigma$= 0.6 keV) FeK$\alpha$ line at 6.4 keV with an EW=0.22 keV is really there. The presence of an accretion disk in addition to the jet were suggested for explaining the origin of the X-ray emission. Chiaberge et al. (2003) modelled the spectral energy distribution from $\Gamma$-ray to radio frequencies and found that it was consistent with a synchrotron self-compton model with an unexpected high resemblance to blazar-like objects. This model, together with the dispute over the existence of FeK$\alpha$, lead Evans et al. (2005) to favor the relativistic jet emission as the main component of the observed emission. We report here the analysis of the spectra extracted from 25.4 ksec *Chandra* and 41 ksec *XMM-Newton* data. *Chandra* results are consistent with our previous analysis (GM+06). *XMM-Newton* data is better reproduced by ME2PL. The hard X-ray luminosity calculated from *XMM-Newton* data is one order of magnitude brighter than that obtained from *Chandra* data, which cannot be interpreted as due to an aperture effect (Table 14).

**NGC 6240 (IC 4625, UGC 10592, 4C 02.44).** This is a very well-known ultraluminous infrared merger remnant (see Appendix F, Fig. F.13) with a strong nonthermal radio excess and two nuclei separated by ≈2". Carral et al. (1990) report a compact radio source at 15 GHz. Making use of the same dataset we analyse here, Komossa et al. (2003) discovered a binary AGN in the galaxy coincident with the optical nucleus. They appear as compact-unresolved at energies between 2.5-8 keV. With the same dataset we use here, Satyapal et al. (2004) classed it as an object that reveals a hard nuclear point source embedded in soft diffuse emission. The spectroscopic analysis shows a very hard radiation for both nuclei, with $\Gamma$=0.2 for the one to the South and 0.9 for the one to the Northeast. The FeK$\alpha$ emission line is present in both nuclei. Ptak et al. (2003) pointed out to the complexity of the nuclear spectrum of this galaxy and constructed a more complex model that, in addition to the standard MEKAL and power law components, also included a Gaussian fit for the FeK$\alpha$ and a Compton reflection component with different column densities. To give an idea of the complexity of the source, let us point out that Boller et al. (2003) best-modeled the FeK$\alpha$ line as resolved into 3 narrow lines: a neutral FeK$\alpha$ line at 6.4 keV, an ionized line at 6.7 keV, and a blend of higher ionized lines (FeXXVI and Fe K$\beta$ line) at 7.0 keV. For consistency with the statistical analysis, we modelled the continuum spectrum with the combination of a thermal plus a power law component, without taking the complexity of the FeK$\alpha$ line into account. High absorption was derived for this source from both the spectral fitting and the estimation from color-color diagrams by GM+06. We have fitted the spectra of this source, obtained from both *Chandra* and *XMM-Newton* data, with ME2PL. The hard X-ray luminosity is one order of magnitude brighter in the spectrum from *XMM-Newton* data compared to that from *Chandra* data. This cannot be totally explained considering that the two nuclear sources are included in the *XMM-Newton* aperture 14. A clear FeK$\alpha$ been detected in our data with equivalent width of 378 keV (see Table 11).

**IRAS 17208-0014.** This Ultraluminous Infrared source has an optical morphology characterised by a single nucleus surrounded by a disturbed disk (see Appendix F, Fig. F.13) containing several compact star clusters, with a single tail. Baan and Klockner (2006) detected a compact flat spectrum nuclear radio source. Its X-ray nuclear emission appears to be unresolved at high energies (Fig. C.61). Risaliti et al. (1999) analysed luminous IR galaxies in X-rays with *BeppoSAX* to investigate the 2-10 keV nature of their emission and classified this object as a star forming galaxy with quite a large X-ray luminosity (L(2-10 keV)=1$\times 10^{42}$ erg s$^{-1}$). Franceschini et al. (2003) reported their analysis on *XMM-Newton* data for a sample of 10 ULIRGs and found that for this galaxy the observations are equally consistent with a model of a thermal plasma with a temperature kT=0.75 keV plus a power law component with $\Gamma$=2.26 and N$_H$=1.1$\times 10^{22}$cm$^{-2}$, and a thermal component with a temperature kT=0.74 keV plus a cut-off power law component with $\Gamma$=1.30 and N$_H$=2.6$\times 10^{21}$cm$^{-2}$, leading in both cases to similar luminosities on the order of a few times $10^{41}$erg s$^{-1}$. Based on the lack of FeK$\alpha$ emission line and the close value between the SFR estimated through the far IR emission and the X-ray emission, they suggested that the X-ray emission had a starburst origin. GM+06 did not tried to fit the spectrum extracted from *Chandra* data due to low count rate; from the position in the color-color diagrams, this galaxy seemed to be consistent with high column density and a combined model with a power law index between 1.6 and 2.0 and a temperature in the range 0.6-0.8 keV. Ptak et al. (2003) analysed the same *Chandra* data on this object and found that the best fit to the global spectrum is provided by a combined power law ($\Gamma$=1.68) and thermal (kT=0.35 keV) with N$_H$=0.52$\times 10^{22}$cm$^{-2}$ model. The nuclear luminosity is estimated to be L(2-10 keV)=4.2$\times 10^{41}$ erg s$^{-1}$, a factor of 3 brighter than the value from GM+06. We report here the analysis of the spectra extracted from 14.6 ksec *Chandra* and 14 ksec *XMM-Newton* observations on this source. *Chandra* data are better explained with a PL with a spectral index of 1.6, while *XMM-Newton* data are better described by MEKAL (kT=0.64 keV). We report a luminosity of $1.6 \times 10^{41}$ erg s$^{-1}$ calculated from *Chandra* data, close to what it was reported before. *XMM-Newton* data result in a lower value ($1.7 \times 10^{40}$ erg s$^{-1}$). The low count rate of these observations donot allow to favour any of the results.

**NGC 6482 (UGC 11009).** This galaxy is the brightest member of a fossil group (see Appendix F, Fig. F.13). Based on different *Chandra* observations that those reported here, Khosroshahi et al. (2004) analysed the temperature profile of the group, but not for the individuals. *Chandra* data on this source shows no hard nuclear source (Fig. C.62) associated with the compact radio source detected by Goudfrooij et al. (1994). The spectral analysis shows that the data are consistent with a thermal plasma at kT=0.68 keV in GM+06. The *Chandra* analysis performed here is consistent with the values obtained before. We have also analysed 6.7 ksec *XMM-Newton* data finding that the best-fit is provided by MEKAL with a temperature of 0.7 keV (also consistent with that from *Chandra* data). The hard X-ray luminosity results to be almost one order of magnitude for *XMM-Newton* data. This discrepancy can be attributed to the extended emission around the nucleus. The nuclear spectrum is better-fitted by a single thermal component, maybe due to the contribution of the emission from the galaxy group. In fact, a very similar spectrum is recovered when using a 25" extraction with *Chandra* data.

**NGC 7130 (IC 5135, IRAS 21453-3511),** is a peculiar galaxy that has no close companions (see Appendix F, Fig. F.13), since the closest projected companion AM2145-351 is at z=0.1. It shows a well-defined nuclear source at high X-ray energies (Fig. C.63). Since most of the UV emission is spectrally characteristic of star formation (Thuan, 1984; Gonzalez-Delgado et al., 2004), Levenson et al. (2005) used the same *Chandra* dataset than we use in this paper; they tried to decompose the AGN and Starburst contributions and found that the AGN contribution manifested mainly at higher energies (> 3 keV). They found that the obscuration of the nucleus is *Compton-thick*, which prevents the detection of the intrinsic emission in the *Chandra* bandpass below 8 keV. The spectral fitting is not statistically acceptable for this source in GM+06 but now, with our refined method, we have that ME2PL shows an acceptable fit ($\Gamma$=2.7, kT=0.76 keV, NH2 = $8.6 \times 10^{23}$ cm$^{-2}$). A clear FeK$\alpha$ has been measured with equivalent width 382 eV.

**NGC 7285 (Arp 93).** NGC 7285 is a member of the close interacting pair Arp 93, together with NGC 7284 at 0.5 arcmin (see Appendix F, Fig. F.13) and cz=4681 km/s. No previous X-ray data have been reported. Here we present the 27.2 ksec *XMM-Newton* observations on this source. The spectral analysis gives MEPL as the best fit with: $\Gamma$=1.6, kT=0.13 keV, NH1 = $6.8 \times 10^{21}$ cm$^{-2}$ and NH2 = $8.7 \times 10^{21}$ cm$^{-2}$. An equally good fit is obtained with 2PL, with $\Gamma$=1.69, NH1 = 8.$\times 10^{20}$ cm$^{-2}$ and NH2 = $1.7 \times 10^{22}$ cm$^{-2}$. A clear FeK$\alpha$ has been measured with equivalent width 212 eV.

**NGC 7331 (UGC 12113).** This is a quite isolated object in terms of not having any large companion at similar redshift (see Appendix F, Fig. F.14) although sometimes it has been considered in interaction with a member of the Stephan Quintet (Dumke et al., 1995). Stockdale et al. (1998) and Roberts and Warwick (2000) used *ROSAT* data to point out the AGN nature of this galaxy. The hard X-ray image extracted from the only available *Chandra* dataset does not show any evidence of a nuclear source, being very diffuse at high energies (Fig. C.65). Note that Filho et al. (2004) described this galaxy as hosting a hard (2-10 keV) X-ray nucleus, but Satyapal et al. (2004) classed it as an object exhibiting multiple, hard off-nuclear point sources of comparable brightness to the nuclear source, based on the same data. The parameters estimated by GM+06 from its position in color-color diagrams are consistent with a spectral index 2-2.6 and a temperature 0.7 keV. The estimation of the luminosity by Satyapal et al. (2004) for an intrinsic power slope of 1.8 is in perfect agreement with GM+06. Gallo et al. (2006) presented the *XMM-Newton* data on this source and found that the spectrum is consistent with a thermal component at kT=0.49 keV plus a power law with Γ=1.79, giving a luminosity that is a factor of 10 larger than that in GM+06 and this work. There is only 49 counts in the 0.5-10.0 keV band of the spectrum extracted from 0.8 ksec *XMM-Newton* data. We estimate an X-ray luminosity with *XMM-Newton* data in good agreement with Gallo et al. (2006) results. The diference in luminosities between *Chandra* and *XMM-Newton* is attributed to the off-nuclear point-like sources seen in the *Chandra* hard energy band images (see Fig. C.65).

**IC 1459 (IC 5265).** IC 1459 is a giant elliptical in a loose group with several spiral galaxies, the most conspicuous in Appendix F, Fig. F.14 being IC 6269B (cz=2870 km/s) and IC 5264 (cz=1940 km/s). A variety of indicators suggesting a recent merger are present in this galaxy, as a nuclear dust lane (Sparks et al., 1985), an ionized gas disk and a number of shells (Forbes et al., 1994). At X-ray frequencies, this galaxy presents an unresolved nuclear source on top of a diffuse halo at high energies (Fig. C.66), in agreement with the classification by Satyapal et al. (2004). A compact radio core has been detected (Slee et al., 1994). Fabbiano et al. (2003), based on a different set of data, found that it shows a rather weak (L(2-10 keV)=8.0× $10^{40}$ erg s$^{-1}$) unabsorbed nuclear X-ray source with Γ=1.88 and a faint FeKα line at 6.4 keV. These characteristics correspond to a normal AGN radiating at sub-Eddington luminosities, at 3× $10^{-7}$ below the Eddington limit. They suggest that ADAF solutions can explain the X-ray spectrum, but these models failed to explain the high radio power of its compact source (Drinkwater et al., 1997). The fitting parameters from GM+06 are in remarkably good agreement with theirs (Γ=1.89, kT=0.30 keV and L(2-10 keV)=3.6× $10^{40}$ erg s$^{-1}$). We report here the results of the analysis of the nuclear spectra extracted from 53 ksec *Chandra* and 26.9 ksec *XMM-Newton* data. The former is better fitted by ME2PL with Γ = 2.17, kT = 0.61 keV and column densities NH1 = $2.0 \times 10^{21}$ cm$^{-2}$ and NH2 = $1.3 \times 10^{22}$ cm$^{-2}$. The spectrum extracted from *XMM-Newton* data shows a MEPL best-fit with spectral index and hydrogen column density NH1 consistent with that reported before. The difference in luminosities dissapears when comparing *Chandra* and *XMM-Newton* spectra obtained with the same aperture.

**NPM1G-12.0625 (Abell 2597).** The brightest galaxy in Abell 2597 cluster (see Appendix F, Fig. F.14). Sarazin et al. (1995) found a nuclear radiosource consisting of unresolved nuclear emission and two diffuse lobes. Previous X-ray data on this galaxy referred to the analysis of the extended emission in its parent cluster (Pointecouteau et al., 2005; Morris and Fabian, 2005), but less attention was paid to the nuclear emission. Satyapal et al. (2004) classed its X-ray morphology based on 40 ksecs *Chandra* data, as those of objects revealing a nuclear point source embeded in difuse emission. We report the analysis of the nuclear spectra extracted from 59 ksec *Chandra* and 89.6 ksec *XMM-Newton* data. *Chandra* data are better fitted by MEPL and *XMM-Newton* with ME2PL with a consistent value of the spectral index, although high. The reported temperature is much lower in the case of *Chandra* data (kT=0.31 keV) compared with *XMM-Newton* data (kT=2.7 keV). This higher value in *XMM-Newton* data is an aperture effect since there is a strong hard diffuse component related to the cluster emission (see Tables 14 and 15).

**NGC 7743 (UGC 12759).** No other similar-sized galaxy is seen within 250 kpc (see Appendix F, Fig. F.14). This LINER appears not to have a broad Hα component (Terashima et al., 2000). It is the only object in the sample by Terashima et al. (2002) with no need of a power-law component to fit its *ASCA* spectrum, what is interpreted as a possible *Compton-thick* nature for this object. A clear compact flat spectrum radio core has been detected by Ho and Ulvestad (2001). In fact, it appears as a *Compton-thick* candidate in the study by Panessa et al. (2006). Our *XMM-Newton* spectrum covers up to ≃5 keV, and is better fitted by MEPL with Γ = 3.16, kT=0.26 keV, NH1 = $3.5 \times 10^{21}$ cm$^{-2}$ and NH2 = $1.7 \times 10^{22}$ cm$^{-2}$. We stress that $\chi^2_r$ = 1.76, which is the smallest value we get, but the count number is at the low limit of our requirements and the spectral fit is therefore not reliable.